\documentclass[fleqn,usenatbib]{mnras}

\usepackage{newtxtext,newtxmath}
\usepackage[T1]{fontenc}
\usepackage{ae,aecompl}

\usepackage{graphicx}	% Including figure files
\usepackage{amsmath}	% Advanced maths commands
\usepackage{amssymb}	% Extra maths symbols

\usepackage[normalem]{ulem}
\usepackage{booktabs}   %for table top/mid/bottom rules

\newcommand{\kpc}   {\mbox{$\:{\rm kpc}$}}
\footnotesize{\tiny}
\newcommand{\kms} {km s$^{-1}$}

\newcommand{\FeH} {\mathrm{[Fe/H]}}
\newcommand{\alphaFe} {$\mathrm{[\alpha/Fe]}$}

\newcommand{\Gaia} {\textit{Gaia}}
\newcommand{\G}{{\mathrm{G}}}
\newcommand{\Gbp}{G_{\mathrm{BP}}}
\newcommand{\Grp}{G_{\mathrm{RP}}}

\newcommand{\Ntracks} {126}
\newcommand{\Nunique} {95}
\newcommand{\Nmultiple} {19}
\newcommand{\Npm} {61}
\newcommand{\NpmD} {45} %pm & distance gradient
\newcommand{\Ngc} {22}
\newcommand{\Nvrad} {7}
\newcommand{\Nvradpm} {5}
\newcommand{\galstreams} {\textit{galstreams}}
\newcommand{\astropy} {\textit{astropy}}

\def\neww#1{{\textcolor{black}{#1}}}

\title[galstreams]{galstreams: A Library of Milky Way Stellar Stream Footprints and Tracks}

\author[C. Mateu]{
Cecilia Mateu,$^{1}$\thanks{E-mail: cmateu@fcien.edu.uy}
\\
$^{1}$Departamento de Astronom\'ia, Facultad de Ciencias, Universidad de la Rep\'ublica, Igu\'a 4225, 14000, Montevideo, Uruguay \\
}

\date{Accepted XXX. Received YYY; in original form ZZZ}

\pubyear{2022}

\hypersetup{draft}

\begin{document}
\label{firstpage}
\pagerange{\pageref{firstpage}--\pageref{lastpage}}
\maketitle

\begin{abstract}
Nearly a hundred stellar streams have been found to date around the Milky Way and the number keeps growing at an ever faster pace. Here we present the \galstreams\ library, a compendium of angular position, distance, proper motion and radial velocity track data for nearly a hundred (\Nunique) Galactic stellar streams. The information published in the literature has been collated and homogenised in a consistent format and used to provide a set of features uniformly computed throughout the library: e.g. stream length, end points, mean pole, stream's coordinate frame, polygon footprint, and  pole and angular momentum tracks. We also use the information compiled to analyse the distribution of several observables across the library and to assess where the main deficiencies are found in the characterisation of individual stellar streams, as a resource for future follow-up efforts. The library is intended to facilitate keeping track of new discoveries and to encourage the use of automated methods to characterise and study the ensemble of known stellar streams by serving as a starting point. The \galstreams~library is publicly available as a Python package and served at the \href{https://github.com/cmateu/galstreams}{\galstreams\  GitHub repository}. 
\end{abstract}

\begin{keywords}
software: public release  -- Galaxy: structure -- Galaxy: halo -- Astronomical databases: catalogues
\end{keywords}

%%%%%%%%%%%%%%%%%%%%%%%%%%%%%%%%%%%%%%%%%%%%%%%%%%

%%%%%%%%%%%%%%%%% BODY OF PAPER %%%%%%%%%%%%%%%%%%

\section{Introduction}

The field of stellar streams is currently in a golden era. It has increasingly grown and all but exploded in the last decade \citep[see e.g.][]{GrillmairCarlin2016,Helmi2020}, thanks to deep wide-area photometric surveys \citep[SDSS, PS-1, DES][]{Grillmair2009,Bernard2016,Grillmair2017_south,Shipp2018} and, more recently, to the amazing possibilities opened by the all-sky astrometric information provided by the \Gaia\ mission since its Second Data Release \citep[DR2][]{GaiaCol_2018_DR2_survey,Malhan2018,Ibata2021}. The availability of widespread kinematic information is having a tremendous impact. From the observational standpoint, enabling a large number of new stellar stream discoveries (see Table~\ref{t:super_summary_table} for a complete list), revealing relationships between streams far apart \citep[e.g. Orphan/Chenab, ATLAS/Aliqa Uma][]{Koposov2019,Li2021},  linking known streams (Fimbulthul, Gj\"oll, Fj\"orm) to their globular cluster progenitors \citep[$\omega$~Cen, NGC~3201, M68][]{Ibata2019_OCen,Ibata2021,Palau2019,Palau2021}; while at the same time revealing intriguing features like gaps, "spurs", a "cocoon" and a "blob" in GD-1 some potentially due to dark matter subhalos \citep{PriceWhelanBonaca2018_gd1,Bonaca2019_gd1,Malhan2019,deBoer2020}, and (at first) unexpected features like the misalignment of velocities with stream tracks found first in the Orphan-Chenab stream \citep{Koposov2019} and later observed in several of the Dark Energy Survey streams \citep{Shipp2018}, now thought to be perturbations caused by a recent close passage of the Large Magellanic Cloud \citep{Erkal2019,Shipp2020}. At the same time, from the theoretical standpoint, it is now making it feasible to advance some of the astrophysical questions that have long motivated interest in stellar streams, like reconstructing the assembly history of the Milky Way \citep[e.g.][]{Naidu2020,Bonaca2021,Malhan2022}, inferring the shape and mass of its dark matter halo \citep{Malhan2019_potential,Reino2021,Vasiliev2021b,Cautun2020} and constraining properties of dark matter subhaloes \citep{Erkal2017,Bonaca2019_gd1,Bonaca2020,Banik2021,Malhan2021_dm,Gialluca2021}.

In the past six years alone, the number of new streams reported in the literature increased almost five-fold, going from $\sim$20 reported in \citet{GrillmairCarlin2016} to nearly a hundred compiled in this work (see Figure~\ref{f:full_lib_galactic_aitoff}). This rate of discovery is not yet showing signs of reaching saturation and is even likely to increase as deep photometry from the Legacy Survey of Space and Time \citep[LSST at the Vera Rubin Observatory,][]{Ivezic2019} will allow probing fainter structures and farther reaches of the Galactic halo. Such a rapid growth makes a compilation as challenging as it makes it necessary. Challenging, not only because of the fast pace, but because the growth of the field has been, and continues to be, entirely organic. Because of their very nature as extended structures, the reporting of stellar stream properties \neww{(position in the sky, angular width, kinematics, etc.)} in the literature has been -to put it mildly-  heterogeneous. Necessary, because systematic studies of Galactic stellar streams as a system  require a homogeneous compendium as a starting point.

In \citet{Mateu2018} we published \galstreams, a first compilation of tidal stream footprints implemented as a Python library. It was based on the compilation by \citet[][their Tables 4.1 and 4.2]{GrillmairCarlin2016} and expanded the information available from approximate RA$/$Dec or $l/b$ regions to extended footprints. Its initial aim was to keep track of known stellar streams and overdensities and to homogenise the available information so that it could be parsed in an automated way and would facilitate identifying whether a stream is indeed a new discovery or which streams are present in a given patch of the sky. In that version (v0.1), pre-\Gaia~DR2, the library contained only celestial positions and distance information to represent the area covered by a stream in the sky. Since then, many studies have used \Gaia~DR2 and EDR3 \citep{GaiaCol_2018_DR2_survey,GaiaCol2021_EDR3_survey} to find and characterise the proper motion signature of many stellar streams, complemented with radial velocity data from existing and new dedicated follow-ups for a several of them \citep[e.g.][]{Ibata2021,Li2022}. Hence, a much more rich compilation including kinematics is now possible for a large number of stellar streams. 

Our goal in this paper and with the new version (v1) of the \galstreams\ library is to collate the available information to provide \neww{a homogeneous framework with celestial, distance, proper motion and radial velocity \emph{track}}\footnote{\neww{\galstreams\ does not provide data for individual member stars for any stream.}} \neww{information available in the literature} for the tidal streams found in the Milky Way.
Such a compilation will make it easier to identify when a discovery is indeed a new structure and to find connections between structures scattered across the sky, and will be an essential starting point for future automated systematic searches for new members of known streams. It is also intended as a resource for the community to identify where efforts can be directed to collect missing or insufficient information for the less studied streams, and avoid duplicating efforts when planning new observations. 

The structure of the paper is as follows: In Sec.~\ref{s:the_library} we present an overview of the information compiled in the library and discuss improvements with respect to the previous version of \galstreams. In Sec.~\ref{s:stream_data_realizations} we present the general procedures used to implement the tracks for the streams found in the literature (Sec.~\ref{s:procedure}) and describe for each one the specific information used to implement it (Sec.~\ref{s:stream_tracks_implemented}). In Sec.~\ref{s:excluded} we discuss structures that are \emph{not} included in the library. In Sec.~\ref{s:multiple_tracks} we compare tracks for streams with multiple instances available and discuss the criteria used to decide the track selected as the default in each case.  In Sec.~\ref{s:global_props} we use the library to discuss properties of the system of stellar streams as a whole and conclude with a summary in Sec.~\ref{s:conclusions}.

\section{The galstreams library}\label{s:the_library}

\begin{figure*}
	\includegraphics[width=2\columnwidth]{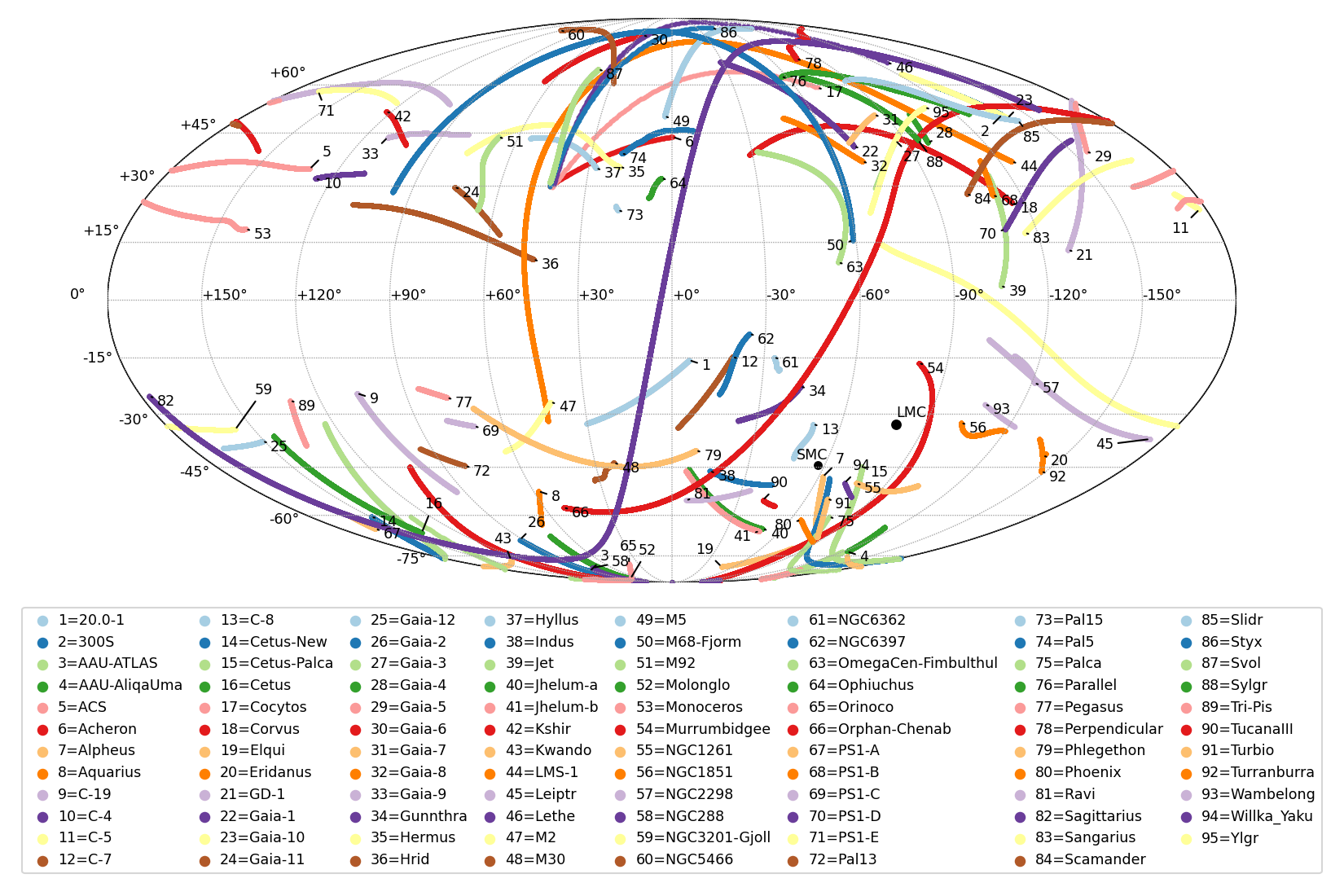}
    \caption{Mollweide projection map in Galactic coordinates of the celestial tracks for the \Nunique~stellar streams implemented in the library. The position of the Large and Small Magellanic Clouds (LMC, SMC) is also shown for reference.}\label{f:full_lib_galactic_aitoff}
\end{figure*} 

\begin{figure*}
	\includegraphics[width=1.9\columnwidth]{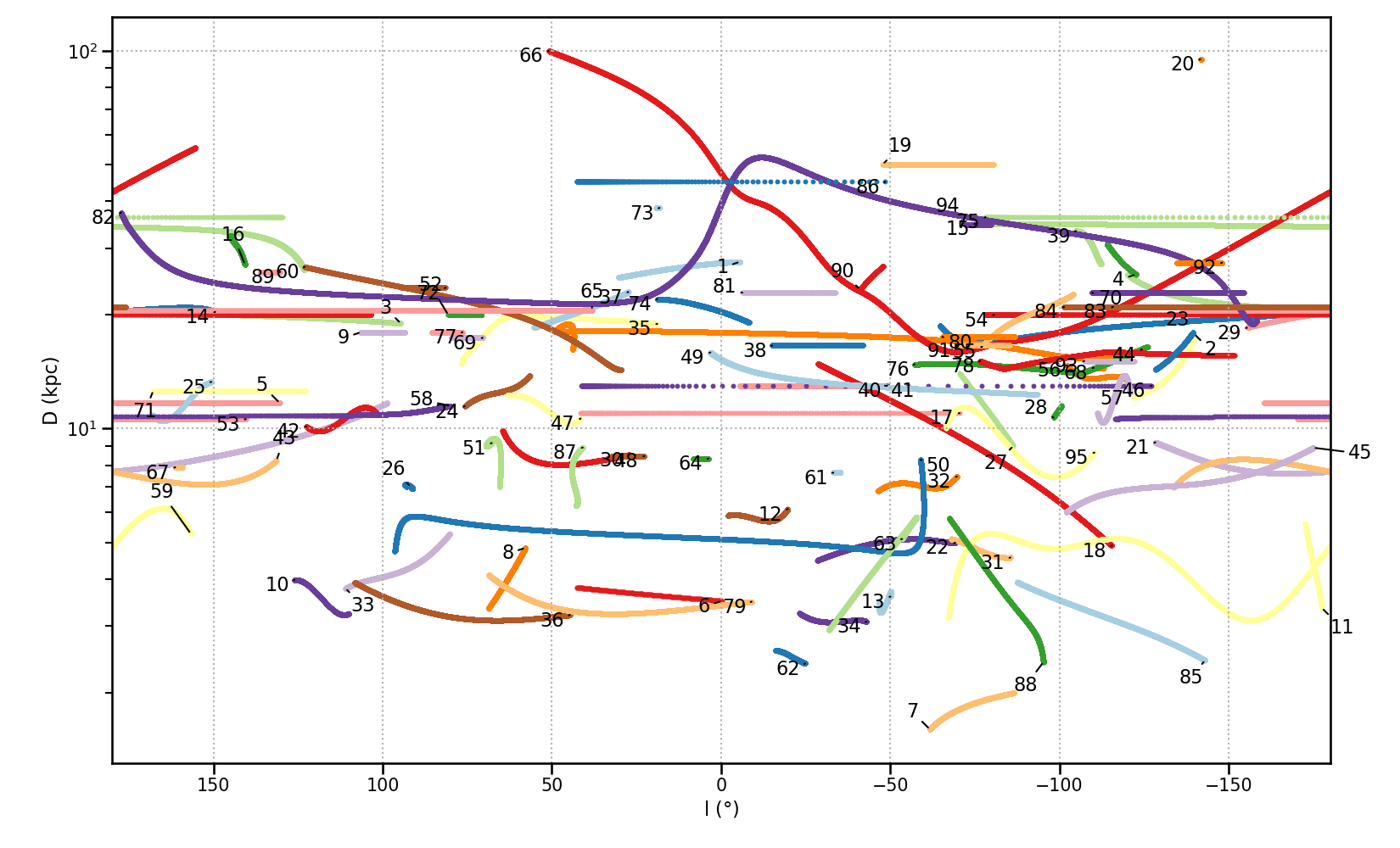}
    \caption{Distance tracks as a function of $l$ for the \Nunique~stellar streams in the library. Colour codes and IDs are the same as in Fig.~\ref{f:full_lib_galactic_aitoff}.} \label{f:full_lib_distance_l}
\end{figure*}

\begin{figure*}
	\includegraphics[width=2\columnwidth]{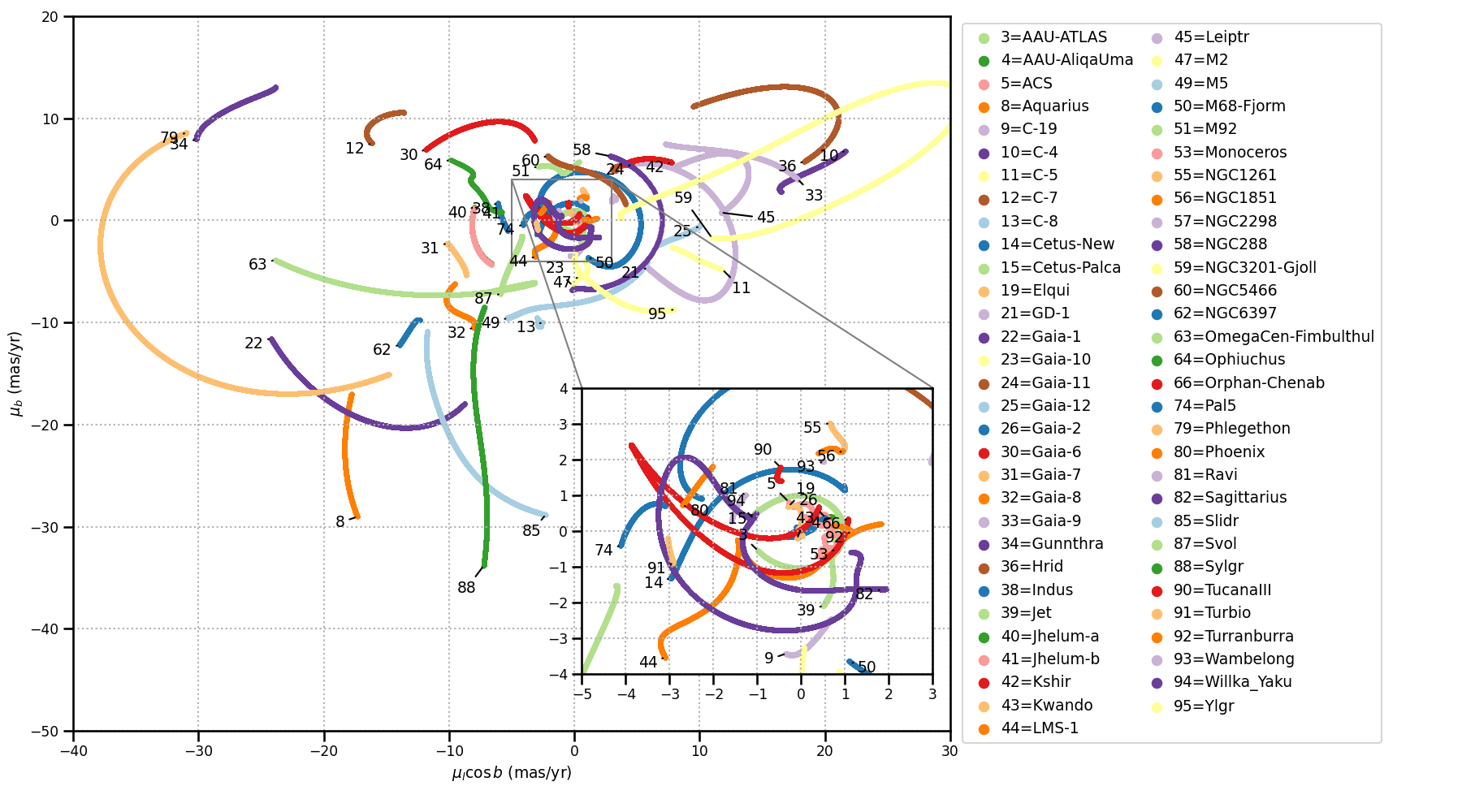}
    \caption{$\mu_b$ versus $\mu_l\cos{b}$ for the \Npm~stellar streams with proper motion tracks in the library.}\label{f:full_lib_pml_pmb}
\end{figure*} 

\begin{figure}
	\includegraphics[width=\columnwidth]{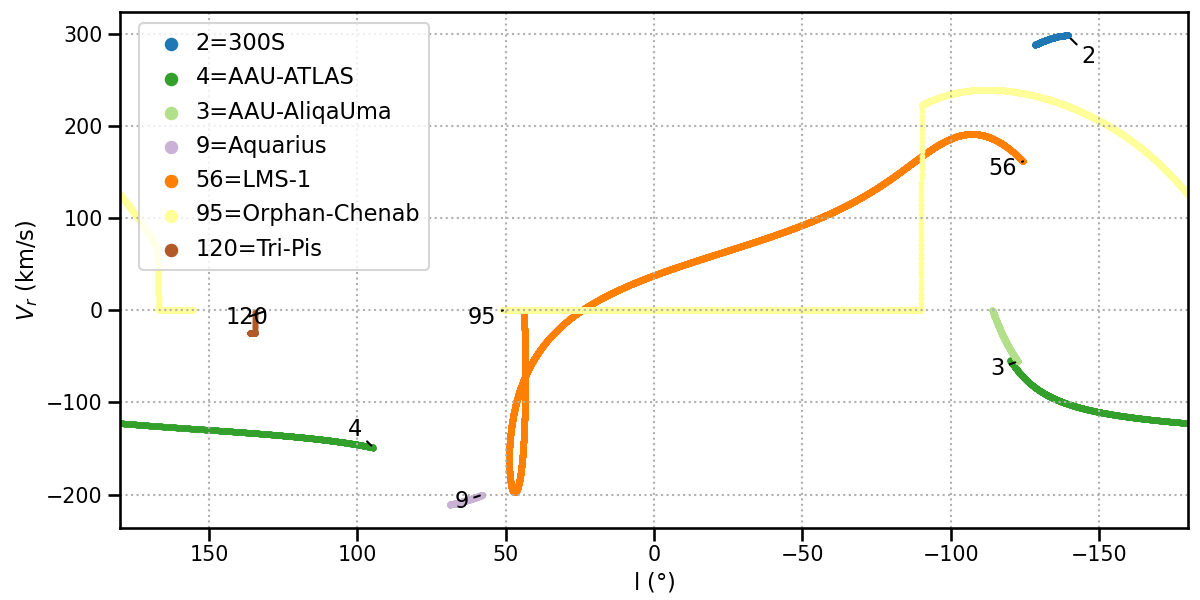}
    \caption{Radial velocity versus Galactic longitude for the \Nvrad~stellar streams with radial velocity tracks in the library.}\label{f:full_lib_vrad_l}
\end{figure} 

The \galstreams~library can be viewed as a compilation of stream's meta-data: it includes information needed to locate a given tidal stream in the sky (v0.1) and its signature in kinematics (v1). \neww{This information is provided as a \emph{track} in up to 6D, each dimension being a one dimensional function describing each component (e.g. distance or proper motion in a given direction) as a function of an angle in the sky. As will be described in Sec.~\ref{s:procedure}, tracks \emph{are not orbital fits}, they are \emph{empirical fits to observed data}, obtained either as interpolations from reported knots or reference points or as polynomial fits to members reported in the literature}.  

The library is implemented in Python, based on \astropy, and served at the \galstreams\ public repository on GitHub\footnote{Available at \url{https://github.com/cmateu/galstreams}}. Figure~\ref{f:full_lib_galactic_aitoff} shows a map of the celestial tracks for the \Nunique~unique stellar streams (i.e. no repetitions), out of the total of \Ntracks\ tracks implemented in the \galstreams\ library \neww{based on information published from different sources in the literature, in which multiple tracks are available for same stellar stream in several instances (see discussion in Sec.~\ref{s:multiple_tracks})}. \neww{Figure~\ref{f:full_lib_distance_l} shows the distance tracks as a function of Galactic longitude for the \Nunique~stellar streams in the library, Figure~\ref{f:full_lib_pml_pmb} the proper motion in Galactic latitude versus the proper motion in Galactic longitude for the \Npm\ streams with proper motion information in the library, and Figure~\ref{f:full_lib_vrad_l} the radial velocity as a function of Galactic longitude for the $\Nvrad$ tracks with this information available in the library. Table~1 summarises the information available for all the stream tracks implemented in the library following the procedure described in Sec.~\ref{s:procedure}.}

The main features of the library in its current version are:

\begin{itemize}
    \item Tracks \neww{in up to 6D}: sky, distance, proper motion and radial velocity tracks for each stream
    \item Stream's (heliocentric) coordinate frame realised as an \astropy\  reference frame
    \item Polygon Footprints
    \item End-points and mid-point celestial coordinates
    \item Pole (at mid point) and pole tracks in the heliocentric and Galactocentric frames for all tracks in the library
    \item Angular momentum track in the heliocentric reference frame at rest with respect to the Galactic centre \neww{(for tracks with proper motion data available)}
    \item Summary object with attributes for the full library: Uniformly reported end points and mid-point, heliocentric and Galactocentric mid-pole, track angular length, track and discovery references and an information flag denoting which of the 6D attributes (sky, distance, proper motions and radial velocity) are available in the track object 
\end{itemize}

Although track information is derived from individual stream members when available in the literature, \textbf{ \galstreams\ does not supply information for individual members}. The 6D tracks and polygon footprints provided should, however, provide a useful tool to conduct automated systematic searches for individual stream members.

\subsection{Improvements and changes with respect to galstreams v0.1}

In its original version (v0.1) the main piece of information provided by \galstreams\ was each stream's Footprint created as a random realisation with uniform area coverage in the sky, \neww{including either mean distance information or distance as a function of position along the footprint, for the relatively few streams where this information was available}. An extension was later added to calculate end points and poles from these random realisations, as a means of providing a quick-and-dirty way of realising each stream's coordinate frame, mainly for plotting purposes. 
%This is why in \citet{Riley2020}, who used and complemented the library to analyse stream's poles, the authors had to recompute end-points for several of the streams reported in \galstreams, so they would exactly match stream's tracks in every case. 

In this new version, the main feature of the library is to provide each stream's track in up to 6D: the sky, distance, proper motions and radial velocity tracks are reported for streams with data available in the literature. Because of the potential use in qualitative plots, a method to produce random realisations of the tracks is provided.

The representation as a celestial track, rather than a randomly populated footprint, is more suited to the nature of the streams we want to represent and allows for a better record keeping of useful stream attributes and their dependencies, such as correlations of distance, radial velocity or proper motion with celestial coordinates. In future versions, other attributes could also be supported in this representation, e.g. width and error tracks or other attributes relating to the stellar population like age and metallicity. New Polygon Footprints, also provided, are constructed based on the tracks. These provide an easy way to select a catalogue's points inside a stream's footprint and to create a similar off-stream selection area for background subtraction. 

The stream's track is stored as an \astropy\  SkyCoord object and, as of this version, distance information is required for a stream to be included in the library. This way 
the track has all coordinate transformation capabilities available in \astropy\  and making the distance a mandatory attribute allows for the track to be easily transformed into any non-heliocentric reference frame, in particular, the Local or Galactocentric Standards of Rest (LSR and GSR, respectively). Although a distance estimate was not mandatory in the previous version, in practice all streams in \galstreams~v0.1 had it, except for the WG1-4 streams from \citet{Agnello2017}. Since there is no further available information for these candidates (proper motions, radial velocity or any information about the stellar population), we have decided not to include WG1-4 in the library and to make a distance estimate (however rough) a mandatory attribute. 

Up to the previous version we had also included footprint realisations for the cloud-like features compiled in Table 4.2 of \citet{GrillmairCarlin2016}: Tri-And I and II, Hercules-Aquila and the Virgo and Pisces Overdensities (see also Sec.~\ref{s:excluded}, about the Virgo Stellar Stream, VSS). We are aiming now at providing more descriptive information that is specific to streams and not meaningful for clouds, so we have decided not to include these or other diffuse structures in \galstreams. The original motivation to include them was for the user to be aware of any known overdensities in a given region of the sky. The discovery of \emph{localised} overdensities does not seem to have proliferated since the publication of \Gaia~DR2; in contrast, major accretion events have been identified, such as Gaia-Enceladus-Sausage or Gaia-Sausage Merger  \citep{Belokurov2018_sausage,Helmi2018_enceladus}, Thamnos  \citep{Koppelman2019}, Sequoia \citep{Myeong2019}; Arjuna, Sequoia and I'itoi \citep{Naidu2020}; Heracles \citep{Horta2021} and Pontus \citep{Malhan2022_Pontus}. These features are not localised, spanning vast regions (if not all) of the celestial sphere. Hence, we have decided to keep the library focused only on stream-like overdensities and not to include any of these structures (see Sec.~\ref{s:excluded} for more details on excluded structures). 

The library has been updated to include newly discovered streams (2020--2022), as well as streams published previously in the literature but missing in the first version (e.g. 300S, Aquarius, Parallel and Perpendicular). To the best of our knowledge, \galstreams\ is up to date as of mid-March 2022 and will be kept updated and documented in the GitHub repository, following the format and procedures described here. 

\section{Streams' data and track realisations}\label{s:stream_data_realizations}

\subsection{Procedure}\label{s:procedure}

The data for stellar streams and the format in which it is reported in the literature is very heterogeneous: it can be comprised of anything from a pole or end-point coordinates, a list of stream members, polynomial equations or spline knots with an arbitrary celestial coordinate acting as free parameter, to a stream track marked or tidal tails shown in a sky projection plot. The initial data for any given stream inevitably needs to be collected and parsed manually into a data set that \galstreams\ can then handle uniformly and implement as a track with other associated attributes computed automatically for each one.

The first step after the initial data collection is to create a set of points to represent the track. Overall, the data available for each stream can be classified into one of these categories: (i) End points, (ii) Pole, mid point and length and (iii) Custom. The first two are closely related and easy to implement automatically. In the third case, which comprises the majority of streams in the literature, typically a polynomial interpolation or fit will be used to produce a track from the collected data (described for each stream in detail in Sec.~\ref{s:stream_tracks_implemented}).  In this case, because of each stream's different orientation in space, the celestial coordinate that can act as the independent variable will be different. However, the natural reference frame to realise each stream's track automatically is its own frame, i.e. that in which the stream lies approximately in the equator and in it, the natural independent variable is the longitude or azimuthal variable usually named $\phi_1$. In most cases, this frame is not known a~priori and needs to be inferred from the data itself.  To homogenise this starting data set from which the track attributes are computed, \galstreams\ uses the following internal procedure:

\begin{itemize}

\item For each stream a set of densely populated reference points or knots along the track is produced. We describe what data is available and how this is done, for each individual stream in Sec.~\ref{s:stream_tracks_implemented}. The set of knots created for each stream is dense enough so that simple linear interpolation can be performed later to evaluate the track at any given point and the stability of the interpolation is ensured. This set of knots will define the stream's track.

\item In case i), when end point's coordinates are provided,  the initial set of points is realised as uniformly spaced points along the ?(shortest) great circle arc connecting them.  

\item Similarly, in case ii), when the pole, mid point and length are provided, the end points are computed and an initial set of points is realised along the great circle as described in the previous point.

\item When an initial track or set of candidate points is reported and there is no existing information either on the end-points, pole or directly the reference frame's rotation matrix, this is guessed automatically based on the data collected from the literature, as follows:
    \begin{itemize}
        \item The geometric average of the first and last point is used as an initial guess for the mid-point and pole. This guess is provisional as this average is, in general, not along the stream track and --unless the input is already a track-- these points are not necessarily representative of the track's end points. An initial stream frame is realised with these parameters and the points rotated to this reference frame. In this initial reference frame $\phi_1$ can be used as an independent variable and the knots, defined as a set of points uniformly spaced in $\phi_1$, will lie reasonably close to the frame's equator ($\phi_2\sim0\degr$). 
        \item In this initial frame, the new reference point's  coordinates $(\phi_1,\phi_2)$ are found by interpolating the knots' $\phi_2$ at the mean $\phi_1$. This point will lie along the track by construction and is a reasonable approximation of a mid-point.
        \item Next, two points near the reference point (within $0\fdg5$) are used to compute a new pole. The stream's reference frame is now computed using the new mid-point and pole. This reference frame ensures the mid-point is at the origin and the track stays close to $\phi_2\sim0\degr$ around it.  
    \end{itemize}
    
\item The literature data points are transformed to the stream's reference frame and are either interpolated (when an equation or knots are provided) or fitted with a polynomial (when individual stream members are) to reduce the points to a track. Once the interpolator or polynomial fitting object is built in the stream's frame, in each of the available (up to 6D) coordinates the track knots to be stored are realised by populating the track between the identified end points with an arbitrary uniform spacing of $0\fdg01$. This resolution was found to be high enough so that in any later step linear interpolation in the track is sufficient and stable. The stream's knots and reference frame are stored internally. 

\item \neww{ Heliocentric and galactocentric pole tracks are provided for all streams in the library, as are the mid-pole (pole coordinates at the stream's mid-point) in both reference frames. In a given frame, the pole track is computed as the vector product between the position vectors of each pair of consecutive points along a stream. This provides a normal or \emph{pole} vector that may change along the track if the stream is not perfectly planar. Because of the difference in point of view between a heliocentric and Galactocentric observer, pole tracks will depend on the reference frame, hence, pole tracks are reported for both reference frames. }

\item \neww{Angular momentum tracks in a heliocentric reference frame at rest with the GSR are provided for all streams with proper motion information available in the library (see discussion in Sec~\ref{s:pm_misalignment}). This angular momentum is computed as $\mathbf{L_{\mathrm{hel}}} = \mathbf{r_{\mathrm{hel}}} \times \mathbf{v}_{\mathrm{GSR}}$, where $\mathbf{r}_{\mathrm{hel}}$ is the heliocentric position and $\mathbf{v}_{\mathrm{GSR}}$ is the velocity with respect to the GSR. It is convenient to use this reference frame because the radial velocity of the stream has no contribution to the angular momentum and this component is only available for a small number of tracks in the library ($\Nvrad$ out of $\Ntracks$).  }

\end{itemize}

The library's policy is to compile data and summary statistics computed from a homogeneous set of \emph{direct observables}. In some papers \emph{predicted} proper motion or radial velocity tracks are reported, these are not included in the library. Also, solar-reflex corrected proper motion tracks and GSR radial velocities are commonly reported in the literature. Since these are dependent on the position and velocity assumed for the Sun and the LSR, we choose to report instead the \emph{observed} heliocentric proper motions and radial velocity tracks, i.e. \emph{without} having subtracted the solar reflex motion. The solar reflex correction can be easily applied by the user consistently across the library with their preferred solar and LSR parameters\footnote{e.g. in Python using  the reflex\_correct method in \emph{gala} \citep{gala}}.

For now, the purpose of \galstreams\ is to allow comparison of new features with know ones and to allow visualising the system of stellar streams in the Galaxy as a whole.  This is, of course, an incomplete view. Information about the stellar population, e.g. age, metallicity or \alphaFe, is not included. 
No judgement is made as to the confidence or robustness of the detection of the streams listed in the library, some having much more available information than others thanks to dedicated follow-up studies. The positional and kinematic information and data flags provided are intended to serve as a tool for users to form their own criteria as to which streams they consider as better characterised or feel more confident in. Therefore, user discretion is advised.

\subsection{Individual stream track implementations} \label{s:stream_tracks_implemented}

Table~\ref{t:super_summary_table} presents a summary for the \Ntracks\ tracks implemented from different sources for the \Nunique\ (unique) streams currently available in \galstreams. 
\neww{How multiple tracks for a given stream are dealt with and} the criteria on which this is decided is discussed for each case with multiple tracks in Sec~\ref{s:multiple_tracks}. \neww{The naming convention followed for streams with multiple tracks that have different names in the literature, but have been robustly identified as being part of the same stream (e.g. Orphan and Chenab), is that each track is labelled with the original name used by each author (Orphan-K19 and Chenab-S19 tracks) and all are ascribed to the same stream with a compound name (Orphan-Chenab stream).} 
Depending on the data available for each stream its 6D track was implemented following one of the three methods described in Sec.~\ref{s:procedure}: (i) Streams realised as great circle arcs from reported end points, (ii) Streams realised as great circle arcs defined by their reported poles and (iii) Streams realised from custom data. \neww{The method used to implement the track is indicated in Table~\ref{t:super_summary_table} in the Imp (implementation) column for each of the categories described in the previous section (ep=end points, po=pole, st=individual stars or knots). } The InfoFlags column indicates which of the 6D data is available for each stream and keeps track of any assumptions made in each track implementation: the first character is 0 if the stream is assumed to be a great circle and 1 if not; the second, third and fourth characters indicate whether distance, proper motions and radial velocity tracks are available (1) or not (0). Whenever the distance, proper motion or radial velocity  measurements are available but there is a caveat, it is indicated with `2' (see e.g. M5-G19 and Ravi). In these cases, see the documentation for details on each particular stream. The `On' column indicates which of the tracks is set as the default for a given stream. The stream's length along the track, equatorial coordinates and distances for the end points are also reported in the table. Finally, the TRefs and DRefs columns indicate with a unique code the references for the track implementation and the stream's discovery, respectively. The reference  corresponding to each code is found in Table~\ref{t:ref_summary_table}.  

In what follows we briefly describe the specific information used to implement each attribute (position, distance, proper motion, radial velocity) of a stream's track and its individual provenance. \neww{As will become apparent, for many streams the data needed to implement the tracks was only available from figures, not tables or quantitative data provided in the article's text. In these cases, the data has been read-off the relevant plots cited using the WebPlotDigitizer tool.}
\footnote{Available at \url{https://apps.automeris.io/wpd/}}

%---- CAREFUL: DO NOT EDIT THE indiv.tex file. This is created automatically (offline) from the individual documentation index card for each stream in the library. Any manual edits made to this file will be lost.

\paragraph*{20.0-1-M18}

The celestial track was implemented as the great circle arc (of minimum length) with galactocentric pole, mid-point coordinates and length as reported by \citet{Mateu2018}. The mean galactocentric distance reported by the authors was adopted for the full track. The coordinates were transformed back to the heliocentric frame assuming $R_\odot=-8.5$~kpc, as adopted by the authors.

\paragraph*{300S-F18}

The celestial, distance and radial velocity tracks were read-off the fit 
shown by \citet{Fu2018} in their Figure 10. Note that the stream members observed and confirmed spectroscopically by \citet{Fu2018} and, thus, their orbital fit, produce a sky track that differs slightly from the one reported in an earlier follow-up by \citet{Bernard2016}. At both ends the \citet{Bernard2016}  track is slightly south of the one in \citet{Fu2018}. 
They also note their reported track was restricted to the area where the stream is most prominent, 
therefore being shorter than that in \citet{Bernard2016}.

\paragraph*{AAU-ATLAS-L21 and AAU-AliqaUMa-L21}

The ATLAS-AliqaUma (AAU) stream is argued by \citet{Li2021} to be a single feature that includes the previously
identified ATLAS \citep{Koposov2014} and AliqaUma streams \citep{Shipp2018}. The S5 spectroscopic survey of the region
shows it is discontinuous in the sky, but continuous in distance, proper motion and radial velocity.

Because of the sharp discontinuity in the sky, we have implemented it as two separate tracks: AAU-ATLAS and AAU-AliqaUMA.
The sky tracks are given by Eq. 3 by \citet{Li2021}, the distance modulus track is given by their Eq.~2 and the radial velocity and proper motion tracks are given in their Eq. 1. The radial velocity was converted back from GSR to LSR assuming the solar and LSR 
used by the authors and provided in their Sec.1. The proper motions given by their Eq.~1 do not include the solar reflex motion correction.

\neww{
The coordinate frame for the stream (for the two branches) is implemented using the rotation matrix provided by \citet{Li2021},
but note the resulting $\phi_1$ is flipped with respect to \citet{Li2021} (i.e. $\phi_1=-\phi_1^{L21}$).
}

\paragraph*{ACS-R21}

The ACS (AntiCentre Stream) proper motion tracks implemented are the median tracks in Fig. 5 of \citet[][data provided by P. Ramos priv. comm.]{Ramos2021}. The celestial track corresponds to the smoothed spline that better represents the mean Galactic latitude of the HEALpix where these structures are detected, as a function of galactic longitude. The authors report a single mean distance of 11.7~kpc adopted here for the full track. 

Although initially thought to be a tidal stream, like in the case of Monoceros (see Monoceros-R21), a fair consensus has been reached that ACS is most likely a feature produced by stars perturbed out of the Galactic disc \citep[e.g.][and references therein]{Laporte2019a,Laporte2019b,Ramos2021}. As for Monoceros, we have chosen to keep it in the library given that its signature is localised in both the sky and proper motion spaces, and well represented
by a simple track in each.

\paragraph*{ATLAS-I21}

The stream's celestial, distance and proper motions tracks were implemented by fitting a seventh degree polynomial to the stream members reported by \citet{Ibata2021} in their Table~1. The distance track was implemented using the distance computed by the authors and readily provided in the table.

\paragraph*{Acheron-G09}

The celestial track was implemented as the great circle arc (of minimum length) with end points reported by \citet{Grillmair2009}. The distance track was implemented by linearly interpolating the distances reported by the authors for the end points.  

\paragraph*{Alpheus-G13}

The celestial track was implemented from the polynomial fit provided by \citet{Grillmair2013} in their Eq.~1:
\begin{eqnarray*}
\alpha = 32.116-0.00256\delta-0.00225\delta^2
\end{eqnarray*}

with $\delta \in [-69^\circ,-45^\circ]$. The authors report mean heliocentric distances of 2 and 1.6~kpc respectively for the southern and northern parts of the stream. We assume these distances to correspond to the ends of the stream and use linear interpolation to give a first approximation to the distance gradient.

\paragraph*{Aquarius-W11}

The celestial, distance, proper motion and radial velocity tracks for the Aquarius stream were implemented by fitting second degree polynomials to the stream members reported by \citet{Williams2011} in their Tables~1 and 3. The $d_R$ distances were used, as recommended by the authors. These were derived using Reduced Proper Motions and assuming a tangential velocity of 250~km/s for the stream stars. Although proper motions from PPMXL are reported by the authors in their Table~1, for consistency among the library, we have used Gaia EDR3 proper motions retrieved by matching the reported members to Gaia EDR3 with a $0\fdg5$ tolerance. For the radial velocity track we used the line-of-sight heliocentric velocity $V_{los}$ reported by the authors in their Table~1.

\paragraph*{C-19-I21}

The celestial and proper motion tracks for the C-19 stream are implemented by fitting a second degree polynomial to the potential member stars reported in Table 1 of \citet{Martin2022}. The mean distance of 18~kpc adopted by the authors in their analysis is assumed here for the full stream, as Gaia EDR3 parallaxes are not informative enough at such large distances. A mean heliocentric radial velocity of is assumed for the full stream, computed from the stars reported in Table 2 of \citet{Martin2022}.

\paragraph*{C-4-I21}

The stream's celestial, distance and proper motions tracks were implemented by fitting a seventh degree polynomial to the stream members reported by \citet{Ibata2021} in their Table~1. The distance track was implemented using the distance computed by the authors and readily provided in the table.

\paragraph*{C-5-I21}

The stream's celestial, distance and proper motions tracks were implemented by fitting a seventh degree polynomial to the stream members reported by \citet{Ibata2021} in their Table~1. The distance track was implemented using the distance computed by the authors and readily provided in the table.

\paragraph*{C-7-I21}

The stream's celestial, distance and proper motions tracks were implemented by fitting a seventh degree polynomial to the stream members reported by \citet{Ibata2021} in their Table~1. The distance track was implemented using the distance computed by the authors and readily provided in the table.

\paragraph*{C-8-I21}

The stream's celestial, distance and proper motions tracks were implemented by fitting a seventh degree polynomial to the stream members reported by \citet{Ibata2021} in their Table~1. The distance track was implemented using the distance computed by the authors and readily provided in the table.

\paragraph*{Cetus-New-Y21}

The celestial, distance and proper motion tracks were obtained by fitting third order polynomials to stars in the Cetus-New wrap identified by \citet[][data provided by Z. Yuan priv. comm.]{Yuan2021}. We have restricted the fit to members with $\delta<0\degr$ to avoid the sharp discontinuity in the track introduced by a few members at $\delta\sim20\degr$, clearly separated from the rest in declination, and which would require a much higher order polynomial to fit the track and introduce seemingly unphysical wiggles.  We therefore caution the track is not representative of the stream members at $\delta\sim20\degr$ reported by \citet{Yuan2021}.

\paragraph*{Cetus-Palca-T21}

The celestial, distance and proper motion tracks were obtained by fitting a third order polynomial to the Blue Horizontal Branch star members identified by \citet[][data provided by G. Thomas  priv. comm.]{Thomas2021}. The distances provided for these stars are photometric standard-candle distances, computed by the authors as described in Sec. 4.2.1. of \citet{Thomas2021}. 
The stream's reference frame is an auto-computed great-circle frame, with origin at $(\alpha,\delta)=(22\fdg 11454259,-6\fdg 7038421)$ as recommended in their Sec. 4.1. We have chosen this for consistency along the library, since all refence frames used are great-circle ones. We do caution that \citet{Thomas2021} report their $(\phi_1,\phi_2)$ coordinates in a small-circle frame, in which the stream lies at $\phi_2\sim 0\degr$, corresponding to a plane offset by $14\fdg 36$ from the great-circle plane with the same pole $(\alpha,\delta)=(125\fdg 1809832,15\fdg 91290743)$.

\paragraph*{Cetus-Palca-Y21}

The celestial, distance and proper motion tracks were obtained by fitting a fifth order polynomial to the Blue Horizontal Branch stars, K giants and Cetus-Palca wrap samples from \citet[][data provided by Z. Yuan priv. comm.]{Yuan2021}. The fit was restricted to $\alpha<200\degr$ to ensure it's stability, since stars at larger right ascension make the track multi-valued in the stream's reference frame. Therefore, we caution the reader the stream track may extend beyond $\alpha\sim 200\degr$. The authors also identified a new wrap of the Cetus stream, dubbed by the authors as the Cetus-New wrap, we implement this separately under that name.

\paragraph*{Cetus-Y13}

The Cetus stream (or Cetus Polar stream) celestial, distance and radial velocity tracks are implemented from the 
reference points reported by \citet{Yam2013} in their Table~1, taking the mean of the galactic latitude range reported in each row. 

\citet{Yam2013} report radial velocities in the GSR frame. The solar parameters used to convert the observed radial velocities to the GSR are not explicitly reported, hence, to revert back to the heliocentric frame and compute the observed radial velocity we assume a solar peculiar velocity of $(U,V,W)_\odot=(11.1,12.24,7.25)$~km/s with respecto to the LSR from \citet{Schoenrich2010} and $V_{LSR}=220$~km/s from \citet{DehnenBinney1998}, available and widely used at the time.

\paragraph*{Chenab-S19}

The stream's celestial and proper motions tracks were implemented by fitting a second degree polynomial using the ICRS data for the stream members reported by \citet{Shipp2019} in their Table~7 (Appendix E). The distance track was implemented using the mean distance of 39.8~kpc from \citet{Shipp2018} for the full track. The stream's coordinate frame is implemented from the coefficients for the rotation matrix reported by \citet{Shipp2019} in their Table~5 (Appendix C).

\paragraph*{Cocytos-G09}

The celestial track was implemented as the great circle arc (of minimum length) with end points reported by \citet{Grillmair2009}. The distance track was implemented by linearly interpolating the distances reported by the authors for the end points.  

\paragraph*{Corvus-M18}

The celestial track was implemented as the great circle arc (of minimum length) with galactocentric pole, mid-point coordinates and length as reported by \citet{Mateu2018}. The mean galactocentric distance reported by the authors was adopted for the full track. The coordinates were transformed back to the heliocentric frame assuming $R_\odot=-8.5$~kpc, as adopted by the authors.

\paragraph*{Elqui-S19}

The stream's celestial and proper motions tracks were implemented by fitting a second degree polynomial using the ICRS data for the stream members reported by \citet{Shipp2019} in their Table~7 (Appendix E). The distance track was implemented using the mean distance of 50.1~kpc from \citet{Shipp2018} for the full track. The stream's coordinate frame is implemented from the coefficients for the rotation matrix reported by \citet{Shipp2019} in their Table~5 (Appendix C).

\paragraph*{Eridanus-M17}

The end points for the tidal tails were computed from the position angle ($\mathrm{PA}$) and length $l$ of the tails reported in \citet{Myeong2017}. Equatorial coordinates $(\alpha_i,\delta_i)$ for the end points were computed as:

\begin{eqnarray*}
\Delta\alpha_i & = & l_i\sin{(\mathrm{PA}_i)}/\cos{\delta_c} \\
\Delta\delta_i & = & l_i\cos{(\mathrm{PA}_i)}
\end{eqnarray*}

where $(\alpha_c,\delta_c)=(66\fdg1854,-21\fdg 1869)$ are the cluster's central coordinates, from the \citet{Harris1996} catalogue. 

We realised the track as a linear interpolation of the end points and cluster coordinates. This is a good approximation given the small extent of the tails (18' and 11') and their linear appearance in Fig.~1 of \citet{Myeong2017}. 
The authors do not estimate a distance gradient, we assume a mean heliocentric distance for the track of 80.8~kpc as cited by the authors from the \citet{Harris1996} compilation.

\paragraph*{Fimbulthul-I21}

The stream's celestial, distance and proper motions tracks were implemented by fitting a seventh degree polynomial to the stream members reported by \citet{Ibata2021} in their Table~1. The distance track was implemented using the distance computed by the authors and readily provided in the table. \citet{Ibata2019_OCen} argue that Fimbulthul is part of a stream generated by the $\omega$ Centauri globular cluster. Note that the reported track does not link up to the cluster itself.

\paragraph*{Fjorm-I21}

The stream's celestial, distance and proper motions tracks were implemented by fitting a seventh degree polynomial to the stream members reported by \citet{Ibata2021} in their Table~1. The distance track was implemented using the distance computed by the authors and readily provided in the table. 
 The radial velocity track was implemented by fitting a polynomial to the radial velocities of stars reported as probable Fj\"orm members in Table~1 of \citet{Ibata2019}. The radial velocity is set to zero outside the range spanned by the member stars. The last InfoFlag bit is set t
o `2' to reflect that the radial velocity is available but does not span the full length of the track.
This track corresponds to the stream referred to as Fj\"orm in \citet{Ibata2019,Ibata2021}, since \citet{Ibata2021} and \citet{Palau2019} show the Fj\"orm stream to be associated to the M68 cluster, it is named M68-Fjorm in the library.

\paragraph*{GD-1-I21}

This version of the GD-1 stream's celestial, distance and proper motions tracks was implemented by fitting a seventh degree polynomial to the stream members reported by \citet{Ibata2021} in their Table~1. The distance track was implemented using the distance computed by the authors and readily provided in the table.

\paragraph*{GD-1-PB18}

This version of the GD-1 track is based on \citet{PriceWhelanBonaca2018_gd1}. The sky and proper motion tracks are found by fitting a fifth degree polynomial to the stream members selected by \citet{PriceWhelanBonaca2018_gd1} using Gaia DR2 and PanSTARRS-1\footnote{Data available at \url{https://doi.org/10.5281/zenodo.1295543}}, and using the color-magnitud diagram, proper motion and stream track masks provided by the authors.

For the distance track, as the stream is too distant for Gaia~DR2 parallaxes to be useful, we have assumed the distance gradient proposed by the authors:
\begin{eqnarray*}
d (\kpc) = 0.05\phi_1(^\circ) +10
\end{eqnarray*}

with $\phi_1$ being the along-stream coordinate in the GD-1 coordinate frame from \citet{Koposov2010}, which we adopt here
as the stream's reference frame.

\paragraph*{Gaia-1-I21}

The stream's celestial, distance and proper motions tracks were implemented by fitting a seventh degree polynomial to the stream members reported by \citet{Ibata2021} in their Table~1. The distance track was implemented using the distance computed by the authors and readily provided in the table.

\paragraph*{Gaia-10-I21}

The stream's celestial, distance and proper motions tracks were implemented by fitting third degree polynomials to stream members extracted from Figure~13 in \citet{Ibata2021}. 

\paragraph*{Gaia-11-I21}

The stream's celestial, distance and proper motions tracks were implemented by fitting a seventh degree polynomial to the stream members reported by \citet{Ibata2021} in their Table~1. The distance track was implemented using the distance computed by the authors and readily provided in the table.

\paragraph*{Gaia-12-I21}

The stream's celestial, distance and proper motions tracks were implemented by fitting third degree polynomials to stream members extracted from Figure~13 in \citet{Ibata2021}. 

\paragraph*{Gaia-2-I21}

The stream's celestial, distance and proper motions tracks were implemented by fitting a seventh degree polynomial to the stream members reported by \citet{Ibata2021} in their Table~1. The distance track was implemented using the distance computed by the authors and readily provided in the table.

\paragraph*{Gaia-3-M18}

The celestial track was implemented as the great circle arc (of minimum length) with end points reported by \citet{Malhan2018}. The distance track was implemented by linearly interpolating the distances reported by the authors for the end points.  

\paragraph*{Gaia-4-M18}

The celestial track was implemented as the great circle arc (of minimum length) with end points reported by \citet{Malhan2018}. The distance track was implemented by linearly interpolating the distances reported by the authors for the end points.  

\paragraph*{Gaia-5-M18}

The celestial track was implemented as the great circle arc (of minimum length) with end points reported by \citet{Malhan2018}. The distance track was implemented by linearly interpolating the distances reported by the authors for the end points.  

\paragraph*{Gaia-6-I21}

The stream's celestial, distance and proper motions tracks were implemented by fitting third degree polynomials to stream members extracted from Figure~13 in \citet{Ibata2021}. 

\paragraph*{Gaia-7-I21}

The stream's celestial, distance and proper motions tracks were implemented by fitting third degree polynomials to stream members extracted from Figure~13 in \citet{Ibata2021}. 

\paragraph*{Gaia-8-I21}

The stream's celestial, distance and proper motions tracks were implemented by fitting a seventh degree polynomial to the stream members reported by \citet{Ibata2021} in their Table~1. The distance track was implemented using the distance computed by the authors and readily provided in the table.

\paragraph*{Gaia-9-I21}

The stream's celestial, distance and proper motions tracks were implemented by fitting a seventh degree polynomial to the stream members reported by \citet{Ibata2021} in their Table~1. The distance track was implemented using the distance computed by the authors and readily provided in the table.

\paragraph*{Gjoll-I21}

The stream's celestial, distance and proper motions tracks were implemented by fitting a seventh degree polynomial to the stream members reported by \citet{Ibata2021} in their Table~1. The distance track was implemented using the distance computed by the authors and readily provided in the table. The authors show this stream is associated to the NGC~3201 cluster.

\paragraph*{Gunnthra-I21}

The stream's celestial, distance and proper motions tracks were implemented by fitting a seventh degree polynomial to the stream members reported by \citet{Ibata2021} in their Table~1. The distance track was implemented using the distance computed by the authors and readily provided in the table.

\paragraph*{Hermus-G14}

The celestial track for Hermus was implemented from the polynomial fit provided by \citet{Grillmair2014} in their Eq.~1 :
\begin{eqnarray*}
\alpha &=& 241.571 + 1.37841\delta - 0.148870\delta^2 + 0.00589502\delta^3 \\
       & &- 1.03927\times 10^{-4}\delta^4 + 7.28133\times 10^{-7}\delta^5
\end{eqnarray*}

with $\delta \in [5^\circ, 50^\circ ]$ reported as the ends of the stream in their Sec. 3.1. The authors report mean heliocentric distances of 15, 20 and 19~kpc respectively for the northern ($\delta =50^\circ$), central ($\delta =40^\circ$) and southern parts ($\delta =5^\circ$) of the stream. We assume these distances to correspond to the mid-point and ends of the stream and use polynomial interpolation in between, with a high enough order to avoid the kink due to the abrupt change at the mid-point. 

\paragraph*{Hrid-I21}

The stream's celestial, distance and proper motions tracks were implemented by fitting a seventh degree polynomial to the stream members reported by \citet{Ibata2021} in their Table~1. The distance track was implemented using the distance computed by the authors and readily provided in the table.

\paragraph*{Hyllus-G14}

The celestial track for Hyllus was implemented from the polynomial fit provided by \citet{Grillmair2014} in their Eq.~1:
\begin{eqnarray*}
\alpha = 255.8150 - 0.78364\delta + 0.01532\delta^2
\end{eqnarray*}

with $\delta \in [11^\circ,34^\circ]$. These limits in declination are not given explicitly in \citet{Grillmair2014}, so they were taken from the compilation in Table~4.1 of \citet{GrillmairCarlin2016}. The authors report mean heliocentric distances of 18.5 and 23~kpc respectively for the northern and southern ends of the stream. We assume these distances to correspond to the ends of the stream and use linear interpolation in between.

\paragraph*{Indus-S19}

The stream's celestial and proper motions tracks were implemented by fitting a second degree polynomial using the ICRS data for the stream members reported by \citet{Shipp2019} in their Table~7 (Appendix E). The distance track was implemented using the mean distance of 16.6~kpc from \citet{Shipp2018} for the full track. The stream's coordinate frame is implemented from the coefficients for the rotation matrix reported by \citet{Shipp2019} in their Table~5 (Appendix C).

\paragraph*{Jet-F22}

The stream's celestial, distance and proper motions tracks were implemented by fitting third degree polynomials to the ICRS data for the stream members reported by \citet{Ferguson2022} in their Table~3. As noted by the authors, distances reported in the table were computed according to their Eq.~3, based on the distance modulus gradient observed for Blue Horizontal Branch stars identified in the stream. The stream's coordinate frame is implemented using the rotation matrix provided in their Eq.~1, coinciding with the frame defined in \citet{Jethwa2018}.

\paragraph*{Jet-J18}

The celestial track was implemented as the great circle arc (of minimum length) with end points reported by \citet{Jethwa2018}. The distance track was implemented by linearly interpolating the distances reported by the authors for the end points.  

\paragraph*{Jhelum-I21}

The stream's celestial, distance and proper motions tracks were implemented by fitting a seventh degree polynomial to the stream members reported by \citet{Ibata2021} in their Table~1. The distance track was implemented using the distance computed by the authors and readily provided in the table.

\paragraph*{Jhelum-a and Jhelum-b (B19)}

This realisation of the Jhelum stream was implemented in two separate branches, Jhelum-a and Jhelum-b, based on the sky tracks provided by \citet{Bonaca2019}. The main component's track (Jhelum-a) is given by their Eq.~1:
\begin{eqnarray*}
\phi_2^a = 0.000546\phi_1^2 -0.00217\phi_1 + 0.583
\end{eqnarray*}

with $\phi_1 \in [-5\degr,+25\degr]$. The secondary component's (Jhelum-b) track is described 
by: 
\begin{eqnarray*}
\phi_2^b = \phi_2^a - 0\fdg9
\end{eqnarray*}

The two tracks are implemented using the same coordinate frame defined by the rotation matrix provided in \citet[][their Sec.~2]{Bonaca2019}.

The proper motion tracks were implemented by fitting a polynomial to points read-off of their Fig.~4 in the stream's coordinate frame. These proper motions have not been corrected for the solar reflex motion. \citet{Bonaca2019} note that despite the two components having a systematic and constant offset in the sky, their proper motions are very similar, being 'kinematically indistinguishable' at the current precision.

\paragraph*{Jhelum-a and Jhelum-b (S19)}

The stream's celestial and proper motions tracks for Jhelum-a and Jhelum-b were implemented by fitting a first degree polynomial to the stream members reported by \citet{Shipp2019} in their Table~7 (Appendix E). The distance track was implemented using the mean distance of 13.2~kpc from \citet{Shipp2018} for the full track for both components. The same stream's coordinate frame is implemented for both branches, from the coefficients for the rotation matrix reported for Jhelum by \citet{Shipp2019} in their Table~5 (Appendix C).

\paragraph*{Kshir-I21}

The stream's celestial, distance and proper motions tracks were implemented by fitting a seventh degree polynomial to the stream members reported by \citet{Ibata2021} in their Table~1. The distance track was implemented using the distance computed by the authors and readily provided in the table.

\paragraph*{Kwando-G17}

The celestial track for Kwando was implemented from the polynomial fit provided by \citet{Grillmair2017_south} in their Eq. 5:
\begin{eqnarray*}
\delta = -7.817 -2.354\alpha +0.1202\alpha^2 -0.00215\alpha^3 
\end{eqnarray*}

with $\alpha \in [+19^\circ,+31^\circ]$, as explicitly reported by the authors. The authors report a FWHM of 22 arcmin corresponding to a physical width of 130~pc, which corresponds to a heliocentric distance of $\approx$20~kpc. This mean distance was adopted for the full track.

\paragraph*{Kwando-I21}

The stream's celestial, distance and proper motions tracks were implemented by fitting a seventh degree polynomial to the stream members reported by \citet{Ibata2021} in their Table~1. The distance track was implemented using the distance computed by the authors and readily provided in the table.

\paragraph*{LMS1-M21}

The celestial and proper motion tracks for the LMS-1 stream were implemented by fitting a fifth degree polynomial to the stream members reported by \citet{Malhan2021} in their Table 1, with proper motions retrieved directly from a cross-match to Gaia EDR3 with $0.5$ arcsec tolerance. For the full stream we have assumed the mean distance of 19~kpc  calculated  by \citet{Malhan2021} from the Gaia EDR3 uncertainty weighed mean parallax, since at these large distances Gaia EDR3 parallaxes are too uncertain to be informative. Note there are no common stars between these and the RR Lyrae and Blue Horizontal Branch stars from \citet{Yuan2020} used to implement the LMS1-Y20 track. 

\paragraph*{LMS1-Y20}

The stream's celestial, distance and proper motion tracks were implemented by fitting a five degree polynomial to the stream RR Lyrae and Blue Horizontal Branch members reported by \citet{Yuan2020} (private communication). Note these have no stars in common with those reported by \citet{Malhan2021} used to implement the LMS1-M21 track. We have chosen to implement these two tracks separately as they derived from different stellar population tracers.

\paragraph*{Leiptr-I21}

The stream's celestial, distance and proper motions tracks were implemented by fitting a seventh degree polynomial to the stream members reported by \citet{Ibata2021} in their Table~1. The distance track was implemented using the distance computed by the authors and readily provided in the table.

\paragraph*{Lethe-G09}

The celestial track was implemented as the great circle arc (of minimum length) with end points reported by \citet{Grillmair2009}. The distance track was implemented by linearly interpolating the distances reported by the authors for the end points.  

\paragraph*{M2-G22}

The celestial and proper motion tracks were obtained by fitting fifth order polynomials to stream members reported by \citet{Grillmair2022} in their Tables 1 and 2, limited to members with weights larger than 0.2 to avoid an apparent bifurcation of the stream at $\alpha\sim 330\degr$. The tables do not include distance estimates, and this information could not be extracted from their Fig. 4, which does seem to show a strong distance gradient ranging from $\sim20$ to $7$~kpc from east to west along the stream. We have adopted the mean distance to the cluster (11.693~kpc) from \citet{Baumgardt2021} as cited by the author, but caution this should \emph{not} provide a good approximation. The InfoFlag for the distance in this case is thus set to 0 accordingly.

\paragraph*{M2-I21}

The stream's celestial, distance and proper motions tracks were implemented by fitting third degree polynomials to stream members extracted from Figure~14 in \citet{Ibata2021}. 

\paragraph*{M30-S20}

The stream's celestial track was implemented by fitting a sixth degree polynomial to knots along the tail detected by \citet[][data provided by A. Sollima in priv. comm]{Sollima2020}. The author's methodology used Gaia DR2 parallaxes in their inference, but the distances are not provided explicitly in the paper. Here the track was implemented using the most recent cluster distance from \citet{Baumgardt2021}.

\paragraph*{M5-G19}

The M5 stream's celestial track was implemented from the polynomial fit provided by \citet{Grillmair2019} in their Eq. 1:
\begin{eqnarray*}
\delta = 37.4026 + 0.2096\alpha -0.001578\alpha^2
\end{eqnarray*}

with $\alpha \in [190^\circ,225^\circ]$, as explicitly reported by the author. 

The proper motion tracks were obtained by fitting a third order polynomial to the 50 highest weigthed candidates provided in their Table 1. Neither the table nor any of the figures include distance \emph{measurements}, which were not necessary given the methodology used. Setting the mean distance to the cluster (7.5~kpc) for the full length of the track should \emph{not} provide a good approximation. The orbit prediction, shown in their Fig.~1, is that the heliocentric distance increases from  7.3 kpc at $\alpha\sim217^\circ$ to $\sim15$ kpc at $\alpha\sim134^\circ$. Since the distance track is a required attribute in the library, we use linear interpolation between these values from the orbit prediction and caution the users that they \emph{do not correspond to observed values}. The InfoFlag for the distance in this case is thus set to 0 to reflect the \emph{observed} distance track is not available.

\paragraph*{M5-I21}

The stream's celestial, distance and proper motions tracks were implemented by fitting a seventh degree polynomial to the stream members reported by \citet{Ibata2021} in their Table~1. The distance track was implemented using the distance computed by the authors and readily provided in the table.

\paragraph*{M5-S20}

The stream's celestial track was implemented by fitting a fifth degree polynomial to knots along the tail detected by \citet[][data provided by A. Sollima in priv. comm]{Sollima2020}. The author's methodology used Gaia DR2 parallaxes in their inference, but the distances are not provided explicitly in the paper. Here the track was implemented using the most recent cluster distance from \citet{Baumgardt2021}.

\paragraph*{M68-I21}

The stream's celestial, distance and proper motions tracks were implemented by fitting a seventh degree polynomial to the stream members reported by \citet{Ibata2021} in their Table~1. The distance track was implemented using the distance computed by the authors and readily provided in the table.

\paragraph*{M68-P19}

This version of the M68 stream's celestial, distance and proper motion tracks were implemented using the candidate stars reported by \citet{Palau2019} in their Table E1 and fitting a third degree polynomial to the data in each coordinate. For the computation of the distance track we assumed the reciprocal of the parallax reported in the table as a distance estimator and excised stars with negative parallaxes and parallaxes $<0.05$~mas, which are clear outliers. We have set the distance InfoFlag to '2' to reflect this estimate should be taken with caution.

\paragraph*{M92-I21}

The stream's celestial, distance and proper motions tracks were implemented by fitting a seventh degree polynomial to the stream members reported by \citet{Ibata2021} in their Table~1. The distance track was implemented using the distance computed by the authors and readily provided in the table.

\paragraph*{M92-S20}

The stream's celestial track was implemented by fitting a sixth degree polynomial to knots along the tail detected by \citet[][data provided by A. Sollima in priv. comm]{Sollima2020}. The author's methodology used Gaia DR2 parallaxes in their inference, but the distances are not provided explicitly in the paper. Here the track was implemented using the most recent cluster distance from \citet{Baumgardt2021}.

\paragraph*{M92-T20}

The M92 celestial track was implemented using the polynomial fit provided by \citet{Thomas2020} in tangent plane coordinates:

\[ \eta = -0.134 +0.041\xi -0.056\xi^2 +0.001\xi^3 \] 

where $-7^\circ<\xi<+9\fdg5$ and $\eta$ are given in degrees, and point West and North following the usual convention. The tangent plane transformation assumes the cluster as the center of projection. We convert $(\xi,\eta)$ to equatorial coordinates following standard procedure \citep[e.g. see Chapter 9 in][]{BB2005}. Finally, for the distance track we assume for the whole track the mean distance of $8.3$~kpc, reported by the authors in their Table~1, as no distance gradient is reported.

\paragraph*{Molonglo-G17}

The celestial track for Molonglo was implemented from the polynomial fit provided by \citet{Grillmair2017_south} in their 
Eq.~2:
\begin{eqnarray*}
\alpha = 345.017 - 0.5843\delta + 0.0182\delta^2
\end{eqnarray*}

with $\delta \in [-24\fdg5,-12^\circ]$, as explicitly reported by the authors. The authors report a mean heliocentric distance of 20~kpc for the stream and FWHM of 30 arcmin, we assume this mean value for the whole distance track. 

\paragraph*{Monoceros-R21}

The Monoceros proper motion tracks correspond to the median tracks in Fig. 5 of \citet[][data provided by P. Ramos private communication]{Ramos2021}. The celestial track corresponds to the smoothed splined that better represents the mean Galactic latitude of the HEALpix where these structures are detected, as a function of galactic longitude. 

Monoceros extends further towards $l>200^\circ$, but here we limit the tracks to the data provided in the blind identification conducted by \citet{Ramos2021}. The authors report a single mean distance of 10.6~kpc, which we adopt here for the full track.

A fair consensus seems to have been reached in the literature that Monoceros is not a tidal stream formed by an accreted galaxy \citep[see review by][]{Yanny2016}, as originally thought \citep{Yanny2003}, but rather a feature excited or perturbed from the disc \citep{Kazantzidis2009,Laporte2019a,Laporte2019b}. In spite of this, we have chosen to keep it in the library given that its signature is localised in both the sky and proper motion spaces, and well represented by a simple track in each.

\paragraph*{Murrumbidgee-G17}

The celestial track for Murrumbidgee was implemented from the polynomial fit provided by \citet{Grillmair2017_south}
in their Eq.~3:
\begin{eqnarray*}
\alpha &=& 367.893 -0.4647\delta - 0.00862\delta^2 + 0.000118\delta^3 \\
       & & +1.2347\times 10^{-6}\delta^4 - 1.13758\times 10^{-7}\delta^5
\end{eqnarray*}

with $\delta \in [-65^\circ,+16^\circ]$. The declination range for the full stream is not explicitly reported in \citet{Grillmair2017_south}. 
The author reports the portion of the stream with $b \in [-65^\circ,-30^\circ]$ is detected at a 
$6\sigma$ significance, but do not explicitly provide the full galactic latitude (or declination) range for the stream. 
Also, the fiducial point reported in Table 1 and used for orbital fitting is not included in that range. 
Since the authors report the stream to be $95^\circ$ long, we take the declination range to be 
$\delta \in [-65^\circ,+16^\circ]$ in order to reproduce this length and for the track to contain the fiducial point
$(\alpha,\delta)=(358\fdg614, 16\fdg274)$.

The authors report a mean heliocentric distance of 20~kpc, adopted here for the full stream's distance track.

\paragraph*{NGC1261-I21}

The stream's celestial, distance and proper motions tracks were implemented by fitting third degree polynomials to stream members extracted from Figure~14 in \citet{Ibata2021}. 

\paragraph*{NGC1851-I21}

The stream's celestial, distance and proper motions tracks were implemented by fitting third degree polynomials to stream members extracted from Figure~14 in \citet{Ibata2021}. 

\paragraph*{NGC2298-I21}

The stream's celestial, distance and proper motions tracks were implemented by fitting third degree polynomials to stream members extracted from Figure~14 in \citet{Ibata2021}. 

\paragraph*{NGC2298-S20}

The stream's celestial track was implemented by fitting a sixth degree polynomial to knots along the tail detected by \citet[][data provided by A. Sollima in priv. comm]{Sollima2020}. The author's methodology used Gaia DR2 parallaxes in their inference, but the distances are not provided explicitly in the paper. Here the track was implemented using the most recent cluster distance from \citet{Baumgardt2021}.

\paragraph*{NGC2808-I21}

The stream's celestial, distance and proper motions tracks were implemented by fitting third degree polynomials to stream members extracted from Figure~14 in \citet{Ibata2021}. 

\paragraph*{NGC288-I21}

The stream's celestial, distance and proper motions tracks were implemented by fitting third degree polynomials to stream members extracted from Figure~14 in \citet{Ibata2021}. 

\paragraph*{NGC288-S20}

The stream's celestial track was implemented by fitting a ninth degree polynomial to knots along the tail detected by \citet[][data provided by A. Sollima in priv. comm]{Sollima2020}. The author's methodology used Gaia DR2 parallaxes in their inference, but the distances are not provided explicitly in the paper. Here the track was implemented using the most recent cluster distance from \citet{Baumgardt2021}.

\paragraph*{NGC3201-I21}

The stream's celestial, distance and proper motions tracks were implemented by fitting a seventh degree polynomial to the stream members reported by \citet{Ibata2021} in their Table~1. The distance track was implemented using the distance computed by the authors and readily provided in the table.

The detection presented in \citet{Ibata2021} corresponds to the stellar stream's detection around the cluster position. The authors argue the Gj\"oll stream is the continuation of the cluster's tails further towards the Galactic anti-center, based on the agreement between the NGC3201 stream and Gj\"oll detections with an orbital fit based on two stars of the cluster's stream. 

\paragraph*{NGC3201-P21}

The NGC3201 stream's celestial, distance and proper motion tracks are implemented using the candidate stars reported by \citet{Palau2021} in their Table~C1 and fitting a seventh degree polynomial to the data in each coordinate. The distance track was computed assuming the reciprocal of the parallax as a distance estimator and excising stars with negative parallaxes and 
parallaxes $<0.05$~mas, which are clear outliers.

\paragraph*{NGC5466-G06}

In the previous version of \galstreams, the NGC~5466 stream track was realised by interpolating between the stream's end points
reported by \citet{Grillmair2006_5466} in their  Fig.~1 caption and using the cluster's position from the
\citet[][2010 edition]{Harris1996} catalogue as a central point.

For this realization of the NGC~5466 stream's celestial track, we read off points along the dot-dashed lined in Figure 2 of \citet{Grillmair2006_5466}. The authors report a width of $1\fdg4$ and a mean heliocentric distance of 16.6~kpc adopted here for the full stream.

It is interesting to note that \citet{Weiss2018} report three detections nearly parallel to the stream but about $\sim 5^\circ$ south of the NGC~5466 stream reported by \citet{Grillmair2006_5466}. These are not included here.

\paragraph*{NGC5466-I21}

The stream's celestial, distance and proper motions tracks were implemented by fitting third degree polynomials to stream members extracted from Figure~14 in \citet{Ibata2021}. 

\paragraph*{NGC5466-J21}

The stream's celestial, distance and proper motions tracks were implemented by fitting second degree polynomials to stream members  
from Table~2 of \citet{Jensen2021}. The authors also report radial velocities for six stars, but these all correspond to cluster members (and one contaminant), hence, radial velocity information has not been included for this track. 

\paragraph*{NGC6101-I21}

The stream's celestial, distance and proper motions tracks were implemented by fitting third degree polynomials to stream members extracted from Figure~14 in \citet{Ibata2021}. 

\paragraph*{NGC6362-S20}

The stream's celestial track was implemented by fitting a sixth degree polynomial to knots along the tail detected by \citet[][data provided by A. Sollima in priv. comm]{Sollima2020}. The author's methodology used Gaia DR2 parallaxes in their inference, but the distances are not provided explicitly in the paper. Here the track was implemented using the most recent cluster distance from \citet{Baumgardt2021}.

\paragraph*{NGC6397-I21}

The stream's celestial, distance and proper motions tracks were implemented by fitting a seventh degree polynomial to the stream members reported by \citet{Ibata2021} in their Table~1. The distance track was implemented using the distance computed by the authors and readily provided in the table.

\paragraph*{OmegaCen-I21}

The stream's celestial, distance and proper motions tracks were implemented by fitting third degree polynomials to stream members extracted from Figure~14 in \citet{Ibata2021}. 

\paragraph*{OmegaCen-S20}

The stream's celestial track was implemented by fitting a fifth degree polynomial to knots along the tail detected by \citet[][data provided by A. Sollima in priv. comm]{Sollima2020}. The author's methodology used Gaia DR2 parallaxes in their inference, but the distances are not provided explicitly in the paper. Here the track was implemented using the most recent cluster distance from \citet{Baumgardt2021}.

\paragraph*{Ophiuchus-B14}

The celestial track was implemented as the great circle arc (of minimum length) with heliocentric pole, mid-point coordinates and length as reported by \citet{Bernard2014}.   The mean distance reported by the authors was adopted for the full track. 

\paragraph*{Ophiuchus-C20}

The celestial and proper motion tracks were implemented by fitting a fifth degree polynomial to the members published in Table~2 of \citet{Caldwell2020} with membership probabilities $P_\mathrm{mem}>0.5$. The distance track was implemented using the mean distance for the full stream, calculated as the reciprocal of the mean weighted parallax of $0.12 \pm 0.01$~mas obtained by the authors for high probability members ($P_\mathrm{mem}>0.9$) with parallax errors $<0.5$~mas. We caution that there probably is a significant distance gradient in the stream, since the authors note \citet{Sesar2015} already observed a distance gradient of $\sim1.5$~kpc over $\sim2\degr$, consistent with their observed color-magnitud diagrams. 

\paragraph*{Orinoco-G17}

Orinoco's celestial track was implemented from the polynomial fit provided by \citet{Grillmair2017_south} in their Eq. 4:
\begin{eqnarray*}
\delta &=& -25.5146 + 0.1672\alpha + -0.003827\alpha^2 -0.0002835\alpha^3 \\
       & & -5.3133\times 10^{-6}\alpha^4
\end{eqnarray*}

with $\alpha>324^\circ$ or $\alpha<23^\circ$. This range in right ascension is not explicitly reported by the authors,
it was inferred from their Fig.~1 to match the extent of the stream shown (A. Drlica-Wagner private comm.).
The authors report a FWHM of 40 arcmin corresponding to a physical width of 240~pc, which corresponds to a heliocentric distance of 20.6~kpc, adopted here for the full track.

\citet{Grillmair2017_south} also mention a putative western extension of Orinoco that is not well approximated by their Eq.~4, but no further information is provided so this is not included in the implemented track.

\paragraph*{Orphan-I21}

The stream's celestial, distance and proper motions tracks were implemented by fitting a seventh degree polynomial to the stream members reported by \citet{Ibata2021} in their Table~1. The distance track was implemented using the distance computed by the authors and readily provided in the table.

\paragraph*{Orphan-K19}

The sky, distance, $\mu_{\phi_1}$ and radial velocity tracks for the Orphan stream are implemented from the knots
reported in Tables C1-C3 and Table 4 of \citet{Koposov2019}, respectively. 
The coordinate frame adopted and supplied with \galstreams\ is that provided by \citet{Koposov2019} in their Appendix B.
The authors report the solar-reflex corrected $\mu_{\phi_1}$ proper motion and radial velocity in the GSR frame,
for which the Sun's peculiar velocity $V_{LSR}=240$\kms from \citet{Schoenrich2010} and position $R_\odot=8.34$kpc 
from \citet{Reid2014} were adopted. We use these values to add back the solar reflex contribution and report all quantities in the heliocentric frame. Radial velocities reported by the authors are limited to $50^\circ<\phi_1<120^\circ$ (corresponding to 
$141\fdg21<\alpha<167\fdg75$), outside this range we have set the radial velocity track to zero.

The $\mu_{\phi_2}$ proper motion track was obtained from the RR Lyrae stream members provided in Table~5 of \citet{Koposov2019}. Their Gaia~DR2 \citep{GaiaCol_2018_DR2_survey} proper motions were retrieved from the Gaia Archive, converted into the stream's coordinate frame and the 
$\mu_{\phi_2}$ track obtained by fitting a 10-degree polynomial. 

\paragraph*{PS1-A-B16}

The celestial track was implemented as the great circle arc (of minimum length) with heliocentric pole, mid-point coordinates and length as reported by \citet{Bernard2016}.   The mean distance reported by the authors was adopted for the full track. 

\paragraph*{PS1-B-B16}

The celestial track was implemented as the great circle arc (of minimum length) with heliocentric pole, mid-point coordinates and length as reported by \citet{Bernard2016}.   The mean distance reported by the authors was adopted for the full track. 

\paragraph*{PS1-C-B16}

The celestial track was implemented as the great circle arc (of minimum length) with heliocentric pole, mid-point coordinates and length as reported by \citet{Bernard2016}.   The mean distance reported by the authors was adopted for the full track. 

\paragraph*{PS1-D-B16}

The celestial track was implemented as the great circle arc (of minimum length) with heliocentric pole, mid-point coordinates and length as reported by \citet{Bernard2016}.   The mean distance reported by the authors was adopted for the full track. 

\paragraph*{PS1-E-B16}

The celestial track was implemented as the great circle arc (of minimum length) with heliocentric pole, mid-point coordinates and length as reported by \citet{Bernard2016}.   The mean distance reported by the authors was adopted for the full track. 

\paragraph*{Pal13-S20}

The celestial track was implemented as the great circle arc (of minimum length) with end points reported by \citet{Shipp2020}    . The distance track was implemented by linearly interpolating the distances reported by the authors for the end points.  

\paragraph*{Pal15-M17}

The celestial track was implemented using end points for the tidal tails computed from the position angle ($\mathrm{PA}$) and length $l$ of the tails reported in \citet{Myeong2017}. Equatorial coordinates $(\alpha_i,\delta_i)$ for the end points were computed as:
\begin{eqnarray*}
\Delta\alpha_i &=& l_i\sin{\mathrm{PA}_i}/\cos{\delta_c} \\
\Delta\delta_i &=& l_i\cos{\mathrm{PA}_i} 
\end{eqnarray*}

where $(\alpha_c,\delta_c)=(255\fdg 01,-0\fdg 5419)$ are the cluster's central coordinates, from the \citet{Harris1996} catalogue.

We implement the track as a linear interpolation of the end points and cluster coordinates. This is a good approximation given the small extent of the tails (59' and 29') and their linear appearance in Fig.~2 of \citet{Myeong2017}.
The authors do not provide a distance or distance gradient estimate, so we adopt a mean distance for the track of 43.5~kpc as cited by the authors from the \citet{Harris1996} compilation.

\paragraph*{Pal5-I21}

The stream's celestial, distance and proper motions tracks were implemented by fitting a seventh degree polynomial to the stream members reported by \citet{Ibata2021} in their Table~1. The distance track was implemented using the distance computed by the authors and readily provided in the table.

\paragraph*{Pal5-PW19}

The Pal~5 stream's proper motion and distance tracks are implemented from the 2D polynomial coefficientes provided by \citet{PriceWhelan2019_pal5} in their Table 1. The celestial track is taken from \citet{Bonaca2020}.

\paragraph*{Pal5-S20}

Only the celestial track is implemented for the stream in this case, based on the anchor points (black circles) shown in Fig.~7 of \citet{Starkman2020}. Since there is no distance gradient information used in this study, we have set the mean distance of 22.5~kpc for the full stream. Although there is 5D information available for the Pal~5 stream from previous studies \citep{PriceWhelan2019_pal5,Ibata2021}, we include it since this study has traced the leading tail by $\sim7\degr$ beyond previously known limits.

\paragraph*{Palca-S18}

The celestial track was implemented as the great circle arc (of minimum length) with end points reported by \citet{Shipp2018}. The distance track was implemented by linearly interpolating the distances reported by the authors for the end points.  

\paragraph*{Parallel-W18}

The Parallel stream celestial and distance tracks are implemented by fitting third degree polynomials to the reference points reported by \citet{Weiss2018} in their Table 2 and using linear interpolation in between.

\paragraph*{Pegasus-P19}

The Pegasus stream celestial track was implemented by fitting a third degree polynomial to the end points reported 
by \citet{Perottoni2019} in their Table 1, plus a few points read-off of their Figure 2, in order to avoid assuming the track is well aproximated by a great circle. The authors report a heliocentric distance of 18~kpc for the full stream, which we adopt here for the distance track. 

\paragraph*{Perpendicular-W18}

The Perpendicular stream celestial and distance tracks are implemented using the reference points reported by \citet{Weiss2018} in their Table 2 and using linear interpolation in between.

\paragraph*{Phlegethon-I21}

The stream's celestial, distance and proper motions tracks were implemented by fitting a seventh degree polynomial to the stream members reported by \citet{Ibata2021} in their Table~1. The distance track was implemented using the distance computed by the authors and readily provided in the table.

\paragraph*{Phoenix-S19}

The stream's celestial and proper motions tracks were implemented by fitting a second degree polynomial using the ICRS data for the stream members reported by \citet{Shipp2019} in their Table~7 (Appendix E). The distance track was implemented using the mean distance of 17.5~kpc from \citet{Balbinot2016} for the full track. The stream's coordinate frame is implemented from the coefficients for the rotation matrix reported by \citet{Shipp2019} in their Table~5 (Appendix C).

\paragraph*{Ravi-S18}

The celestial track was implemented as the great circle arc (of minimum length) with end points reported by \citet{Shipp2018}. The distance track was implemented by linearly interpolating the distances reported by the authors for the end points.  The stream's coordinate  frame is implemented from the coefficients for the rotation matrix reported by \citet{Shipp2019} in their Table~5 (Appendix C). The proper motion track was implemented using the mean by-eye proper motion measurement for the stream in observed ICRS coordinates, reported in \citet{Shipp2019} in their Table 3.  
The InfoFlag for the proper motion in this case is thus set to 2 to reflect the proper motion track is available but is an approximation.

\paragraph*{Sagittarius-A20}

The Sagittarius stream's celestial and proper motion tracks implemented are those derived by \citet{Antoja2020} coupled with the distance track from \citet{Ramos2020} corresponding to their RR Lyrae stars' Strip sample. To implement these we have used the polynomial interpolators provided by the authors in the GitHub repository Brugalada\footnote{\url{https://github.com/brugalada/Sagittarius}}.

In the previous version of \galstreams\ the Sagittarius stream footprint had been implemented by supplying a realisation of the \citet{LawMajewski2010} model in a spherical potential. We have chosen to implement the new track based on \citet{Antoja2020} and \citet{Ramos2020} as these correspond to nearly all-sky (except for the Galactic disc crossing) blind detections made with direct observables with no prior Sagittarius model information. This way, in terms of track implementation, Sagittarius stands on equal footing as the rest of the streams.

\paragraph*{Sangarius-G17}

The celestial track was implemented as the great circle arc (of minimum length) with heliocentric pole, mid-point coordinates and length as reported by \citet{Grillmair2017}. The mean distance reported by the authors was adopted for the full track. 

\paragraph*{Scamander-G17}

The celestial track was implemented as the great circle arc (of minimum length) with heliocentric pole, mid-point coordinates and length as reported by \citet{Grillmair2017}. The mean distance reported by the authors was adopted for the full track. 

\paragraph*{Slidr-I21}

The stream's celestial, distance and proper motions tracks were implemented by fitting a seventh degree polynomial to the stream members reported by \citet{Ibata2021} in their Table~1. The distance track was implemented using the distance computed by the authors and readily provided in the table. The radial velocity track was implemented by fitting a polynomial to the radial velocities of stars reported as probable members in Table~1 of \citet{Ibata2019}. The radial velocity is set to zero outside the range spanned by the member stars. The last InfoFlag bit is set to `2' to reflect that the radial velocity is available but does not span the full length of the track.

\paragraph*{Styx-G09}

The celestial track was implemented as the great circle arc (of minimum length) with end points reported by \citet{Grillmair2009}. The distance track was implemented by linearly interpolating the distances reported by the authors for the end points.  

\paragraph*{Svol-I21}

The stream's celestial, distance and proper motions tracks were implemented by fitting a seventh degree polynomial to the stream members reported by \citet{Ibata2021} in their Table~1. The distance track was implemented using the distance computed by the authors and readily provided in the table.

\paragraph*{Sylgr-I21}

The stream's celestial, distance and proper motions tracks were implemented by fitting a seventh degree polynomial to the stream members reported by \citet{Ibata2021} in their Table~1. The distance track was implemented using the distance computed by the authors and readily provided in the table.  The radial velocity track was implemented by fitting a polynomial to the radial velocities of stars reported as probable members in Table~1 of \citet{Ibata2019}. The radial velocity is set to zero outside the range spanned by the member stars. The last InfoFlag bit is set to `2' to reflect that the radial velocity is available but does not span the full length of the track.

\paragraph*{Tri-Pis-B12}

The celestial track for the stream was implemented from the polynomial fit provided by \citet{Bonaca2012} for the Triangulum stream in their Eq. 1:
\begin{eqnarray*}
\delta = -4.4\alpha + 128\fdg5
\end{eqnarray*}

with $\alpha \in [+21^\circ,+24^\circ]$, as explicitly reported by the authors. \citet{Bonaca2012} report a mean heliocentric 
distance of 26~kpc for the stream and a width of $0\fdg2$. 

This coincides with the feature named 'stream a' in \citet{Grillmair2012}.\footnote{Since this reference is a conference proceedings and not a full length paper, we have chosen not to cite it as a discovery reference.}
Soon after the discovery by \citet{Bonaca2012}, \citet{Martin2013} reported the independent discovery of the same structure, based on radial velocity data, naming it the Pisces stream. This detection spans the easternmost $\sim1^\circ$ of the $\sim13^\circ$ track detected by \citet{Bonaca2012}. Based on their spectroscopic metallicity measurement of $\FeH=-2.2$, \citet{Martin2013} find a distance of 35~kpc to the stream, much larger than the 26~kpc found by Bonaca et al's, based on a significantly larger metallicity of $\FeH\sim -1$ estimated from isochrone fitting. Here we will adopt the larger distance estimate of 35~kpc for the full track, as it is based in the more reliable spectroscopic measurement of the metallicity.

The radial velocity track was implemented with the mean of the radial velocities from \citet{Martin2013} because the available data is too noisy and its along-stream span too short to justify higher order fitting. We set this mean value as the radial velocity for track in the range $23\fdg2<\alpha<24\fdg2$ spanned by the observations; outside this range we have set the radial velocity track to zero. The radial velocities reported by \citet{Martin2013} are in the GSR. To revert back to the heliocentric frame and compute the observed radial velocity we have assumed  a solar peculiar velocity with respect to the LSR $(U,V,W)_\odot=(11.1,12.24,7.25)$~km/s \citep{Schoenrich2010} and  $V_{LSR}=220$~km/s \citep{DehnenBinney1998}, since the solar parameters used to convert to the GSR were not reported by the authors.

Since the previous version of galstreams this stream has been referred to as Triangulum-Pisces (in short Tri-Pis), following \citet{GrillmairCarlin2016}. We have kept this naming convention to account for the two independent discoveries.

\paragraph*{TucanaIII-S19}

The stream's celestial and proper motions tracks were implemented by fitting a second degree polynomial using the ICRS data for the stream members reported by \citet{Shipp2019} in their Table~7 (Appendix E). The distance track was implemented by a linear interpolation between the end points and distances reported in \citet{Shipp2018}.The stream's coordinate frame is implemented from the coefficients for the rotation matrix reported by \citet{Li2018}, which makes the origin of the coordinate frame centred on the stream's progenitor.

\paragraph*{Turbio-S18}

The celestial track was implemented as the great circle arc (of minimum length) with end points reported by \citet{Shipp2018}. The distance track was implemented by linearly interpolating the distances reported by the authors for the end points. The stream's coordinate frame is implemented from the coefficients for the rotation matrix reported by \citet{Shipp2019} in their Table~5 (Appendix C).
The proper motion track was implemented using the mean by-eye proper motion measurement for the stream in observed ICRS coordinates reported in \citet{Shipp2019} in their Table 3.
The InfoFlag for the proper motion in this case is thus set to 2 to reflect the proper motion track is available but is an approximation.

\paragraph*{Turranburra-S19}

The stream's celestial and proper motions tracks were implemented by fitting a second degree polynomial using the ICRS data for the stream members reported by \citet{Shipp2019} in their Table~7 (Appendix E). The distance track was implemented using the mean distance of 27.5~kpc from \citet{Shipp2018} for the full track. The stream's coordinate frame is implemented from the coefficients for the rotation matrix reported by \citet{Shipp2019} in their Table~5 (Appendix C).

\paragraph*{Wambelong-S18}

The celestial track was implemented as the great circle arc (of minimum length) with end points reported by \citet{Shipp2018}. The distance track was implemented by linearly interpolating the distances reported by the authors for the end points. The stream's coordinate frame is implemented from the coefficients for the rotation matrix reported by \citet{Shipp2019} in their Table~5 (Appendix C).
The proper motion track was implemented using the mean by-eye proper motion measurement for the stream  in observed ICRS coordinates reported in \citet{Shipp2019} in their Table 3.
The InfoFlag for the proper motion in this case is thus set to 2 to reflect the proper motion track is available but is an approximation.

\paragraph*{Willka Yaku-S18}

The celestial track was implemented as the great circle arc (of minimum length) with end points reported by \citet{Shipp2018}. The distance track was implemented by linearly interpolating the distances reported by the authors for the end points.  The stream's coordinate frame is implemented from the coefficients for the rotation matrix reported by \citet{Shipp2019} in their Table~5 (Appendix C).
The proper motion track was implemented using the mean by-eye proper motion measurement for the stream in observed ICRS coordinates reported in \citet{Shipp2019} in their Table 3.
The InfoFlag for the proper motion in this case is thus set to 2 to reflect the proper motion track is available but is an approximation.

\paragraph*{Ylgr-I21}

The stream's celestial, distance and proper motions tracks were implemented by fitting a seventh degree polynomial to the stream members reported by \citet{Ibata2021} in their Table~1. The distance track was implemented using the distance computed by the authors and readily provided in the table.

\label{s:indiv}

\subsection{Excluded clouds and other structures}\label{s:excluded}

Some structures reported or described in the literature as `streams' are not included in \galstreams: in particular the Helmi streams \citep{Helmi1999,Helmi2017}, S1-S4 \citep{Myeong2018}, Nyx \citep{Necib2020} and Icarus \citep{ReFiorentin2021}. Although correctly named streams due to their coherence in velocity, these structures are either very close to the Sun ($\lesssim 2$~kpc) or even permeate the solar neighbourhood and do not produce localised signatures in the sky suitable to be represented by well-defined celestial or proper motion tracks. Hence, they are not included in the library. Similarly, early accretion events that are now at a high phase-mixing stage such as Gaia-Sausage-Enceladus \citep{Belokurov2018_sausage,Helmi2018_enceladus}; Thamnos \citep{Koppelman2019}; Sequoia \citep{Myeong2019}; Aleph, Arjuna, I'toi and Wukong \citep{Naidu2020}, are also left out. 

Extra-tidal features and incipient tidal tails have been reported in the literature for many globular clusters \citep[see e.g.][]{NiedersteOstholt2010,Balbinot2011,Navarrete2017,CarballoBello2018,Piatti2020,Piatti2020_ngc7099,Kundu2021}. We have chosen to report here only features clearly extending several degrees beyond the tidal radius. For more details and relevant earlier references about tidal tails and extra-tidal features in globular clusters, see discussion in \citet{Sollima2020} and \citet{Ibata2021}. 

Finally, other excluded structures are the Virgo Stellar Stream (VSS) and Virgo Overdensity (VOD). First identified by \citet{Vivas2001,Newberg2002} and \citet{Duffau2006}, there has been a long standing debate about their nature and possible mutual association.
The VOD has a cloud like morphology; while the VSS, although originally thought to be a tidal stream, is shown by \citet{Vivas2016} to have a cloud-like morphology, based on the kinematic identification of new RR Lyrae members of the VSS and new moving groups found in the region. More recently, \citet{Donlon2020} argue the VSS and VOD, together with the Perpendicular and Parallel streams \citep{Weiss2018} and other moving groups reported in the literature \citep[see][for a review]{Donlon2020} are related and were formed by a single event, which they call the Virgo Radial Merger (VRM). Because of their cloud-like morphology the track representation is unsuitable for the VOD and VSS, so they have been excluded from the library.  We have kept the Parallel and Perpendicular streams in the library, with their original names, because for these the stream track representation is adequate.  

A census of cloud-like or non-localised structures would be a useful contribution, but will be better served with a different representation. This is out of the scope of the present work, but worth considering for a separate package.

\section{Comparison of streams with multiple tracks}\label{s:multiple_tracks}

Several stellar streams in the library (\Nmultiple) have multiple realisations of their tracks,  based on discovery or follow-up with either different techniques, tracers and/or surveys, and  with varying degrees of available information. Before discussing global properties of the stream's compilation, we compare here different  tracks available for each stream with multiple realisations and select a `default' track to be shown for a given stream in the library, so that when summary or global visualisations or statistics are made only one instance of each stream is considered. The setting of the default track is done by means of the `On' attribute of each stream track and can be changed by the \galstreams\ user at will. 

\subsection{Streams without known progenitors}

\begin{figure*}
	\includegraphics[width=2.1\columnwidth]{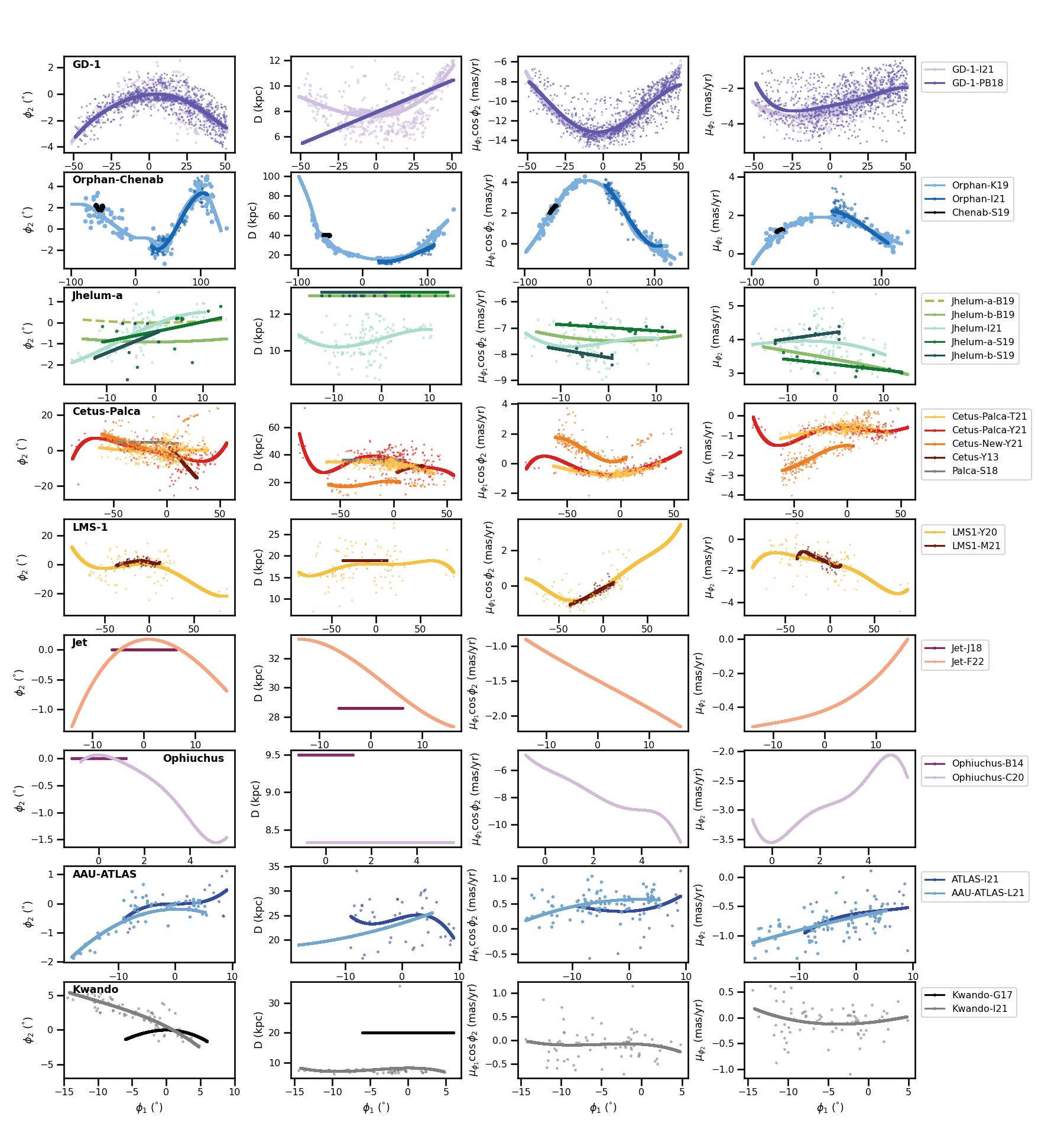}
    \caption{Celestial, distance and proper motion tracks (left to right) for stellar streams with multiple tracks implemented in the library and without known progenitors. From top to bottom: GD-1, Orphan-Chenab, Jhelum, Cetus or Cetus-Palca, LMS-1, Jet, Ophiuchus, AAU-ATLAS and Kwando. The tracks in each row are shown in the stream's reference frame for the top reference listed in that row. Where available, the stars used in each case to define the track implementation are shown in the corresponding colour, as summarised in the legend and referenced in Table~\ref{t:super_summary_table} and Section~\ref{s:indiv}.}
    \label{f:track_compare_notclusters}
\end{figure*}

Figure~\ref{f:track_compare_notclusters} shows the celestial, distance and proper motion tracks for nine streams --GD-1, Orphan-Chenab, Jhelum, Cetus/Cetus-Palca, LMS-1, Jet, Ophiuchus, AAU and Kwando-- without known progenitors. The stars on which the tracks are based are also shown when available from the literature (but these are not included in the library).

\subsubsection{GD-1}

For the GD-1 stream (top row) there is remarkable agreement in the celestial and $\mu_{\phi_1}$ tracks. Minor differences, e.g. in proper motion are clearly much smaller than the average dispersion of the stars they're based on. The most significant difference is observed in the distance track. We have assumed for the PB18 track the linear distance gradient the authors proposed since the stream is too distant for \Gaia~DR2 parallaxes to be useful. The I21 track, on the other hand, includes distances \emph{inferred} by the STREAMFINDER algorithm \citet{Malhan2018} using the observed $\G$, $\Gbp$\ and $\Grp$\ magnitudes from \Gaia, and show a parabolic distance gradient. Given its more detailed inferred (rather than assumed) distance gradient track, we set the GD-1-I21 track as the default for the GD-1 stream.

\subsubsection{Orphan-Chenab}

For the Orphan-Chenab stream (second row) the agreement between the tracks is remarkable, only very minor differences are visible at $\phi_1\sim100\degr$ in distance with the I21 track is $\sim6$~kpc below the K19 one and at $\phi_1\sim20\degr$ in $\mu_{\phi_2}$ with the I21 track having slightly larger proper motion than K19. The agreement in distance is particularly noteworthy, as K19 is derived from photometric distances of RR Lyrae stars and I21 are distances inferred by STREAMFINDER based on the star's $\mathrm{G}$ magnitude, $\mathrm{G_{BP}-G_{RP}}$ colour and a simple stellar population model \citep{Malhan2018}. Given its much larger extent we set the Orphan-K19 track as the default for the Orphan-Chenab stream. 

We have also included Chenab in this comparison, recognised by  \citet{Koposov2019} to be a part of Orphan's southern tail, as clearly illustrated by the agreement of Chenab's and Orphan's celestial, distance and proper motion tracks.  Hence, we have adopted the Orphan-Chenab name for the stream and take Chenab to be a part of it, set as `Off' in the library.

\subsubsection{Jhelum}

Jhelum is quite a complex case. \citet{Bonaca2019} first recognised two separate branches in the Jhelum stream by using deep DES photometry combined with \Gaia~DR2 proper motions. The data revealed the two branches as separate overdensities in the sky, but with indistinguishable kinematics. Soon after, also using DES and \Gaia~DR2, \citet{Shipp2019} identified two components in Jhelum, this time in proper motion space and with little difference in the spatial distribution of the two components. Exactly the reverse as \citet{Bonaca2019}. More recently, \citet{Ibata2021} recover a stream they identify as Jhelum using STREAMFINDER with \Gaia~EDR3.

Figure~\ref{f:track_compare_notclusters} (third row) shows this quite clearly: the two tracks from S19 (auto-computed by \galstreams\ via polynomial fitting of the member stars) almost coincide in the sky and are clearly separated in both proper motion components. The two tracks from B19 are separated by nearly a degree in the sky and overlap completely in proper motion, so only one line is visible in the last two panels of the figure. In both components, the B19 proper motion track lies approximately in between the two S19 tracks. In the sky, the two nearly parallel S19 tracks cross the B19 ones at a slight angle (the aspect ratio of the figure exaggerates the inclination).   The I21 celestial track agrees well with the S19 Jhelum-a/b tracks, while the proper motion tracks agrees with B19.  Another aspect to keep in mind is the role of errors and the intrinsic dispersion of the stream in proper motion. B19 note the dispersion in proper motion for both Jhelum components to be relatively large and comparable to the proper motion uncertainty $\sim0.7$ to $1.2$~$\mathrm{mas\,yr^{-1}}$; the I21 members in Figure~\ref{f:track_compare_notclusters} also show a dispersion of $\sim1\,\mathrm{mas\,yr^{-1}}$.  

Given these two scenarios, it seems likely that if the two Jhelum components shared a single but long and sinuous track (the B19 case) in the proper motion plane (reflex corrected, Figure~\ref{f:track_jhelum_pm1pm2}), this could be mistaken by a Gaussian Mixture Model as two (or more) proper motion components, which statistically would be a random mixture of the two spatial components (the S19 case). If this were the case, it would be akin to the bifurcation of the Sagittarius stream, in which two spatially distinct components appear to have no difference in their proper motion tracks \citep[see e.g.]{Ramos2022_bif,Ramos2021}, \neww{as mentioned also by \citep{Shipp2019}}. \neww{The addition of radial velocity information as well as a}  further increase in the precision of proper motions in the next \Gaia~Data Release will probably elucidate this. In the mean time, as we find the B19 scenario more likely to explain the three sets of observations (B19, S19 and I21), we will set these as the default tracks for the Jhelum-a/b  components of the stream, but warn the reader to take this with a grain of salt. 

\subsubsection{Cetus, Palca and Cetus-Palca}

\begin{figure}
	\includegraphics[width=\columnwidth]{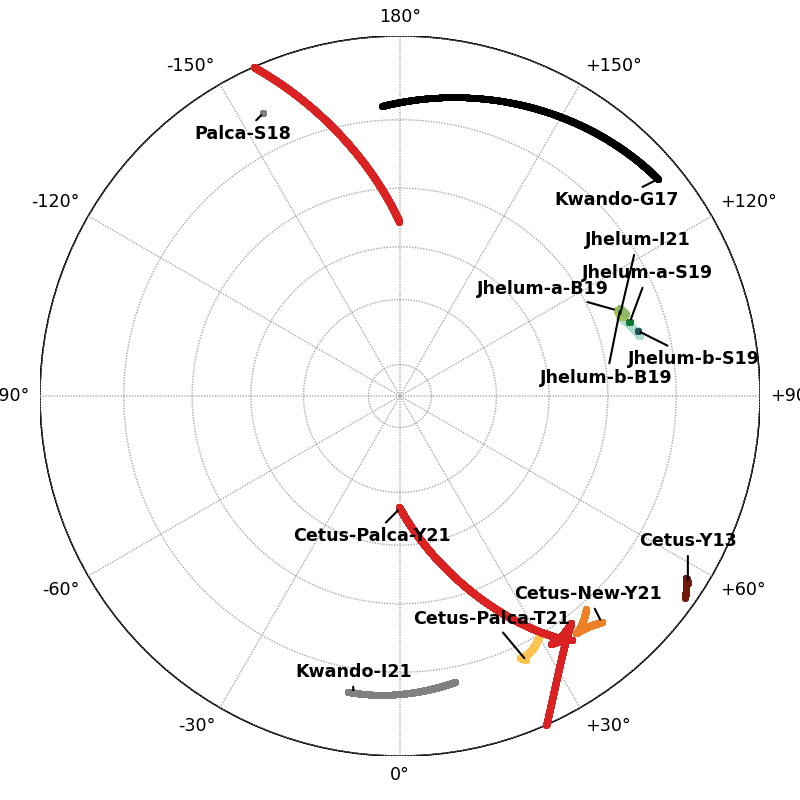}
    \caption{Heliocentric pole tracks for the multiple tracks available for the Kwando, Jhelum and Cetus/Cetus-Palca streams, in a north polar azimuthal projection in Galactic coordinates. The colour coding is the same as in Figure~\ref{f:track_compare_notclusters}.}
    \label{f:poletrack_compare_notclusters}
\end{figure}

\begin{figure}
	\includegraphics[width=0.8\columnwidth]{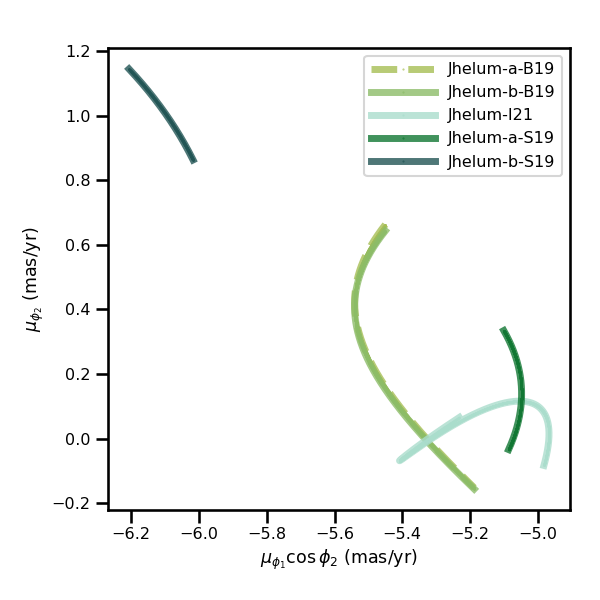}
    \caption{\neww{Proper motion $\mu_{\phi_2}$ vs $\mu_{\phi_1}\cos{\phi_1}$ (vector-point  diagram)} in the Jhelum-a-B19 coordinate frame for the multiple tracks available for the Jhelum stream components. \neww{Proper motions have been corrected for the solar reflex motion in order to ensure the solar contribution is taken into account and not responsible for the observed differences (or the lack thereof).} The colour coding is the same as in Figure~\ref{f:poletrack_compare_notclusters}}
    \label{f:track_jhelum_pm1pm2}
\end{figure}

For the Cetus stream, Figure~\ref{f:track_compare_notclusters} (fourth row) shows four different tracks: the original discovery track Y13, Cetus-Palca-T21, Cetus-Palca-Y21 and Cetus-New, a new branch identified by \citet{Yuan2021}. The track for Palca is also shown, as T21 and Y21 claim the Palca and Cetus stream reported in their studies to be related.
Their corresponding pole tracks are also shown in Figure~\ref{f:poletrack_compare_notclusters}. 
The comparison between the celestial tracks shows a complex scenario. Both the $\phi_1-\phi_2$ plot and the pole tracks show that the Y21, T21 and Palca tracks roughly cross the same are of the sky, but there are systematic differences between the two. The T21 track is parallel to Palca, separated by $\sim 5\degr$. The Y21 track is inclined with respect to both T21 and Palca, also shown  by the fact the I21 pole track --despite being extensive-- does not contain either the Palca pole or the T21 pole track, which it only barely overlaps. The better part of these track's poles (Palca, T21, Y21 and Cetus-New), however, occupy a reasonably well-defined locus in Figure~\ref{f:poletrack_compare_notclusters}, confined to an area with $\sim 10\degr$-radius. The overall good agreement observed between the Y21 and T21 distance and proper motion's tracks might suggest they are indeed tracing the same feature, and the differences in their celestial (and pole) tracks might have to do with sampling issues and probable contaminants at the ends of the streams introducing spurious oscillations in the Y21 tracks. The case of Cetus-New seems more clear, as the track is evidently distinct in distance and both proper motions. Finally, The Cetus-Y13 celestial track looks highly inclined with respect to the others and with its pole separated by more than $30\degr$ from the area where the Cetus-Palca poles cluster. 
Kinematic data could have helped disentangle this, but Cetus-Y13 has radial velocity information but no proper motion data, while the Cetus-Palca Y21 and T21 tracks have proper motion but no radial velocity data. 
Since a definitive association between several of these tracks seems unclear, we will keep the Cetus-Y13 track as the default for Cetus; the T21 as the default for the main branch of Cetus-Palca, as it is the most stable track with the best behaved pole track; and Cetus-New as a separate branch under that name. The Palca track is also kept as a separate stream.

\subsubsection{LMS-1}

\begin{figure}
	\includegraphics[width=\columnwidth]{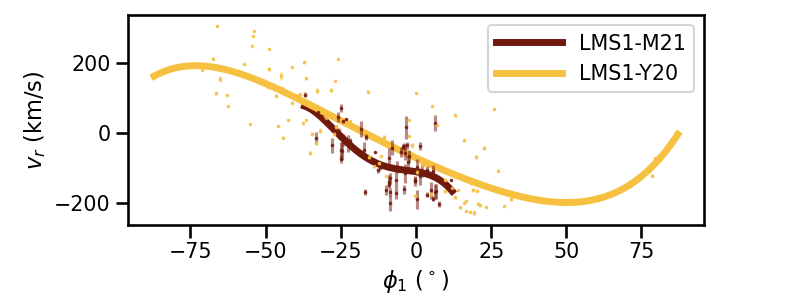}
    \caption{Radial velocity tracks for the two implementations available for the LMS-1 stream, in the LMS-1-Y20 stream reference frame. The colour coding is the same as in Figure~\ref{f:track_compare_notclusters}. Individual members from M21 and Y20 are shown, including error bars in radial velocity, with the same colour coding.}
    \label{f:vrad_compare_lms1}
\end{figure}

For LMS-1 there are two available tracks, the discovery track LMS1-Y20 and the follow-up by M21, shown in Figure~\ref{f:track_compare_notclusters} (penultimate row). There is also radial velocity information for both, shown in Figure~\ref{f:vrad_compare_lms1}. Both tracks show excellent agreement in every coordinate, sky, distance, proper motions and radial velocities and, although there is overlap in $\phi_1$, there are no objects in common between the two studies. Since the Y20 track spans a much larger range of $\phi_1$, we set this as the default for the LMS-1 stream.

\subsubsection{Jet}

For Jet there are two available tracks, the discovery track Jet-J18 and the follow-up by F22, shown in Figure~\ref{f:track_compare_notclusters} (sixth row) which includes proper motion data and extends the stream's track by $\sim20\degr$ in length. Although the new F22 track differs from the great-circle discovery track, this discrepancy is only by $\sim0\fdg1$, so there is good agreement. There is, however, a significant discrepancy in the distance, with the F22 track being systematically more distant by up to $\sim4$~kpc. Given the much larger extent and proper motion data available in the F22 track, we set this as the default for the Jet stream.  

\subsubsection{Ophiuchus}

\neww{
Similarly to Jet, there are two available tracks for Ophiuchus: the discovery track Ophiuchus-B14 and the follow-up by C20, shown in Figure~\ref{f:track_compare_notclusters} (seventh row) which includes proper motion data and extends the stream's track by $\sim10\degr$ in length. The celestial tracks coincide very well. In the distance tracks there is a $\sim1$~kpc discrepancy, with the B14 track having the more distant estimate. Given the much larger extent and proper motion data available in the C20 track, we set this as the default for the Ophiuchus stream.  
}

\subsubsection{AAU-ATLAS}

\neww{
The two available tracks for the ATLAS branch of the ATLAS-AliqaUma stream are shown in Figure~\ref{f:track_compare_notclusters} (penultimate row), together with members from I21 and spectroscopically confirmed members from \citet{Li2021}. The two tracks show excellent agreement in the sky, distance and proper motion spaces. The AAU-ATLAS-L21 track extends $\sim5\degr$ further than the ATLAS-I21 track\footnote{Note that the ATLAS-I21 track had initially been dubbed V\'id in the preprint version of I21.} and includes published radial velocities, hence it is set as the default for the ATLAS branch of the AAU stream. }

\subsubsection{Kwando}

\neww{
The Kwando-G17 track from the \citet{Grillmair2017_south} discovery paper is compared against the Kwando-I21 track in the last row of Figure~\ref{f:track_compare_notclusters}. Their heliocentric pole tracks are compared in Figure~\ref{f:poletrack_compare_notclusters}. There is significant disagreement between the two celestial and distance tracks. In the sky, the I21 track intersects the G17 one at a significant inclination, confirmed by the difference in their (heliocentric) pole tracks of well over $20\degr$ in pole latitudes. In distance, the I21 track is located at half the distance of the G17 track. Kwando is one of several streams found using photometry alone pre-GaiaDR2 \citep[e.g. streams from, but not limited to,][]{Bernard2016,Grillmair2017,Grillmair2017_south,Mateu2018} that have lacked further follow-up (see discussion in Sec.~\ref{s:gprops_info}). Further studies will be required to clarify the two (separate) issues of robustness of the Kwando stream and whether the association of Kwando-I21 to the Kwando stream is correct or if it corresponds to a new stellar stream\footnote{The Kwando-I21 track had initially been dubbed C-9 in the preprint version of I21.}. In the mean time, we have kept the G17 as the default for the Kwando stream, because the Kwando-I21 association with it is unclear. 
}

%}

\subsection{Streams associated with surviving globular clusters}

The celestial, distance and proper motion tracks for stellar streams that are associated with surviving globular clusters and have multiple track realisations, are shown in Figures~\ref{f:track_compare_clusters1} and \ref{f:track_compare_clusters2}. Figure~\ref{f:track_compare_clusters1} shows, from top to bottom, the multiple tracks for clusters Pal~5, NGC~3201-Gj\"oll, M92, M5 and M68-Fj\"orm and Figure~\ref{f:track_compare_clusters2} shows the tracks for NGC~288, NGC~2298, NGC~5466, M2 and $\omega$~Cen-Fimbulthul. 
%Although extra-tidal features have been reported in the literature for many globular clusters \citep[see e.g.][]{Leon2000,CarballoBello2018,Piatti2020}, we have chosen to report here only features clearly extending several degrees beyond the tidal radius. For more details and relevant references about this, see discussion in \citet{Sollima2020} and \citet{Ibata2021}.  

\subsubsection{Pal~5}

For Pal~5, the PW19 and I21 tracks have very similar extents in $\phi_1$ and coincide remarkably well in the sky. The S21 track extends the cluster's leading tail by $\sim7\degr$ compared to the previous ones, however, it lacks distance and proper motion information. In distance and proper motions there is overall agreement between the PW19 and I21 tracks around the position of the cluster ($\phi_1\sim0\degr$) and the leading tail ($\phi_1>0\degr$), but they start to differ in proper motions in the trailing tail at $\phi_1\lesssim-5\degr$. 
The PW19 is based on RR Lyrae stars whose distances are more precise and require no inference involving the orbit, as do the STREAMFINDER distances in I21. The PW19 data is also more constraining as revealed by the lower dispersion in the distance and proper motion tracks. Hence, we set the PW19 track as the default for Pal~5. We note, however, that since the two studies are based on different stellar populations these differences could be physically meaningful.

\subsubsection{NGC~3201-Gj\"oll}

For NGC~3201 the two available tracks (P21 and I21, second row of Fig.~\ref{f:track_compare_clusters1}) coincide very well, in the sky, distance and in both proper motions. The I21 track's extent is of about $10\degr$ and centred around the cluster, compared to $>100\degr$ length for the P21 track. In their work, \citet{Palau2021} recognised the Gj\"oll stream \citep{Ibata2019} as part of the NGC~3201 tidal tail identified with their algorithm. An excellent agreement is clearly seen between the two celestial, distance and proper motion tracks in the figure.  We will set the much longer P21 track as the default for the NGC~3201-Gj\"oll stream. The Gj\"oll track is set as `Off' and ascribed to NGC~3201-Gj\"oll.

\subsubsection{M92}

For M92 the S20, T20 and I21 tracks (third row in Fig.~\ref{f:track_compare_clusters1}) intersect the cluster position in the sky. I21 and S20 coincide at  positive $\phi_1$, but the S20 track departs from the I21 at negative $\phi_1$. Still, it is a minor deviation considering it is reasonably within the dispersion of the I21 track stars. The T20 track does not agree with either I21 or S20 --even considering the dispersion of I21 members-- with the largest mismatch again at negative $\phi_1$. Since the S20 and T20 tracks contain no kinematic information, the I21 is set as the default for M92.

\subsubsection{M5}

For M5 (fourth row in Fig.~\ref{f:track_compare_clusters1}) there are three available tracks: S20 limited to the celestial track and intersecting the cluster position; and G19 and I21, both with full sky, distance and proper motion data but neither crossing the cluster's position. The celestial I21 track agrees with G19 for the most part, with the angle changing somewhat at the positive $\phi_1$ end; but note that the aspect ratio of the figure is probably exaggerating the differences.  The S20 interestingly could be the extension of the stream at negative $\phi_1$. To better judge the differences between the three celestial tracks, we compare their corresponding (heliocentric) pole tracks, as shown in Figure~\ref{f:poletrack_compare_clusters} in a north polar azimuthal projection. This shows the I21 and G19 poles overlap for the most part, but the S20 does not, indicating a completely different --and highly varying-- orbital plane. This casts doubt as to the likelihood of the S20 track truly being associated to the M5 stream. In the proper motion tracks there seem to be some differences between I21 and G19, but being so much shorter in comparison to G19 the more meaningful comparison would be of its mean proper motion which does seem consistent.
The distance tracks, however, are entirely inconsistent with I21 located a nearly half the distance of the G19 track, \neww{but, although separated more than $30\degr$ away from the cluster, much closer to the \citet{Baumgardt2021} distance for the cluster used in the S20 track}. Given its much larger extent we set the G19 track as the default for the M5 cluster, but recommend to use its distance information with caution.

\subsubsection{M68-Fj\"orm}

\begin{figure*}
	\includegraphics[width=2.1\columnwidth]{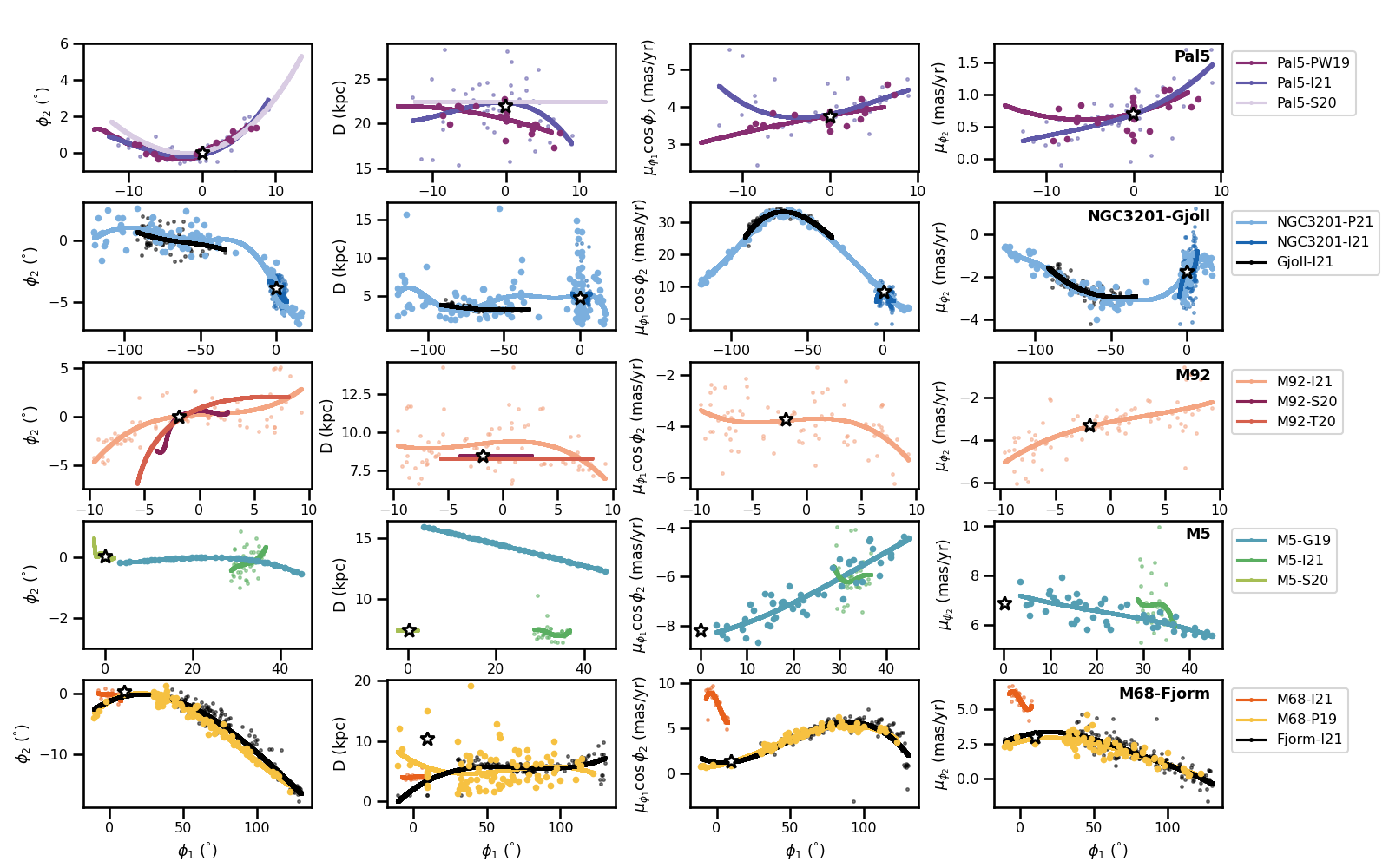}
    \caption{Celestial, distance and proper motion tracks (left to right) for stellar streams with multiple tracks implemented in the library, that are associated with globular clusters. From top to bottom: Pal~5, NGC~3201-Gj\"ol, M92, M5 and M68-Fj\"orm. The tracks are shown in the stream's reference frame for the top reference listed for each row. Globular cluster positions and proper motions from \citet{Vasiliev2021} and \neww{ (mean) distances from \citet{Baumgardt2021}} are shown with a black star in each panel. Where available, the stars used in each case to define the track implementation are shown in the corresponding colour, as summarised in the legend and referenced in Table~\ref{t:super_summary_table} and Section~\ref{s:indiv}.} 
    \label{f:track_compare_clusters1}
\end{figure*}

For M68 (last row in Fig.~\ref{f:track_compare_clusters1}),
the association of Fj\"orm and M68,  already  pointed out by \citet{Palau2019}, is also seen here in the overall excellent agreement between the tracks \neww{in the sky and both proper motion components. However, note the disagreement between the two distance tracks at $\phi_1\lesssim30\degr$, where the I21 predicts a much shorter distance than P19. We caution that in both samples, there is a sharp decrease in the number of member stars at that $\phi_1$ range precisely. It is also worth noting the disagreement between both tracks and the cluster distance from \citet{Baumgardt2021}. On the other hand, at $\phi_1\gtrsim30\degr$ the two distance tracks agree well, despite the caveat mentioned in Sec.~\ref{s:indiv} about P19 distances having been computed from the reciprocal of the reported parallax. }

\begin{figure}
	\includegraphics[width=0.9\columnwidth]{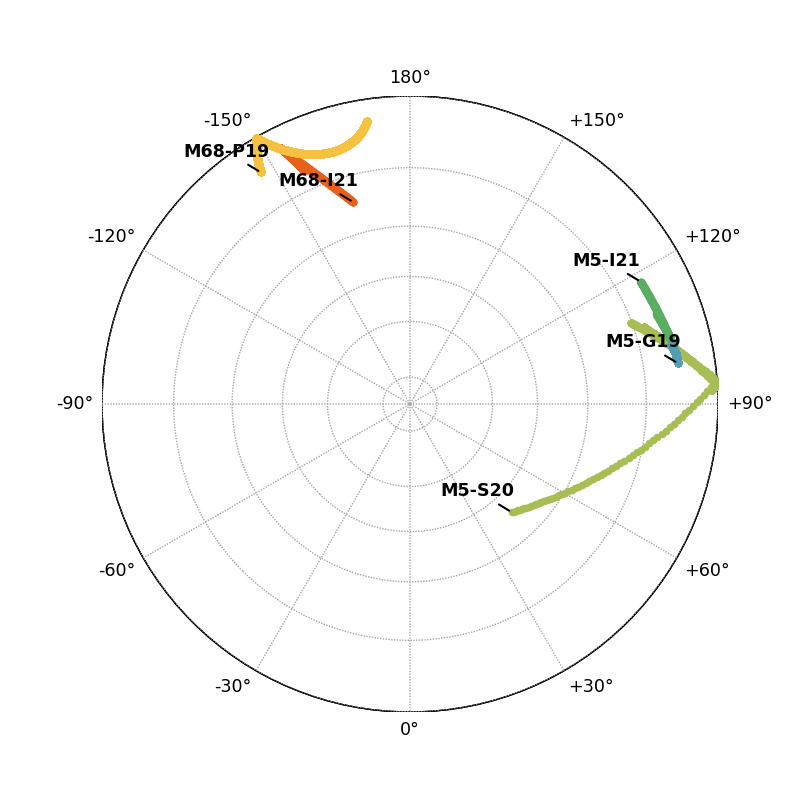}
    \caption{Heliocentric pole tracks for M5 and M68, in a north polar azimuthal projection in Galactic coordinates. The colour coding is the same as in Figure~\ref{f:track_compare_clusters1}. }
    \label{f:poletrack_compare_clusters}
\end{figure}

When compared to the M68-I21 track, there is significant disagreement. The Fj\"orm-I21 celestial track (based on the candidates reported in their Table 1) and the M68-P19 track both intersect the cluster position in the sky, but M68-I21  crosses the P19 track at an angle. Again, to avoid being mislead by the aspect ratio of the plot, we compare the corresponding pole tracks in Figure~\ref{f:poletrack_compare_clusters}. This shows the two tracks have very minor overlap, indicative of significantly different orbital planes, further supported by systematic differences observed between the two tracks in both proper motion components. The M68-I21 track has a significantly larger proper motion (over 4 times larger) than the cluster even at similar $\phi_1$, where some overlap would be expected. 

These discrepancies cast doubt into the association of the M68-I21 track with the M68 cluster and could be pointing to this feature being a distinct new stream. \neww{Radial velocity information would help settle the matter, but is currently only available for the Fj\"orm-I21 track, but not for the M68-P19 or the M68-I21 tracks.} The M68-P19 track is set as the default for the M68-Fj\"orm stream. For naming clarity, the Fj\"orm track is set as "Off" and ascribed to the M68-Fj\"orm stream, given its perfect agreement with the M68-P19 track. The M68-I21 track is set as "Off" and not ascribed to M68-Fj\"orm \neww{until further information is available} and the proper motion discrepancy is resolved.

\begin{figure*}
	\includegraphics[width=2.1\columnwidth]{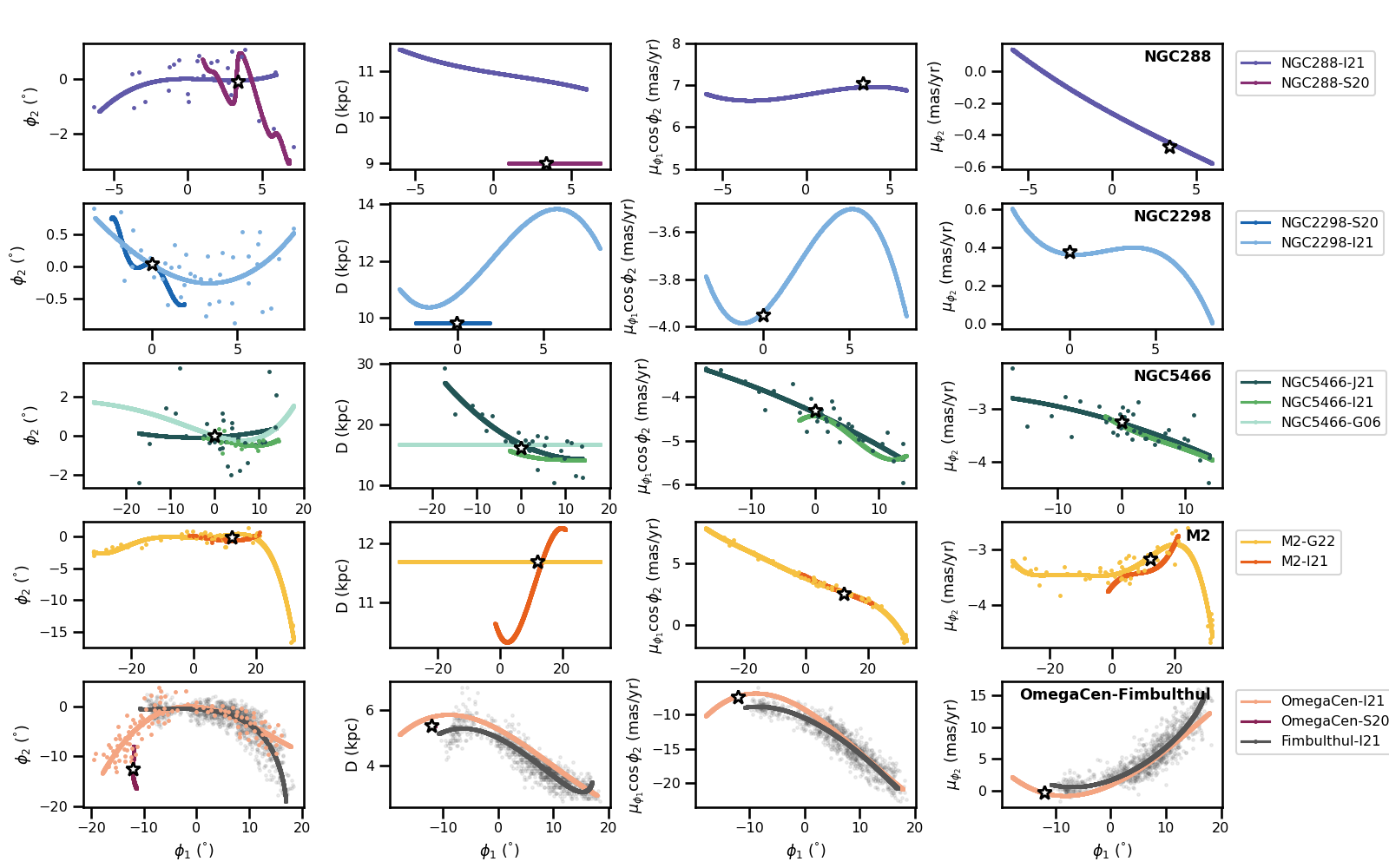}
    \caption{Celestial, distance and proper motion tracks (left to right) for stellar streams with multiple tracks implemented in the library, that are associated with globular clusters. From top to bottom: NGC~288, NGC~2298, NGC~5466, M2 (NGC~7089) and $\omega$~Cen-Fimbulthul (NGC~5139). The tracks are shown in the stream's reference frame for the top reference listed for each row. Globular cluster positions and proper motions from \citet{Vasiliev2021} and \neww{ (mean) distances from \citet{Baumgardt2021}} are shown with a black star in each panel. Where available, the data used in each case to define the track implementation is shown in the corresponding colour, as summarised in the legend and referenced in Table~\ref{t:super_summary_table} and Section~\ref{s:indiv}.}
    \label{f:track_compare_clusters2}
\end{figure*} 

% \subsubsection{NGC~288, NGC~2298 and NGC~5466}

% For these three clusters there are tracks available from I21 and S20, as shown in the first three rows of Figure~\ref{f:track_compare_clusters2}. In all cases the I21 tracks trace the stream for a longer span and include proper motion data; S20 tracks are limited to information in the plane of the sky. The two sets of tracks for NGC~2298 and NGC~5466 show reasonable agreement, within the dispersion. The case of NGC~288 is interesting, the S20 track is much more wiggly and has a drop towards negative $\phi_2$ that seems unphysical at first, but coincides with the I21 stars in that region. The analysis of kinematic data in the S20 overdensity track would help confirm their membership to the cluster's tails. \neww{Note also the significant difference in the NGC288-I21 distance and the \citet{Baumgardt2021} distance estimate for the cluster, of nearly 2~kpc. In the other two cases there is also a slight discrepancy of $<1$~kpc}.

% In these three cases the I21 track is set as the default for each cluster, given the more extensive coverage and availability of proper motion data.

\subsubsection{NGC~288, NGC~2298}

For these two clusters there are tracks available from I21 and S20, as shown in the first two rows of Figure~\ref{f:track_compare_clusters2}. In both cases the I21 tracks trace the stream for a longer span and include proper motion data; S20 tracks are limited to information in the plane of the sky. The two sets of tracks for NGC~2298 show reasonable agreement, within the dispersion. The case of NGC~288 is interesting, the S20 track is much more wiggly and has a drop towards negative $\phi_2$ that seems unphysical at first, but coincides with the I21 stars in that region. The analysis of kinematic data in the S20 overdensity track would help confirm their membership to the cluster's tails. \neww{Note also the significant difference in the NGC288-I21 distance and the \citet{Baumgardt2021} distance estimate for the cluster, of nearly 2~kpc. In the other two cases there is also a slight discrepancy of $<1$~kpc}.

In both cases the I21 track is set as the default for each cluster, given the more extensive coverage and availability of proper motion data.

\subsubsection{NGC~5466}

\neww{There are three different tracks available for NGC~5466: the discovery track from G06, with no distance gradient or proper motion data, and the I21 and J21 tracks both with distance and proper motion data. The I21 and J21 tracks agree very well in all coordinates, with the J21 track having a much larger extension (by $\sim20\degr$, particularly tracking the stream's tail in negative $\phi_1$, not present in I21). The G06 celestial track shows a better agreement with the previous two in $\phi_1>0\degr$, while at $\phi_1<0\degr$ the disagreement is larger but never exceeds $\sim2\degr$. As noted in Sec.~\ref{s:indiv}, the G06 distance was assumed to be the mean cluster distance and therefore, the disagreement is expected. Note also how the I21 track is limited to $<10$~kpc, a distance limit of the STREAMFINDER algorithm, as we will discuss in Sec.~\ref{s:gprops_distance} ). For this stream, due to its length and 5D data availability, the J21 is set as the default.
}

\subsubsection{M2 (NGC~7089)}

As discussed by \citet{Grillmair2022}, there is excellent agreement between the I21 and G22 tracks for M2, as shown in the fourth row of Figure~\ref{f:track_compare_clusters2}. The G22 extends the observed length of the tail by over $30\degr$ and includes proper motion data and is set as the default for this cluster. We caution that at this point it does not include information about the distance gradient, which is predicted by \citet{Grillmair2022} to be significant based on integration of the cluster's orbit. This information \emph{is} available in the I21 track showing a gradient of $\sim 2$~kpc in its $20\degr$ span.

\subsubsection{$\omega$~Cen-Fimbulthul }

The bottom row of Figure~\ref{f:track_compare_clusters2} shows the I21 track for the $\omega$~Centauri (NGC~5139) cluster constructed from members published in \citet{Ibata2021} and compared to the Fimbulthul stellar stream first found by \citet{Ibata2019} and identified by
\citet{Ibata2019_OCen} as the long sought stream of $\omega$~Cen. The agreement between these two tracks is, therefore, expected and is shown here only to illustrate the coverage of the new track (I21) with respect to Fimbulthul and the cluster itself. As the plot shows, the cluster is separated several degrees from the celestial track,  while it does coincide with the track in both proper motion components.  The S20 track, with only celestial data, is also shown in the first panel, it joins with the cluster, by construction as the search in \citet{Sollima2020}  was targeted around globular clusters, and seems disjoint from the I21 track. In this case, again we set the I21 track as the default for the $\omega$~Cen-Fimbulthul stream and set Fimbulthul as `Off' and not considered a separate stream any further.

\section{Global Properties of the System of Stellar Streams}\label{s:global_props}

In this section we will discuss global properties of the \galstreams\ library. Table~\ref{t:super_summary_table} provides a summary of the main characteristics of the total \Ntracks\ stream tracks implemented. As discussed in the previous section, a single track was selected as the default one for each of the \Nunique\ unique stellar streams in the library, these are indicated as `On' in the table.

\subsection{Available Information}\label{s:gprops_info}

\begin{figure}
	\includegraphics[width=\columnwidth]{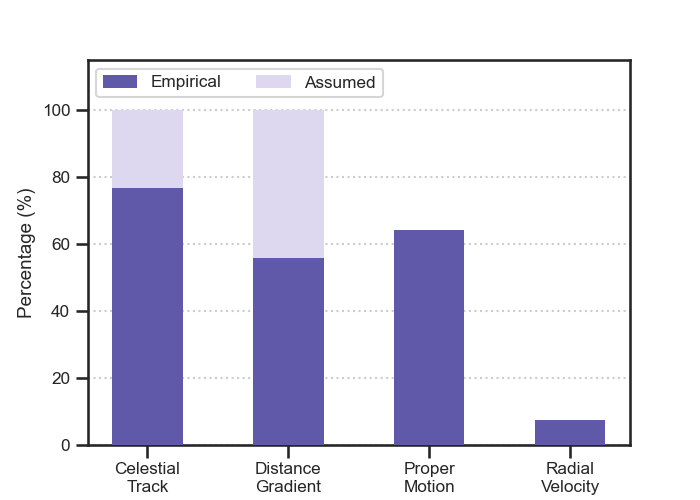}
    \caption{Summary of available information for the \Nunique~stellar streams implemented in the library. The histogram shows, from left to right, the percentage of streams with an empirically determined celestial track (not assumed to be a great-circle), distance gradient information, proper motion and radial velocity tracks.}\label{f:full_lib_info_hist}
\end{figure}

Figure~\ref{f:full_lib_info_hist} summarises the available information for the \Nunique~stellar streams implemented in the library. In order to be included in the current version, a stream must have a celestial track and minimal distance information (e.g. a mean distance or distance at an anchor point). However, the degree of detail with which this is reported varies. In its first version, most of the streams in \galstreams\ were assumed to be great-circles and their tracks implemented under this assumption. In the current version, over three quarters of the streams in the library (77\%) have a detailed celestial track, the remainder are great-circles by construction. 

Somewhat surprisingly, only about half (55\%) the streams have sufficiently detailed distance information to implement a distance gradient in the track, for the remaining 45\% only a mean distance is provided. This last percentage does \emph{not} include cases in which a constant distance is \emph{observed} to be a good approximation for the stream, these cases are indicated in the library as having distance gradient information. This deficit is partly due to a lack of follow-up on many streams (e.g. PS1-A/E, Corvus, Molonglo, Murrumbidgee, etc.) detected before \Gaia~DR2, and can also be justified for some of the shorter streams (e.g. Pal 15, Eridanus), but for the majority it reflects how challenging and observationally demanding it remains to estimate distances for these relatively low-contrast structures when outside the \Gaia-sphere. 

Conversely, well over half the streams (64\%) do have detailed proper motion tracks, an achievement possible thanks to the \Gaia~DR2 and EDR3 releases \citep{GaiaCol_2018_DR2_survey,GaiaCol2021_EDR3_survey}. Most  streams that do not have proper motion information were discovered prior to the \Gaia\ data releases and have not been revisited, so there is a clear opportunity to complement the available information for the majority of those within the reach of \Gaia. Finally, as was natural to expect, only a small portion (7\%) of the streams have radial velocity information available. 

\subsection{Celestial Distribution}\label{s:gprops_spatial}

Figure~\ref{f:full_lib_galactic_aitoff} shows in a Mollweide projection map in Galactic coordinates, the celestial tracks of the \Nunique~unique stellar streams (i.e. no repetition), out of the total of \Ntracks\ tracks implemented in the \galstreams\ library. The map clearly shows the observational bias against the detection of stellar streams near the Galactic plane.  \neww{This bias is anticipated and expected as current detection techniques will struggle to disentangle the faint signature of a stellar stream against the vastly more numerous disc background, and subject to crowding effects and the high and non-uniform extinction affecting photometric depth and completeness}. Only a few streams -Sagittarius, Orphan-Chenab, LMS-1 and NGC~3201- have been unequivocally traced on both sides of the plane, and all cross it almost perpendicularly, a significant off-plane component of motion facilitating detection against the disc background.

\begin{figure*}
	\includegraphics[width=2\columnwidth]{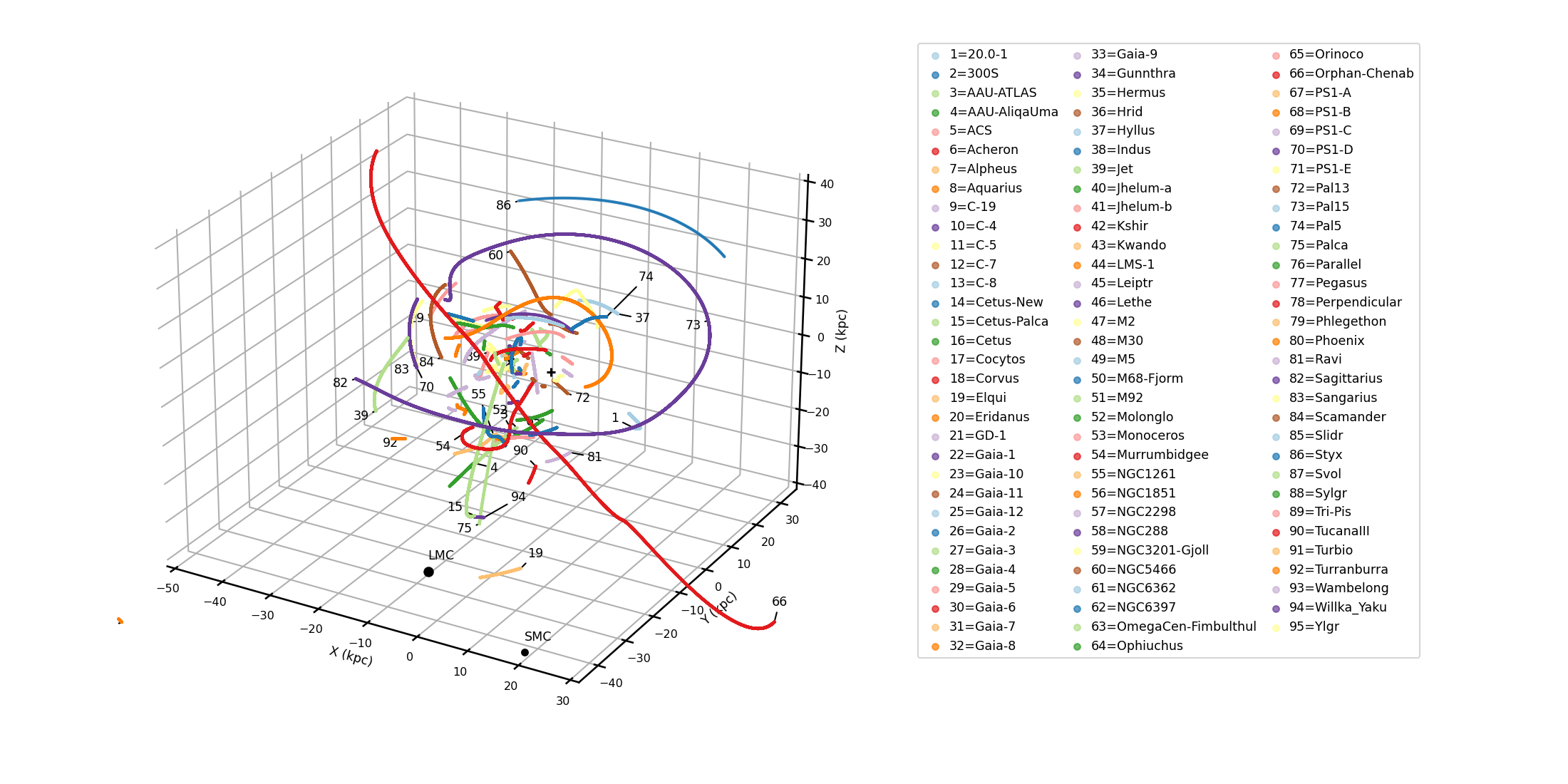}
    \caption{3D view of the stellar streams in \galstreams\ in cartesian  Galactocentric coordinates. The location of the Galactic Centre is indicated with a plus sign. The numeric ID labels are shown in the figure only for streams more distant than 25~kpc. }\label{f:full_lib_galactic_3D}
\end{figure*}

A 3D view in cartesian Galactocentric coordinates is shown in Figure~\ref{f:full_lib_galactic_3D}. The bias in the distribution in favour of nearby streams ($\lesssim20$~kpc) is evident in the clustering observed around the Sun's position (-8.12~kpc). Another apparent clustering of several streams is found in the direction of the Magellanic Clouds, but at shorter distances. The streams in this region are  Vid, NGC1261 \citep{Ibata2021}, Cetus/Cetus-Palca \citep{Yam2013,Thomas2021}, Willka Yaku,   Tucana III, Atlas-AliqaUma, Ravi \citep[DES,][]{Shipp2018}, Murrumbidgee, Molonglo, Kwando, Orinoco \citep[SDSS,][]{Grillmair2017_south}. The group is a mix of streams discovered by different groups, but all except for the first two, discovered with STREAMFINDER, involved an implementation of matched-filtering. It will be interesting to explore to what extent is this clustering real. \citet{Bonaca2021} has already identified groups of streams with a common origin by means of orbital clustering, but only 2 out of the streams in this group are included in that analysis (Atlas-AliqaUma and Ravi, found to be associated with differnt groups) since there is no proper motion information available for the SDSS streams and the data for the rest was published afterwards.

\subsection{Distance Distribution}\label{s:gprops_distance}

\begin{figure}
	\includegraphics[width=\columnwidth]{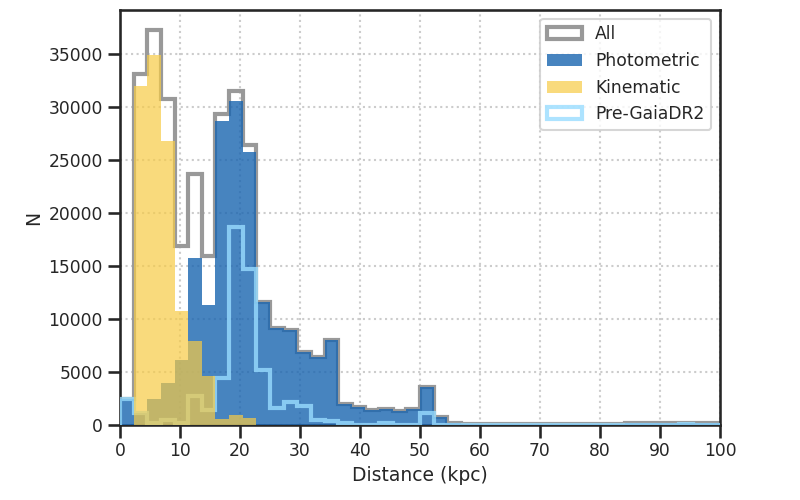}
    \caption{Heliocentric distance histogram for all \Nunique~unique stream tracks in the library \neww{(gray)}, \neww{each stream uniformly populated along its track}. The distributions for streams detected based on kinematic (light yellow) or photometric (dark blue) information are shown separately. The distribution for streams detected prior to Gaia~DR2 is also shown (light blue open histogram) to highlight their bias toward large distances.} \label{f:full_lib_distance_hist}
\end{figure} 

Figure~\ref{f:full_lib_distance_l} showed a plot of all stream's distance tracks in the library as a function of Galactic longitude. This information is also shown summarised in Figure~\ref{f:full_lib_distance_hist}, in which all (unique) celestial tracks have been aggregated into a single (heliocentric) distance distribution, \neww{each having been populated with a uniform spacing along the track. This figure essentially corresponds to a histogram of  Figure~\ref{f:full_lib_distance_l} along the y-axis}.
These two figures show that the vast majority of streams are located within 30~kpc (95), with only five streams (Styx, Cetus-Palca, Elqui, Orphan-Chenab and Sagittarius) having part or all of their track in the 30-40~kpc range and Eridanus being the farthest located at a mean $\sim95$~kpc. 

\neww{Figure~\ref{f:full_lib_distance_hist} presents two clear peaks at $\sim7$~kpc and $\sim20$~kpc. This observed bimodal distribution is produced by the combination of two effects: the real distance distribution of stellar streams plus the effects of observational biases involved in the main two types of methods used in the identification of the streams. First, the distribution of streams in distance is expected to decrease with distance as tidal forces dwindle and systems are less prone to losing stars \citep[see e.g. Figs. 11 and 14 of][]{Mateu2017}. On the other hand, are the selection effects of the methods used in finding the stellar streams. In this case, selection effects are quite different depending on whether or not kinematic information was required by the detection method. This is illustrated in Figure~\ref{f:full_lib_distance_hist}, where the distributions of streams detected based on kinematics \citep[those from][]{Malhan2018,Palau2019,Ibata2019,Grillmair2019,Palau2021, Jensen2021,Ibata2021,Grillmair2022} or on photometry are shown separately (light yellow and dark blue histograms respectively). The figure shows that for streams discovered using kinematic information, the selection function decreases with distance and has a sharp drop-off at $\sim12$~kpc. This drop-off is driven by the limits of the volume in which Gaia's proper motions are precise enough for the methods to work. Combined with the decrease in distance, this produces the first peak at shorter distances ($<10$~kpc).} 

\neww{Conversely, streams detected photometrically clearly dominate at large distances ($\gtrsim 15$~kpc). The distribution shows a peak at $\sim20$~kpc\footnote{This prevalence of streams at $\sim20$~kpc was already noticeable in \galstreams\ v0.1, in which all streams had been detected with photometry alone (pre-dating Gaia DR2).} produced by the combination of an increasing probability of detection at larger distances, due to photometric methods being biased against nearby streams that --in projection-- would be more diffuse and hence more difficult to detect, combined with the declining number of expected streams as a function of distance.   It is clear, then, that the selection functions of streams with/without kinematics, are not only different but almost complementary. This also suggests a sweet-spot at distances of $\sim10-15$~kpc where there are likely more streams to be found, since the two mainstream method families for stream finding have lower detection probabilities at this distance range but the number of expected streams hasn't yet significantly declined.}

\begin{figure}
	\includegraphics[width=\columnwidth]{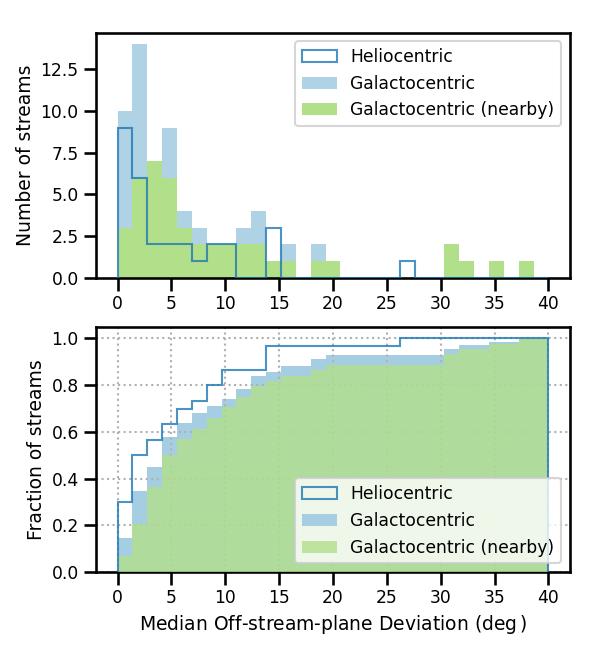}
    \caption{Histogram (top) and cumulative (bottom) distribution of the maximum deviation off each stream's mid-plane for a Heliocentric (open) and Galactocentric (filled blue) observer. In both panels, the distributions are also plotted for nearby streams (heliocentric distance less than 15~kpc, filled green).}
    \label{f:full_lib_offstreamplane_hist}
\end{figure} 

\subsection{Great circle deviations}\label{s:gprops_gc_deviations}

In Section~\ref{s:gprops_info} we showed \Ngc\ (23\%) of the stream tracks in the library are great circles by construction. The remaining streams have enough data available for more detailed tracks to be constructed empirically, which may or may not be well fit by great circles. Here we will assess how well these streams are represented by great circles, in both the heliocentric and galactocentric reference frames.

There is no particular physical reason for the projections of streams to lie along great circles for a heliocentric observer. Instead, for a galactocentric observer, stellar streams formed in a potential with spherical symmetry would have constant angular momentum and, thus, remain confined to a constant plane. \neww{For this observer} such a stream, in projection, would look like a great circle, however complex its radial structure may be \citep[see e.g.][]{Johnston1996,Mateu2017}. Breaking of spherical symmetry will cause the orbital plane to precess and the stream to deviate from a great circle, even for a Galactocentric observer.  In the Milky Way, the Sagittarius stream's orbital plane presents a small rate of precession \citep[$\lesssim 10\degr$][]{Belokurov2014_sgr}, showing that although not perfectly spherical, the (galactocentric) great circle approximation should be a good one for streams in the outer halo ($\gtrsim25$~kpc). In addition, the  difference in the projected view between the heliocentric and Galactocentric observers becomes ever smaller as distance to the stream increases \neww{and the Sun-Galactic-Centre distance becomes negligible.}. 

A large prevalence of great circle streams in the heliocentric frame, particularly for nearby streams, would thus be indicative of an observational bias in the detection techniques \neww{since there is no physical reason to expect it}. Figure~\ref{f:full_lib_offstreamplane_hist} shows the distribution (top panel, cumulative in the bottom panel) of the median deviation of each stream's pole track from its mid-pole (see Sec.~\ref{s:procedure}) for a heliocentric (open) and Galactocentric observer (filled). Note that in these plots each stream contributes only a single point. For a heliocentric observer the distribution (open) shows 60\% of the stream's tracks deviate less than $5\fdg5$ from their mid-plane; about half are less than $3\degr$ wide (on median). For a galactocentric observer the distribution as a whole is similar in shape, but shifted towards slightly larger values indicating  larger deviations from a great circle, 60\% show deviations $\sim8\degr$ and half deviate just under $5\fdg5$. The figure also shows the distribution of Galactocentric deviations for the nearest streams (light green histogram), defined as those with stars  at distances $<15$~kpc. This shows that, as expected, the tail of the distribution towards large deviations is dominated by nearby streams: i.e. nearby streams are less likely to be well represented by great circles from a galactocentric perspective. Vice versa, the complementary distribution for more distant streams (not shown) is dominated by smaller deviations, meaning more distant streams are indeed similarly thin from either point of view. This indicates there is no evident bias towards the detection of (heliocentric) great circle streams.

\subsection{Proper motion misalignment, angular momentum and pole tracks}\label{s:pm_misalignment}

\begin{figure*}
	\includegraphics[width=2.1\columnwidth]{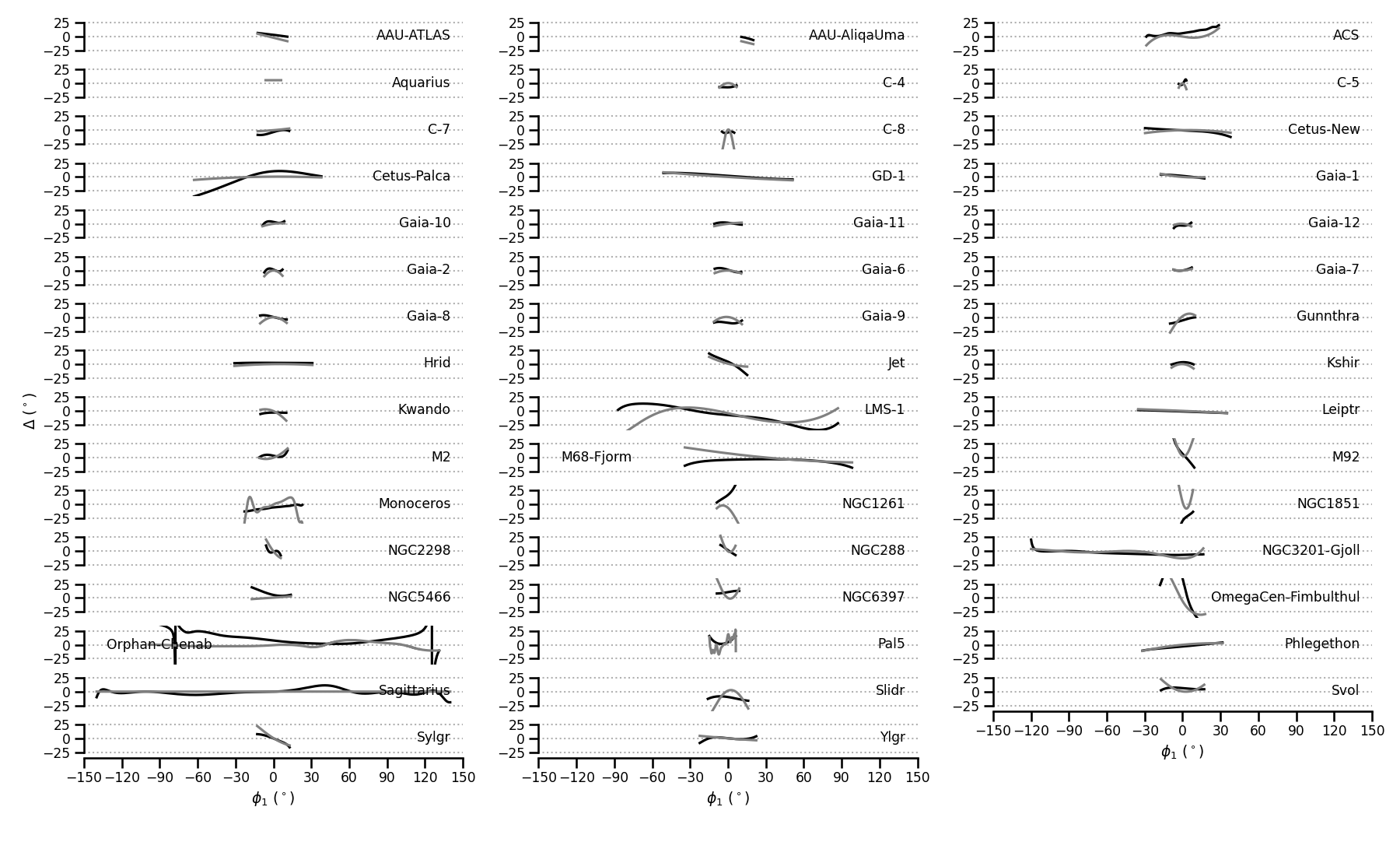}
    \caption{The angle $\Delta$ versus the along-stream coordinate $\phi_1$, computed as ratio of proper motions $(\dfrac{\mu_{\phi_2}}{\mu_{\phi_1}})_\mathrm{corr}=\tan{\Delta}$ corrected by the solar reflex motion (black) and the slope of the stream track tangent $\dfrac{d\phi_2}{d\phi_1}=\tan{\Delta}$ (light grey). In an unperturbed stream proper motions are aligned along the stream track and the two lines coincide. }
    \label{f:full_lib_missalign_track}
\end{figure*} 

\begin{figure*}
	\includegraphics[width=2\columnwidth]{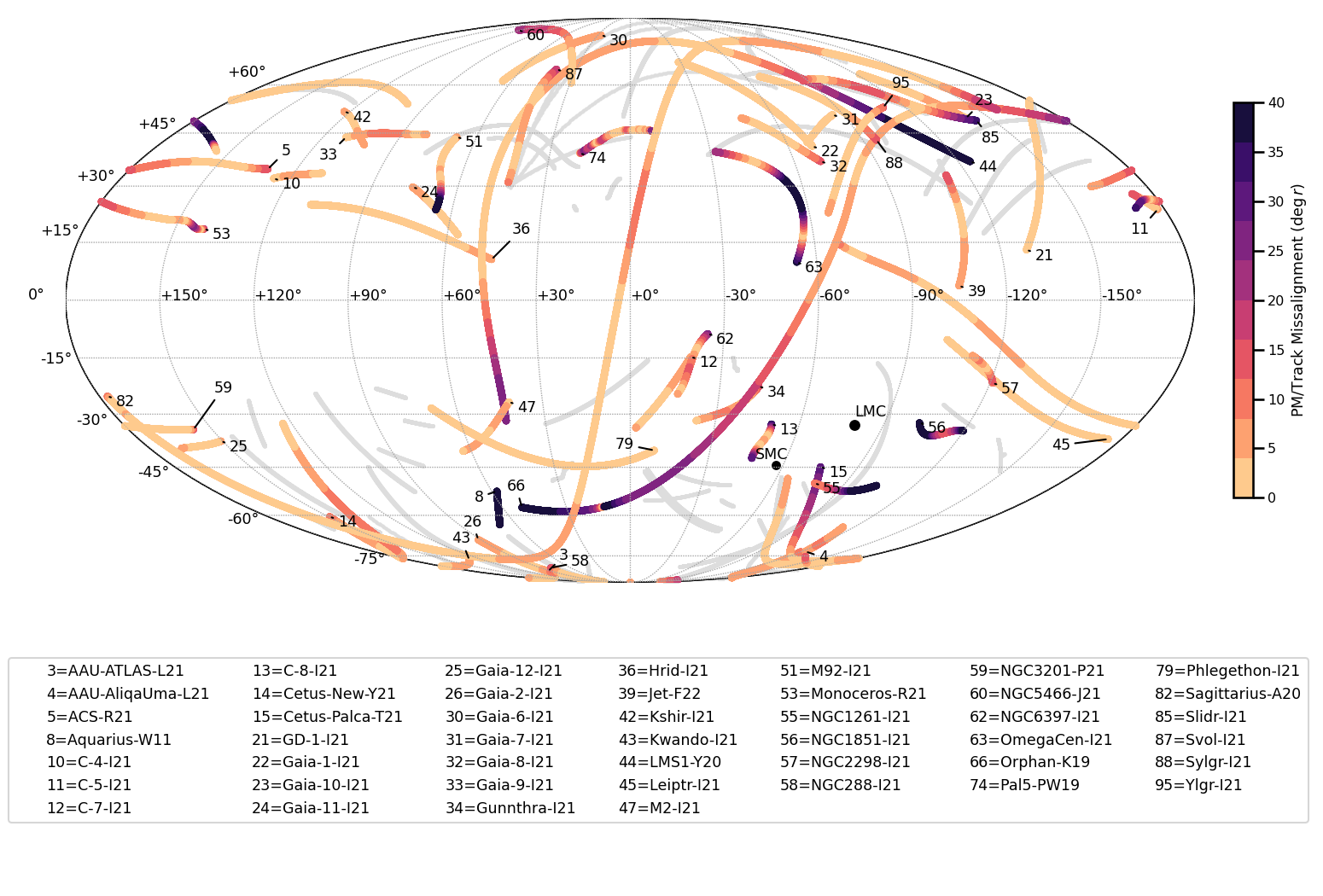}
    \caption{Mollweide projection map in Galactic coordinates of the celestial tracks colour coded by the absolute proper motion to track misalignment. The streams without proper motion information are shown in the background in light grey.}
    \label{f:full_lib_missalign_map}
\end{figure*}  

The availability of proper motion tracks for over half the library makes it possible to compute the angular momenta along the track for a large number of streams, \neww{albeit only in a heliocentric reference frame}. It is convenient at this point to report the angular momentum in a heliocentric frame at rest with respect to the GSR, because in this reference frame the radial velocity has no contribution to the angular momentum and currently only \Nvradpm\ streams have radial velocity information in addition to proper motions, necessary to get the full angular momentum vector in the GSR (see Sec~\ref{s:procedure}).  %\neww{This angular momentum is computed as $\mathbf{L_{\mathrm{hel}}} = \mathbf{r_{\mathrm{hel}}} \times \mathbf{v}_{\mathrm{GSR}}$, where $\mathbf{r}_{\mathrm{hel}}$ is the heliocentric position and $\mathbf{v}_{\mathrm{GSR}}$ is the velocity with respect to the GSR.} This way, the missing radial velocity \neww{of the stream} has no contribution to the angular momentum. 

In an undisturbed stellar stream, stars are expected to be moving mainly along the stream, which approximately traces the orbit of the progenitor \citep{Binney2008}. In such a case, the stream star's velocity, and therefore its proper motion in projection, is expected to be tangent to the stream track itself \footnote{See \citet{Sanders2013} for a discussion of the limitations of this approximation.}. However, analyses of \Gaia~DR2 data have revealed several cases in which the proper motions are significantly misaligned with the stream's track \citep{Erkal2019,Shipp2019}. First observed in the southern tail of the Orphan-Chenab stream \citep{Koposov2019,Erkal2019}, \citeauthor{Erkal2019} showed the proper motion misalignment can be attributed to the dynamical effect of the LMC during a recent (<350 Myr ago) close encounter with the stream. \citet{Shipp2019} have observed a similar effect in several of the DES streams (Indus, Jhelum, and to a lesser extent Aliqa Uma and Turranburra), which they also attribute to the effect of the LMC \citep{Shipp2020}.

For the stellar streams in \galstreams\ that have proper motion and distance track information available, we can provide a systematic survey of proper motion misalignment. This could be visualised in two different but equivalent ways: by comparing the tangent to the stream track $d\phi_2/d\phi_1$ with the ratio of proper motions $\mu_{\phi_2}/\mu_{\phi_1}$ along the track (corrected for the solar reflex motion), as in \citet{Erkal2019}; or by measuring the angular distance between the angular momentum and the pole vector along the track. For an unperturbed stream, in the first case, the ratio of proper motions should coincide with the slope of the tangent along the track $d\phi_2/d\phi_1$; in the second case, the pole and angular momentum vectors should be co-linear. Using the first visualisation, Figure~\ref{f:full_lib_missalign_track} shows the proper motion\footnote{To correct for the solar reflex motion, we assume $(X,Y,Z)_\odot=(8.122,0.0,0.208)$~kpc \citep{GravityCol2018, Bovy2019} and $V_\odot=(12.9, 245.6, 7.78)$~\kms \citep{Drimmel2018,GravityCol2018, Reid2014} for the position and velocity of the Sun in the Galactic Standard of Rest, respectively.} and stream tangent tracks (black and light grey, respectively) for the \NpmD\ streams with proper motion \emph{and} distance gradient data. In this figure all the tracks are shown in the same horizontal and vertical scales (in degrees) to make the comparison easier between different streams. Figure~\ref{f:full_lib_L_pole_sep_tracks} in Appendix~\ref{a:vis} shows the alternative visualisation, the angular distance between the angular momentum and pole vectors, which corresponds to the absolute difference between the tracks in Figure~\ref{f:full_lib_missalign_track}.   Figure~\ref{f:full_lib_L_pole_tracks_map} shows a map in heliocentric Galactic coordinates of the angular momentum and pole tracks for each stream. In these plots, only stellar streams with distance gradient information are shown. Streams with only mean distances available are excluded because proper motions used to judge the misalignment must be corrected for the solar reflex motion. Since this correction is distance dependent, a spurious misalignment can appear if a significant distance gradient is misrepresented by assuming only a mean distance. 

Some very long streams with no misalignment are, e.g., NGC3201-Gj\"oll (P21 track shown), Phlegethon and Leiptr in addition to well-known cases like GD-1 \citep{PriceWhelanBonaca2018_gd1,Grillmair2006_gd1}. On the other hand, the misalignment is evident in the mismatch between the proper motion ratio and stream tangent track in the reported cases of Orphan-Chenab, Jhelum, Indus, Elqui, Turranburra and AAU-AliqaUma \citep{Erkal2019,Shipp2020} as well as in some parts of the Sagittarius track \citep[see]{Vasiliev2021b}, \neww{although it exhibits very little misaligment for the most part}; and also in the Cetus-Palca stream, Slidr and LMS-1 and  M68-Fj\"orm,
previously unreported. 

In the case of Cetus-Palca-T21 and M68-Fj\"orm 
the misalignment happens at one end of the stream (similar to the Orphan-Chenab case). The detection of misaligned proper motions in Cetus-Palca-T21 appears robust, its track is fairly stable and based on members all along the track, the observed misalignment is substantial (similar to that in Orphan-Chenab) and the distance track is robust as it is based on photometric standard-candle distances of Blue Horizontal Branch stars (see details in Cetus-Palca-T21), so it is unlikely to be a distance-related effect. The case of M68-Fj\"orm is also one of apparently high confidence, the two tracks are based on independent detections by \citet{Ibata2021} and \citet{Palau2019}, yet the  misalignment observed in the two cases is nearly identical (only the latter is shown).  For LMS-1, on the contrary, we caution that the misalignment is observed at the ends of the track where there are fewer members to constrain its shape (see Sec.~\ref{s:multiple_tracks}, Fig.~\ref{f:track_compare_notclusters}), compared to the central part ($|\phi_1|<50\degr$) where the two tracks agree very well.

Other, much shorter streams, also show signs of proper motion misalignment at the ends of the track (Gaia-8, NGC~1261, M92, NGC~1851, NGC~6397, Slidr, Sv\"ol, Sylgr, Vid, Ylgr). Some cases should be taken with caution and are worth revising, e.g. C-8, Slidr, Sylgr, in which the misalignment could be due to a poorly constrained celestial (or distance) track if there are few members in the stream as a whole (C-8) or at the end where the misalignment is most prominent (Slidr, Sylgr). 

Figure~\ref{f:full_lib_missalign_map} illustrates the spatial distribution of proper motion misalignment in a Mollweide projection map in Galactic coordinates, where the colour coding is proportional to the absolute difference (in degrees) between track slope computed from the proper motion ratio and tangent along the track. The area in the southern Galactic hemisphere close to the LMC and SMC where many streams (but not all, see e.g. Willka Yaku) are perturbed is evident. The perturbed tail of the Cetus-Palca stream and NGC~1261 are also around this region. In the northern Galactic hemisphere, the perturbed end of M68-Fj\"orm coincides (in projection) with a similarly perturbed end of $\omega Cen$-Fimbulthul; a bit further north, also Gaia-8 shows a perturbed end. The perturbed ends of Slidr, LMS-1 and Sylgr are also in a similar region; bearing in mind the most uncertain end of LMS-1, the one with fewest member candidates, is the end that crosses the Galactic disc, opposite to this one. Several other streams are located in this region, but do not have proper motion information available (shown in grey); these would be interesting to target in a search for any other signs of proper motion misalignment in this area and might provide useful dynamical constraints if used simultaneously in orbital fitting.

\section{Summary}\label{s:conclusions}

Nearly a hundred stellar streams (\Nunique) have been found to date around the Milky Way and the number is rapidly growing. In this work we have combed the literature to collate the angular position, distance, proper motion and radial velocity data available for all published streams and stream candidates, \neww{ for which we provide 3D ($\Ntracks$), 5D ($\Npm$) and 6D ($\Nvrad$) stream tracks where available} in a homogenised format in the \galstreams\ library. The library contains  
 \Ntracks\ tracks with up to 6D information, corresponding to \Nunique~unique streams in our Galaxy, together with a series of computed attributes and utilities, such as stream coordinate frames, end points, pole vector and pole tracks in the heliocentric and Galactocentric reference frames, heliocentric angular momentum tracks and polygon footprints.  

An overview of the latest information available for each stellar stream shows that a remarkable almost two thirds of the streams already have proper motion tracks available, thanks to data from \Gaia\ DR2 and EDR3. \neww{The distance distribution of the streams separated by whether the identification technique used to find them required proper motion data or not (Fig.~\ref{f:full_lib_distance_hist}) showed clearly that kinematically detected streams are biased toward short distances ($\lesssim10$~kpc) with a peak at $\sim7$~kpc, while photometrically detected streams are biased toward larger distances ($\gtrsim12$~kpc) with a peak at $\sim 20$~kpc. The observed peaks in the distance distribution are clearly due to selection effects, not to be confused with real physical features. Further, there is potential to unearth new stellar streams in the distance range $\sim10-15$~kpc so far disfavoured by stream finding techniques with widespread use.}

Perhaps surprisingly, almost half the streams in the library ($45\%$) are lacking enough information to represent a distance gradient and are implemented with only a mean distance estimate for the full stream. In some instances even these are only rough approximations to a distance estimate. This deficit is due to a lack of follow-up on many stream candidates after their first publication, particularly for streams discovered before \Gaia~DR2. As we have shown in Sec.~\ref{s:gprops_distance}, pre-\Gaia~DR2 streams are mostly distant ones ($\gtrsim15$~kpc), beyond the reach of \Gaia\ parallaxes. For these, photometric distances will play a key role in providing distance gradient measurements, either by the use of standard candle tracers -e.g. RR Lyrae or Blue Horizontal Branch stars- or by less precise  isochrone fitting of colour-magnitude diagrams.  Since distance information is required for dynamics and ensemble studies of stellar streams are highly desirable to the Galactic dynamics community, efforts to provide consistent distance gradient measurements for all known streams, ideally based on a common distance scale, will be a highly valuable contribution. 

The \galstreams\ library is served as a Python package publicly available \href{https://github.com/cmateu/galstreams}{via a GitHub repository } and a summary of the streams' properties and information available in the compilation for each track is presented here in Table~\ref{t:super_summary_table}.  

\begin{table*}
\centering
\caption{Summary attributes for the tracks available in \galstreams.}\label{t:super_summary_table}
\tabcolsep=0.1cm
\begin{footnotesize}
\begin{tabular}{llllrrrrrrrrll}
\toprule
   StreamName &         TrackName & InfoFlags & Imp & On & Length & $\alpha_i$ & $\delta_i$ & $D_i$ & $\alpha_f$ & $\delta_f$ & $D_f$ &  TRefs &        DRefs \\
              &                   &           &     &    &($^\circ$)&($^\circ$)&($^\circ$)  & (kpc) &($^\circ$)  &($^\circ$)  & (kpc) &        &              \\                       
\midrule
       20.0-1 &        20.0-1-M18 &      0000 &  po &  1 &   36.6 &      280.2 &      -41.0 &  27.7 &      311.5 &      -16.3 &  25.1 &     47 &           47 \\
         300S &          300S-F18 &      1101 &  st &  1 &   11.1 &      151.8 &       16.0 &  18.0 &      163.2 &       14.4 &  14.4 &     14 &        51,63 \\
 AAU-AliqaUma &  AAU-AliqaUma-L21 &      1111 &  st &  1 &    9.7 &       31.2 &      -32.9 &  25.6 &       40.9 &      -38.5 &  30.8 &     39 &           60 \\
    AAU-ATLAS &     AAU-ATLAS-L21 &      1111 &  st &  1 &   23.6 &        8.9 &      -20.2 &  19.0 &       30.7 &      -33.5 &  25.6 &     39 &           36 \\
    AAU-ATLAS &         ATLAS-I21 &      1110 &  st &  0 &   18.1 &       17.4 &      -25.2 &  24.8 &       35.1 &      -34.9 &  20.4 &     32 &           36 \\
      Acheron &       Acheron-G09 &      0000 &  ep &  1 &   36.5 &      230.0 &       -2.0 &   3.5 &      259.0 &       21.0 &   3.8 &     20 &           20 \\
          ACS &           ACS-R21 &      1110 &  st &  1 &   57.7 &      154.4 &       80.0 &  11.7 &      125.5 &       24.0 &  11.7 &     59 &           17 \\
      Alpheus &       Alpheus-G13 &      1100 &  st &  1 &   24.2 &       21.6 &      -69.0 &   1.6 &       27.7 &      -45.0 &   2.0 &     21 &           21 \\
     Aquarius &      Aquarius-W11 &      1111 &  st &  1 &   12.4 &      338.8 &       -7.5 &   4.8 &      351.1 &       -9.9 &   3.3 &     70 &           70 \\
         C-19 &          C-19-I21 &      1010 &  st &  1 &   29.7 &      354.2 &       36.1 &  18.0 &      355.3 &        6.4 &  18.0 &     44 &           32 \\
          C-4 &           C-4-I21 &      1110 &  st &  1 &   13.6 &      165.2 &       84.3 &   4.0 &      249.6 &       77.1 &   3.2 &     32 &           32 \\
          C-5 &           C-5-I21 &      1110 &  st &  1 &    6.1 &      113.4 &       36.7 &   3.4 &      119.8 &       33.6 &   5.6 &     32 &           32 \\
          C-7 &           C-7-I21 &      1110 &  st &  1 &   24.9 &      270.9 &      -53.1 &   6.1 &      305.0 &      -42.6 &   5.9 &     32 &           32 \\
          C-8 &           C-8-I21 &      1110 &  st &  1 &   10.2 &      327.4 &      -81.8 &   3.6 &      346.6 &      -72.6 &   3.5 &     32 &           32 \\
    Cetus-New &     Cetus-New-Y21 &      1110 &  st &  1 &   67.8 &       25.6 &       -0.2 &  20.4 &       26.6 &      -68.0 &  18.7 &     74 &           74 \\
  Cetus-Palca &   Cetus-Palca-T21 &      1110 &  st &  1 &  100.9 &       47.3 &      -67.1 &  34.9 &       13.0 &       30.4 &  26.3 &     68 &  72,50,60,68 \\
  Cetus-Palca &   Cetus-Palca-Y21 &      1110 &  st &  0 &  152.2 &        0.5 &       46.2 &  25.0 &      127.4 &      -72.1 &  55.6 &     74 &  72,50,60,68 \\
        Cetus &         Cetus-Y13 &      1100 &  st &  1 &   30.1 &       20.0 &       -4.1 &  27.2 &       32.2 &       23.5 &  32.5 &     71 &        72,50 \\
      Cocytos &       Cocytos-G09 &      0000 &  ep &  1 &   75.1 &      186.0 &       -3.0 &  11.0 &      259.0 &       20.0 &  11.0 &     20 &           20 \\
       Corvus &        Corvus-M18 &      0000 &  po &  1 &   73.0 &      140.7 &      -13.0 &   4.9 &      215.9 &      -19.0 &  14.8 &     47 &           47 \\
        Elqui &         Elqui-S19 &      1010 &  st &  1 &   10.9 &       10.8 &      -37.2 &  50.1 &       22.6 &      -43.4 &  50.1 &  61,60 &           60 \\
     Eridanus &      Eridanus-M17 &      1000 &  st &  1 &    0.6 &       66.0 &      -21.4 &  95.0 &       66.4 &      -21.0 &  95.0 &  48,28 &           48 \\
   Fimbulthul &    Fimbulthul-I21 &      1110 &  st &  0 &   38.8 &      188.7 &      -40.0 &   5.1 &      223.6 &      -24.7 &   3.4 &     32 &           30 \\
       Gaia-1 &        Gaia-1-I21 &      1110 &  st &  1 &   34.6 &      183.1 &      -20.8 &   5.0 &      203.3 &        7.5 &   4.5 &     32 &           41 \\
      Gaia-10 &       Gaia-10-I21 &      1110 &  st &  1 &   17.2 &      150.7 &       16.0 &  17.4 &      168.5 &       14.1 &  12.1 &     32 &           30 \\
      Gaia-11 &       Gaia-11-I21 &      1110 &  st &  1 &   21.7 &      267.7 &       48.7 &  11.5 &      277.3 &       28.3 &  13.8 &     32 &           32 \\
      Gaia-12 &       Gaia-12-I21 &      1110 &  st &  1 &   13.7 &       36.0 &       20.3 &  13.3 &       46.7 &       11.2 &  10.9 &     32 &           30 \\
       Gaia-2 &        Gaia-2-I21 &      1110 &  st &  1 &   14.4 &        1.8 &       -8.5 &   7.0 &        9.1 &      -21.0 &   7.0 &     32 &           41 \\
       Gaia-3 &        Gaia-3-M18 &      0000 &  ep &  1 &   18.5 &      171.0 &      -15.0 &   9.0 &      179.0 &      -32.0 &  14.0 &     41 &           41 \\
       Gaia-4 &        Gaia-4-M18 &      0000 &  ep &  1 &    8.9 &      163.0 &      -11.0 &  10.7 &      167.0 &       -3.0 &  11.5 &     41 &           41 \\
       Gaia-5 &        Gaia-5-M18 &      0000 &  ep &  1 &   23.7 &      137.0 &       23.0 &  18.5 &      154.0 &       42.0 &  20.5 &     41 &           41 \\
       Gaia-6 &        Gaia-6-I21 &      1110 &  st &  1 &   21.1 &      204.8 &       25.5 &   8.5 &      224.4 &       38.6 &   9.9 &     32 &           30 \\
       Gaia-7 &        Gaia-7-I21 &      1110 &  st &  1 &   14.6 &      175.0 &       -8.4 &   4.6 &      184.3 &      -19.8 &   5.1 &     32 &           30 \\
       Gaia-8 &        Gaia-8-I21 &      1110 &  st &  1 &   21.1 &      181.7 &      -24.9 &   7.5 &      200.0 &      -12.8 &   6.8 &     32 &           32 \\
       Gaia-9 &        Gaia-9-I21 &      1110 &  st &  1 &   22.0 &      219.9 &       70.8 &   3.8 &      244.2 &       51.8 &   5.3 &     32 &           32 \\
         GD-1 &          GD-1-I21 &      1110 &  st &  1 &  102.5 &      123.3 &      -10.1 &   9.2 &      219.9 &       57.9 &  11.7 &     32 &           18 \\
         GD-1 &         GD-1-PB18 &      1110 &  st &  0 &  100.0 &      124.1 &       -8.0 &   5.5 &      219.6 &       58.2 &  10.5 &  57,35 &           18 \\
     Gunnthra &      Gunnthra-I21 &      1110 &  st &  1 &   20.2 &      265.9 &      -76.4 &   3.1 &      300.1 &      -60.3 &   3.2 &     32 &           32 \\
       Hermus &        Hermus-G14 &      1100 &  st &  1 &   47.6 &      245.4 &        5.0 &  18.9 &      253.2 &       50.0 &  14.9 &     22 &           22 \\
         Hrid &          Hrid-I21 &      1110 &  st &  1 &   61.7 &      278.6 &       15.2 &   3.2 &      288.9 &       76.7 &   3.9 &     32 &           32 \\
       Hyllus &        Hyllus-G14 &      1100 &  st &  1 &   23.5 &      249.1 &       11.0 &  23.0 &      246.9 &       34.0 &  18.5 &  22,27 &           22 \\
        Indus &         Indus-S19 &      1010 &  st &  1 &   18.2 &      324.1 &      -52.0 &  16.6 &      350.3 &      -64.2 &  16.6 &  61,60 &           60 \\
          Jet &           Jet-F22 &      1110 &  st &  1 &   30.3 &      129.0 &      -34.5 &  33.3 &      148.5 &      -10.1 &  27.3 &     13 &           34 \\
          Jet &           Jet-J18 &      0000 &  ep &  0 &   12.5 &      133.8 &      -27.7 &  28.6 &      142.0 &      -17.8 &  28.6 &     34 &           34 \\
     Jhelum-a &      Jhelum-a-B19 &      1010 &  st &  1 &   30.0 &        7.3 &      -52.2 &  13.0 &      321.5 &      -46.2 &  13.0 &     09 &           60 \\
     Jhelum-a &      Jhelum-a-S19 &      1010 &  st &  0 &   24.6 &      323.0 &      -46.9 &  13.2 &        0.5 &      -51.5 &  13.2 &  61,60 &           60 \\
     Jhelum-b &      Jhelum-b-B19 &      1010 &  st &  1 &   30.0 &        7.1 &      -51.3 &  13.0 &      322.1 &      -45.4 &  13.0 &     09 &           09 \\
     Jhelum-b &      Jhelum-b-S19 &      1010 &  st &  0 &   13.4 &      341.9 &      -50.8 &  13.2 &        3.1 &      -50.6 &  13.2 &  61,60 &           09 \\
       Jhelum &        Jhelum-I21 &      1110 &  st &  0 &   27.7 &       10.4 &      -49.9 &  10.9 &      327.4 &      -48.6 &  11.2 &     32 &           60 \\
        Kshir &         Kshir-I21 &      1110 &  st &  1 &   17.4 &      193.4 &       65.6 &  10.1 &      237.7 &       67.3 &  11.0 &     32 &           32 \\
       Kwando &        Kwando-G17 &      1000 &  st &  0 &   12.5 &       19.0 &      -23.9 &  20.0 &       31.0 &      -29.4 &  20.0 &     24 &           24 \\
       Kwando &        Kwando-I21 &      1110 &  st &  1 &   20.6 &       14.9 &      -14.0 &   8.2 &       29.3 &      -29.6 &   7.0 &     32 &           24 \\
       Leiptr &        Leiptr-I21 &      1110 &  st &  1 &   70.4 &       57.9 &        3.4 &   8.9 &      115.8 &      -44.5 &   6.0 &     32 &           30 \\
        Lethe &         Lethe-G09 &      0000 &  ep &  1 &   81.2 &      171.0 &       18.0 &  13.0 &      258.0 &       20.0 &  13.0 &     20 &           20 \\
        LMS-1 &          LMS1-M21 &      1011 &  st &  0 &   49.6 &      195.2 &       13.4 &  19.0 &      245.2 &       31.4 &  19.0 &     42 &           73 \\
        LMS-1 &          LMS1-Y20 &      1111 &  st &  1 &  179.2 &      145.2 &       -0.1 &  16.2 &      315.9 &       -5.6 &  16.2 &     73 &           73 \\
           M2 &            M2-G22 &      1010 &  st &  0 &   73.3 &      281.1 &      -16.2 &  11.7 &      347.4 &       -9.7 &  11.7 &  26,04 &           15 \\
           M2 &            M2-I21 &      1110 &  st &  1 &   22.6 &      310.4 &       -5.0 &  10.6 &      331.6 &        2.6 &  12.2 &     32 &           15 \\
          M30 &           M30-S20 &      1000 &  st &  1 &    8.8 &      320.8 &      -25.4 &   8.5 &      328.3 &      -20.6 &   8.5 &  65,28 &           65 \\
           M5 &            M5-G19 &      1010 &  st &  1 &   41.4 &      226.8 &        3.8 &  15.9 &      188.1 &       21.0 &  12.3 &     25 &           25 \\
           M5 &            M5-I21 &      1110 &  st &  0 &    8.1 &      195.8 &       17.4 &   7.5 &      203.8 &       15.0 &   7.5 &     32 &           25 \\
           M5 &            M5-S20 &      1000 &  st &  0 &    4.9 &      228.0 &        3.0 &   7.5 &      231.7 &        0.3 &   7.5 &  65,28 &           25 \\
          M68 &           M68-I21 &      1110 &  st &  0 &   15.4 &      186.3 &      -43.8 &   4.1 &      191.0 &      -29.0 &   4.1 &     32 &           53 \\   
\bottomrule
\end{tabular}

\end{footnotesize}
\end{table*}

\begin{table*}
\centering
\contcaption{}
\tabcolsep=0.1cm
\begin{footnotesize}
\begin{tabular}{llllrrrrrrrrll}
\toprule
          StreamName &          TrackName & InfoFlags & Imp & On & Length & $\alpha_i$ & $\delta_i$ & $D_i$ & $\alpha_f$ & $\delta_f$ & $D_f$ &     TRefs &        DRefs \\
              &                   &           &     &    &($^\circ$)&($^\circ$)&($^\circ$)  & (kpc) &($^\circ$)  &($^\circ$)  & (kpc) &        &              \\
\midrule
    M68-Fjorm &         Fjorm-I21 &      1110 &  st &  0 &  141.4 &      189.3 &      -47.2 &   0.0 &      292.2 &       65.3 &   7.2 &     32 &        53,30 \\
           M68-Fjorm &            M68-P19 &      1210 &  st &  1 &  133.5 &      190.1 &      -47.3 &   8.3 &      273.5 &       66.5 &   4.7 &        53 &        53,30 \\
                 M92 &            M92-I21 &      1110 &  st &  1 &   21.1 &      246.9 &       43.0 &   9.2 &      273.2 &       38.4 &   7.0 &        32 &     65,67,32 \\
                 M92 &            M92-S20 &      1000 &  st &  0 &    9.3 &      254.4 &       41.4 &   8.5 &      264.6 &       41.0 &   8.5 &     65,28 &     65,67,32 \\
                 M92 &            M92-T20 &      1000 &  st &  0 &   18.2 &      250.3 &       39.3 &   8.3 &      271.3 &       38.6 &   8.3 &        67 &     65,67,32 \\
            Molonglo &       Molonglo-G17 &      1000 &  st &  1 &   19.4 &       10.3 &      -24.5 &  20.0 &      354.6 &      -12.0 &  20.0 &        24 &           24 \\
           Monoceros &      Monoceros-R21 &      1110 &  st &  1 &   46.9 &       78.5 &       71.7 &  10.6 &      116.1 &       30.6 &  10.6 &        59 &           49 \\
        Murrumbidgee &   Murrumbidgee-G17 &      1000 &  st &  1 &  123.6 &      124.3 &      -65.6 &  20.0 &      358.6 &       16.0 &  20.0 &        24 &           24 \\
             NGC1261 &        NGC1261-I21 &      1110 &  st &  1 &   17.4 &       41.7 &      -62.6 &  16.5 &       56.5 &      -47.5 &  22.7 &        32 &        32,60 \\
             NGC1851 &        NGC1851-I21 &      1110 &  st &  1 &   15.8 &       84.0 &      -50.6 &  14.7 &       78.4 &      -36.5 &  13.5 &        32 &        32,60 \\
             NGC2298 &        NGC2298-I21 &      1110 &  st &  1 &   11.8 &       93.3 &      -32.2 &  12.4 &      105.6 &      -38.1 &  11.0 &        32 &     02,65,32 \\
             NGC2298 &        NGC2298-S20 &      1000 &  st &  0 &    4.7 &      104.5 &      -37.7 &   9.8 &      100.6 &      -34.6 &   9.8 &     65,28 &     02,65,32 \\
             NGC2808 &        NGC2808-I21 &      1110 &  st &  0 &   19.5 &      124.2 &      -57.3 &   1.9 &      157.0 &      -71.0 &   6.5 &        32 &           32 \\
              NGC288 &         NGC288-S20 &      1000 &  st &  0 &   10.3 &       12.2 &      -24.2 &   9.0 &       13.2 &      -31.1 &   9.0 &     65,28 &  15,60,65,32 \\
              NGC288 &         NGC288-I21 &      1110 &  st &  1 &   12.1 &        5.9 &      -20.0 &  11.5 &       15.3 &      -28.4 &  10.6 &        32 &  15,60,65,32 \\
             NGC3201 &        NGC3201-I21 &      1110 &  st &  0 &   11.9 &      146.9 &      -45.5 &   5.4 &      164.0 &      -47.2 &   5.4 &        32 &        54,30 \\
       NGC3201-Gjoll &        NGC3201-P21 &      1110 &  st &  1 &  136.9 &       42.3 &       20.8 &   5.3 &      178.4 &      -46.7 &   3.2 &        54 &        54,30 \\
       NGC3201-Gjoll &          Gjoll-I21 &      1110 &  st &  0 &   57.5 &       64.9 &        2.4 &   3.9 &      112.2 &      -34.1 &   3.3 &        32 &        54,30 \\
             NGC5466 &        NGC5466-G06 &      1000 &  st &  0 &   45.3 &      228.8 &       20.5 &  16.6 &      180.9 &       39.6 &  16.6 &        16 &        05,16 \\
             NGC5466 &        NGC5466-I21 &      1110 &  st &  0 &   16.8 &      208.8 &       29.6 &  15.6 &      224.7 &       21.0 &  14.1 &        32 &        05,16 \\
             NGC5466 &        NGC5466-J21 &      1110 &  st &  1 &   30.9 &      192.8 &       35.5 &  26.8 &      224.6 &       21.8 &  14.3 &        33 &        05,16 \\
             NGC6101 &        NGC6101-I21 &      1110 &  st &  0 &   10.5 &      242.9 &      -72.8 &  14.8 &      270.4 &      -67.7 &  16.0 &        32 &           32 \\
             NGC6362 &        NGC6362-S20 &      1000 &  st &  1 &    4.5 &      258.7 &      -64.6 &   7.6 &      264.2 &      -67.9 &   7.6 &     65,28 &           65 \\
             NGC6397 &        NGC6397-I21 &      1110 &  st &  1 &   18.3 &      257.7 &      -54.5 &   2.4 &      288.3 &      -53.1 &   2.6 &        32 &           32 \\
 OmegaCen-Fimbulthul &       OmegaCen-I21 &      1110 &  st &  1 &   43.4 &      198.7 &      -52.9 &   5.1 &      213.2 &      -19.3 &   2.9 &        32 &        31,30 \\
 OmegaCen-Fimbulthul &       OmegaCen-S20 &      1000 &  st &  0 &    8.6 &      196.1 &      -45.2 &   5.4 &      207.4 &      -48.7 &   5.4 &     65,28 &        31,30 \\
           Ophiuchus &      Ophiuchus-B14 &      0000 &  po &  0 &    2.4 &      240.6 &       -7.2 &   9.5 &      243.0 &       -6.7 &   9.5 &        06 &           06 \\
           Ophiuchus &      Ophiuchus-C20 &      1010 &  st &  1 &    6.8 &      241.0 &       -7.2 &   8.3 &      247.6 &       -7.3 &   8.3 &        11 &           06 \\
             Orinoco &        Orinoco-G17 &      1000 &  st &  1 &   21.4 &        0.0 &      -25.5 &  20.6 &       22.7 &      -28.4 &  20.6 &        24 &           24 \\
       Orphan-Chenab &         Orphan-I21 &      1110 &  st &  0 &   86.7 &      145.5 &       40.0 &  30.4 &      177.2 &      -41.7 &  14.5 &        32 &     19,60,37 \\
       Orphan-Chenab &         Orphan-K19 &      1111 &  st &  1 &  230.6 &      341.0 &      -13.7 & 100.0 &      136.4 &       60.3 &  55.4 &        37 &     19,60,37 \\
       Orphan-Chenab &         Chenab-S19 &      1010 &  st &  0 &   10.1 &      322.1 &      -59.0 &  39.8 &      328.6 &      -49.6 &  39.8 &     61,60 &     19,60,37 \\
               PS1-A &          PS1-A-B16 &      0000 &  po &  1 &    4.9 &       28.4 &       -6.5 &   7.9 &       30.1 &       -1.9 &   7.9 &        07 &           07 \\
               PS1-B &          PS1-B-B16 &      0000 &  po &  1 &    9.9 &      145.6 &      -15.3 &  14.5 &      151.1 &       -7.0 &  14.5 &        07 &           07 \\
               PS1-C &          PS1-C-B16 &      0000 &  po &  1 &    7.9 &      330.2 &       11.7 &  17.4 &      334.9 &       18.2 &  17.4 &        07 &           07 \\
               PS1-D &          PS1-D-B16 &      0000 &  po &  1 &   44.9 &      138.7 &      -21.6 &  22.9 &      140.7 &       23.3 &  22.9 &        07 &           07 \\
               PS1-E &          PS1-E-B16 &      0000 &  po &  1 &   24.9 &      160.1 &       46.2 &  12.6 &      193.0 &       62.9 &  12.6 &        07 &           07 \\
               Pal13 &          Pal13-S20 &      0000 &  ep &  1 &   10.9 &      344.3 &        8.9 &  23.6 &      350.2 &       18.2 &  23.6 &        62 &           60 \\
               Pal15 &          Pal15-M17 &      1000 &  st &  1 &    1.5 &      255.3 &       -1.5 &  38.4 &      254.8 &       -0.1 &  38.4 &     48,28 &           48 \\
                Pal5 &           Pal5-S20 &      1000 &  st &  0 &   27.2 &      240.3 &        5.2 &  22.5 &      220.6 &      -12.0 &  22.5 &        66 &           52 \\
                Pal5 &           Pal5-I21 &      1110 &  st &  0 &   22.3 &      223.1 &       -7.5 &  17.8 &      240.2 &        6.1 &  20.4 &        32 &           52 \\
                Pal5 &          Pal5-PW19 &      1110 &  st &  1 &   21.4 &      242.1 &        6.9 &  22.0 &      224.7 &       -4.8 &  19.1 &     56,10 &           52 \\
               Palca &          Palca-S18 &      0000 &  ep &  1 &   57.3 &       30.3 &      -53.7 &  36.3 &       16.2 &        2.4 &  36.3 &        60 &           60 \\
            Parallel &       Parallel-W18 &      1100 &  st &  1 &   37.7 &      192.8 &        0.0 &  14.8 &      156.2 &        8.7 &  16.5 &        69 &           64 \\
             Pegasus &        Pegasus-P19 &      1000 &  st &  1 &    8.6 &      328.3 &       20.9 &  18.0 &      333.4 &       28.1 &  18.0 &        55 &           55 \\
       Perpendicular &  Perpendicular-W18 &      1100 &  st &  1 &   21.3 &      186.0 &        7.5 &  15.1 &      184.3 &       27.5 &  15.6 &        69 &           69 \\
          Phlegethon &     Phlegethon-I21 &      1110 &  st &  1 &   63.6 &      313.7 &      -48.5 &   3.5 &      325.5 &       14.0 &   4.1 &        32 &           30 \\
             Phoenix &        Phoenix-S19 &      1010 &  st &  1 &   11.8 &       19.8 &      -55.1 &  17.5 &       27.0 &      -44.3 &  17.5 &     61,03 &           03 \\
                Ravi &           Ravi-S18 &      0020 &  ep &  1 &   16.6 &      334.8 &      -44.1 &  22.9 &      344.0 &      -59.7 &  22.9 &     60,61 &        60,47 \\
         Sagittarius &    Sagittarius-A20 &      1110 &  st &  1 &  280.0 &       62.8 &       16.0 &  37.3 &      149.8 &       27.8 &  19.4 &     01,58 &  45,46,29,40 \\
           Sangarius &      Sangarius-G17 &      0000 &  po &  1 &   50.1 &      134.0 &      -17.6 &  21.0 &      131.5 &       32.4 &  21.0 &        23 &           23 \\
           Scamander &      Scamander-G17 &      0000 &  po &  1 &   65.2 &      151.6 &      -20.5 &  21.0 &      143.6 &       44.3 &  21.0 &        23 &           23 \\
               Slidr &          Slidr-I21 &      1110 &  st &  1 &   34.2 &      149.2 &       17.6 &   2.4 &      178.9 &        2.0 &   3.9 &        32 &           30 \\
                Styx &           Styx-G09 &      0000 &  ep &  1 &   60.4 &      194.0 &       20.0 &  45.0 &      259.0 &       21.0 &  45.0 &        20 &           20 \\
                Svol &           Svol-I21 &      1110 &  st &  1 &   34.7 &      220.6 &       27.8 &   8.9 &      258.0 &       21.4 &   6.3 &        32 &           30 \\
               Sylgr &          Sylgr-I21 &      1110 &  st &  1 &   26.1 &      165.3 &      -11.6 &   2.4 &      188.1 &        0.4 &   5.8 &        32 &           30 \\
             Tri-Pis &        Tri-Pis-B12 &      1001 &  st &  1 &   13.5 &       21.0 &       36.1 &  26.0 &       24.0 &       22.9 &  26.0 &     08,43 &        08,43 \\
           TucanaIII &      TucanaIII-S19 &      1010 &  st &  1 &    4.3 &      354.6 &      -59.8 &  23.4 &        3.0 &      -59.5 &  26.9 &  61,60,38 &           12 \\
              Turbio &         Turbio-S18 &      0020 &  ep &  1 &   15.0 &       28.0 &      -61.0 &  16.6 &       27.9 &      -46.0 &  16.6 &     60,61 &           60 \\
         Turranburra &    Turranburra-S19 &      1010 &  st &  1 &   13.7 &       59.7 &      -18.6 &  27.5 &       72.6 &      -25.3 &  27.5 &     61,60 &           60 \\
           Wambelong &      Wambelong-S18 &      0020 &  ep &  1 &   14.2 &       90.5 &      -45.6 &  15.1 &       79.3 &      -34.3 &  15.1 &     60,61 &           60 \\
         Willka\_Yaku &    Willka\_Yaku-S18 &      0020 &  ep &  1 &    6.4 &       36.1 &      -64.6 &  34.7 &       38.4 &      -58.3 &  34.7 &     60,61 &           60 \\
                Ylgr &           Ylgr-I21 &      1110 &  st &  1 &   44.8 &      163.8 &        2.4 &   8.6 &      181.7 &      -39.2 &  10.0 &        32 &           30 \\
\bottomrule
\end{tabular}

\end{footnotesize}
\end{table*}

\begin{table}
\caption{Code References for Table~\ref{t:super_summary_table}}\label{t:ref_summary_table}
\tabcolsep=0.1cm
\begin{scriptsize}
\begin{tabular}{ll}
\toprule
Code & Reference\\ \midrule
01 & \citet{Antoja2020} \\ 
 02 & \citet{Balbinot2011} \\ 
 03 & \citet{Balbinot2016} \\ 
 04 & \citet{Baumgardt2021} \\ 
 05 & \citet{Belokurov2006_5466} \\ 
 06 & \citet{Bernard2014} \\ 
 07 & \citet{Bernard2016} \\ 
 08 & \citet{Bonaca2012} \\ 
 09 & \citet{Bonaca2019} \\ 
 10 & \citet{Bonaca2020} \\ 
 11 & \citet{Caldwell2020} \\ 
 12 & \citet{Drlicawagner2015} \\ 
 13 & \citet{Ferguson2022} \\ 
 14 & \citet{Fu2018} \\ 
 15 & \citet{Grillmair1995} \\ 
 16 & \citet{Grillmair2006_5466} \\ 
 17 & \citet{Grillmair2006_acs} \\ 
 18 & \citet{Grillmair2006_gd1} \\ 
 19 & \citet{Grillmair2006_orphan} \\ 
 20 & \citet{Grillmair2009} \\ 
 21 & \citet{Grillmair2013} \\ 
 22 & \citet{Grillmair2014} \\ 
 23 & \citet{Grillmair2017} \\ 
 24 & \citet{Grillmair2017_south} \\ 
 25 & \citet{Grillmair2019} \\ 
 26 & \citet{Grillmair2022} \\ 
 27 & \citet{GrillmairCarlin2016} \\ 
 28 & \citet{Harris1996} \\ 
 29 & \citet{Ibata2001} \\ 
 30 & \citet{Ibata2019} \\ 
 31 & \citet{Ibata2019_OCen} \\ 
 32 & \citet{Ibata2021} \\ 
 33 & \citet{Jensen2021} \\ 
 34 & \citet{Jethwa2018} \\ 
 35 & \citet{Koposov2010} \\ 
 36 & \citet{Koposov2014} \\ 
 37 & \citet{Koposov2019} \\ 
 38 & \citet{Li2018} \\ 
 39 & \citet{Li2021} \\ 
 40 & \citet{Majewski2003} \\ 
 41 & \citet{Malhan2018} \\ 
 42 & \citet{Malhan2021} \\ 
 43 & \citet{Martin2013} \\ 
 44 & \citet{Martin2022} \\ 
 45 & \citet{Mateo1996} \\ 
 46 & \citet{Mateo1998} \\ 
 47 & \citet{Mateu2018} \\ 
 48 & \citet{Myeong2017} \\ 
 49 & \citet{Newberg2002} \\ 
 50 & \citet{Newberg2009} \\ 
 51 & \citet{NiedersteOstholt2009} \\ 
 52 & \citet{Odenkirchen2001} \\ 
 53 & \citet{Palau2019} \\ 
 54 & \citet{Palau2021} \\ 
 55 & \citet{Perottoni2019} \\ 
 56 & \citet{PriceWhelan2019_pal5} \\ 
 57 & \citet{PriceWhelanBonaca2018_gd1} \\ 
 58 & \citet{Ramos2020} \\ 
 59 & \citet{Ramos2021} \\ 
 60 & \citet{Shipp2018} \\ 
 61 & \citet{Shipp2019} \\ 
 62 & \citet{Shipp2020} \\ 
 63 & \citet{Simon2011} \\ 
 64 & \citet{Sohn2016} \\ 
 65 & \citet{Sollima2020} \\ 
 66 & \citet{Starkman2020} \\ 
 67 & \citet{Thomas2020} \\ 
 68 & \citet{Thomas2021} \\ 
 69 & \citet{Weiss2018} \\ 
 70 & \citet{Williams2011} \\ 
 71 & \citet{Yam2013} \\ 
 72 & \citet{Yanny2009} \\ 
 73 & \citet{Yuan2020} \\ 
 74 & \citet{Yuan2021} \\ 
 \bottomrule
\end{tabular}
 
\end{scriptsize}
\end{table}

%%%%%%%%%%%%%%%%%%%%%%%%%%%%%%%%%%%%%%%%%%%%%%%%%%
%--------------------------
\section*{Acknowledgements}
%--------------------------

CM warmly thanks Adrian Price-Whelan, Ana Bonaca, Robyn Sanderson, Pau Ramos and Ting Li for their helpful comments, feature requests and encouragement; Mauro Cabrera for help with data extraction, as well as Alex Drlica-Wagner and Nora Shipp for feature corrections on DES streams and Antonio Sollima, Zhen Yuan, Khyati Malhan, Guillaume Thomas and Rodrigo Ibata for kindly sharing supplementary data that allowed for the implementation of several stream tracks. This project was developed in part at the 2019 Santa Barbara \Gaia\ Sprint, hosted by the Kavli Institute for Theoretical Physics at the University of California, Santa Barbara. This research was supported in part at KITP by the Heising-Simons Foundation and the National Science Foundation under Grant No. NSF PHY-1748958.
This research was partly supported by funding from the MIA program, CSIC project C120-347 at Universidad de la Rep\'ublica, Uruguay and project FCE\_1\_2021\_1\_167524 from the Fondo Clemente Estable of the Agencia Nacional de Innovaci\'on e Investigaci\'on (ANII).

%CSIC-C120-347


\emph{Software:}
    astropy \citep{astropy2018},
    gala \citep{gala},
    matplotlib \citep{mpl},
    numpy \citep{numpy},
    scipy \citep{scipy2001},
    jupyter \citep{jupyter2016}, and 
    TOPCAT \citep{Topcat2005,Stilts2006}
\section*{Data Availability}

The streams' tracks discussed in this work are available at the \galstreams\ Python package served \href{https://github.com/cmateu/galstreams}{at this  GitHub repository}.

\bibliographystyle{mnras}

%%%%%%%%%%%%%%%%%%%%%%%%%%%%%%%%%%%%%%%%%%%%%%%%%%

%%%%%%%%%%%%%%%%% APPENDICES %%%%%%%%%%%%%%%%%%%%%

\appendix

\section{Other Library Visualisations}\label{a:vis}

\begin{figure*}
	\includegraphics[width=2.\columnwidth]{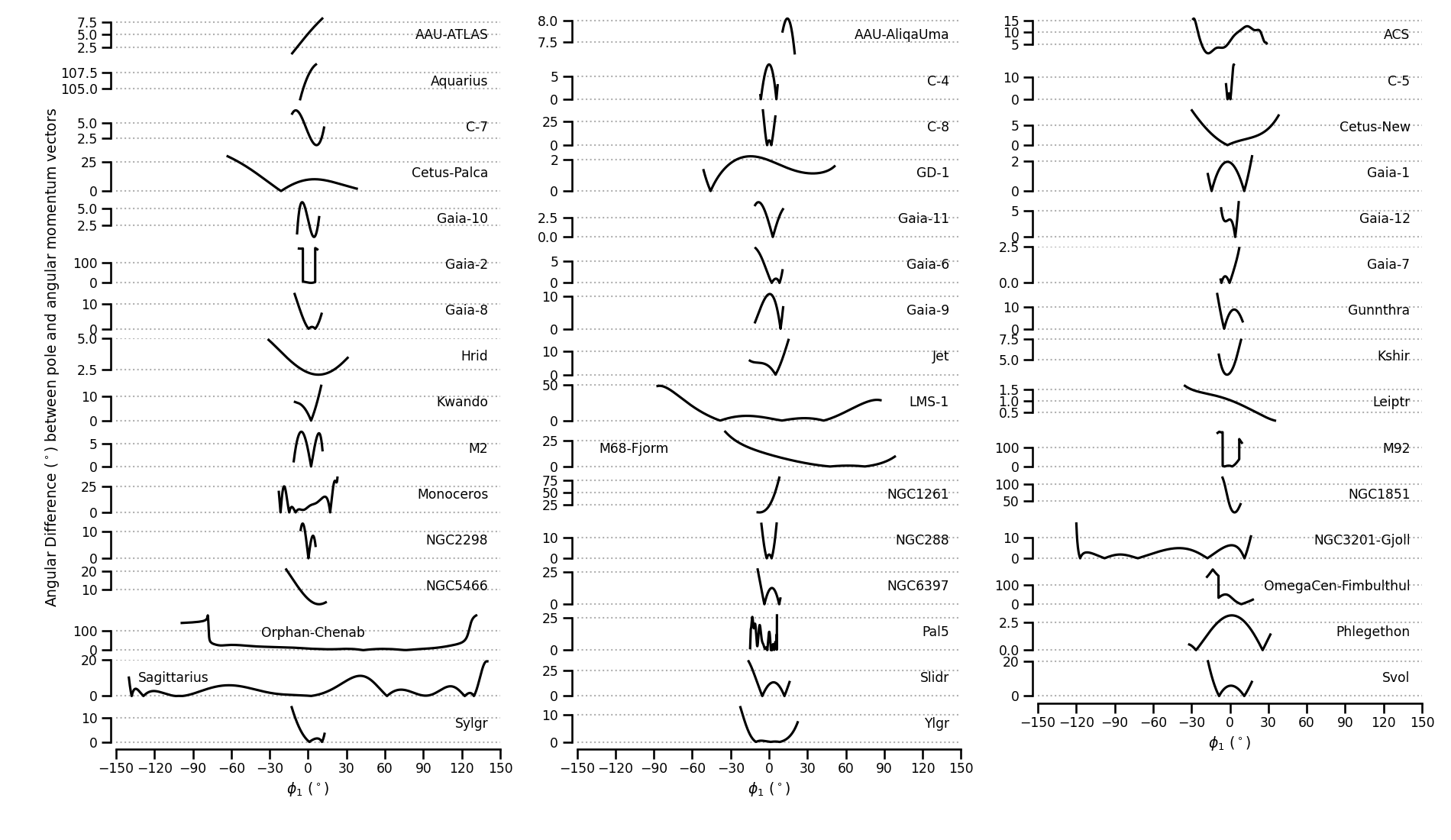}
    \caption{Angular separation (in degrees) between the angular momentum and pole vector for each stream as a function of the along-stream coordinate $\phi_1$, for the streams with available distance gradient and proper motion data. Note the different scale in the y-axis for the different panels. Proper motion misalignment shown in Figure~\ref{f:full_lib_missalign_track} corresponds to non-zero separations in these plots. }
    \label{f:full_lib_L_pole_sep_tracks}
\end{figure*}

\begin{figure*}
	\includegraphics[width=2.\columnwidth]{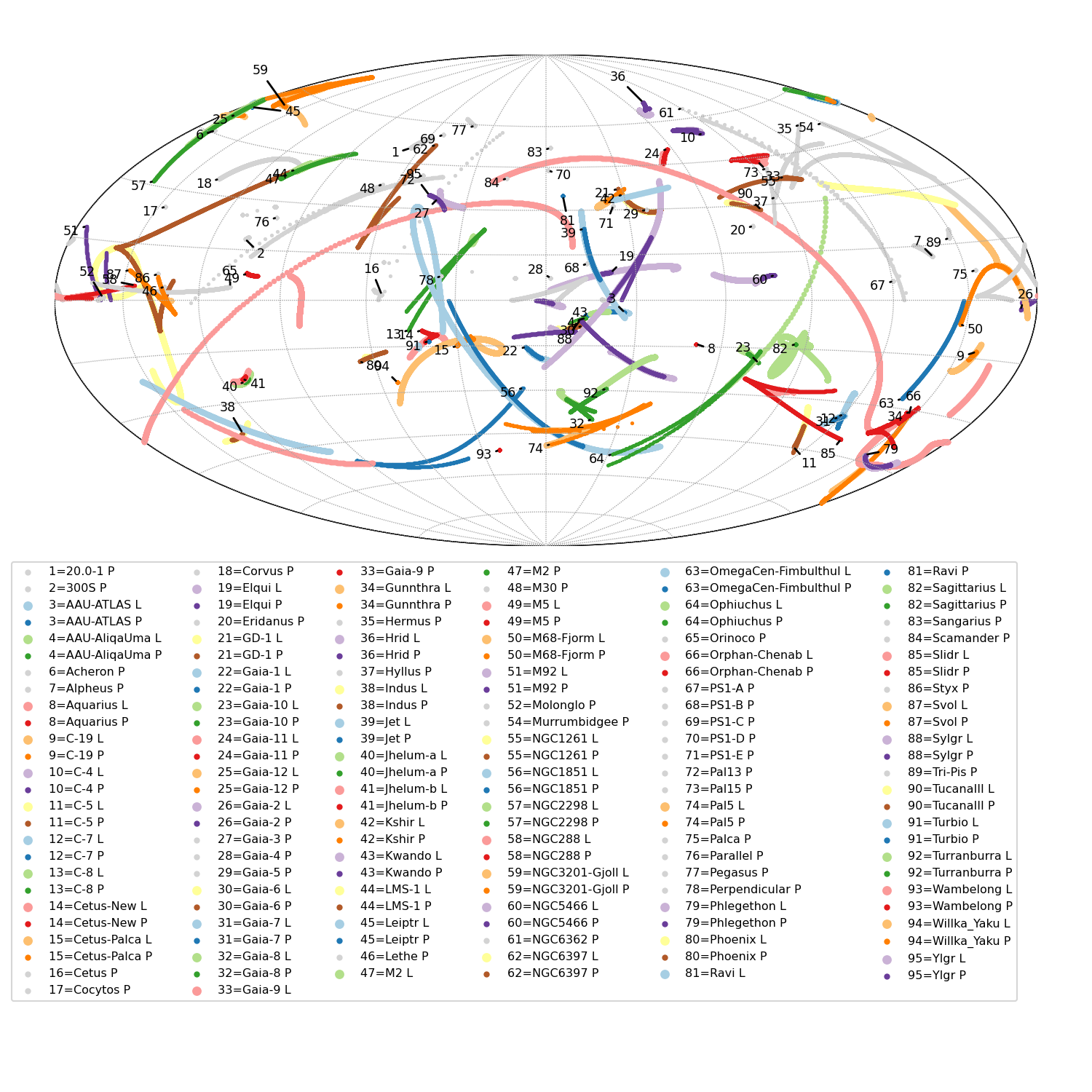}
    \caption{Heliocentric pole (P, dark colour) and angular momentum (L, light colour) tracks for the streams in the library, in a Mollweide projection in Galactic coordinates. Only the pole tracks are shown for the streams for which there is no proper motion data available (grey).}
    \label{f:full_lib_L_pole_tracks_map}
\end{figure*}

This Appendix includes additional library visualisations. Figure~\ref{f:full_lib_L_pole_sep_tracks} shows the angular separation between the angular momentum (heliocentric) and pole vectors for each stellar stream in the library, as discussed in Sec.~\ref{s:pm_misalignment}. A zero separation indicates the track's pole coincides with it's angular momentum, indicative of an unperturbed stream. 
Figure~\ref{f:full_lib_L_pole_tracks_map} shows the individual pole and angular momentum tracks in a map in Galactocentric coordinates, the point-by-point angular difference between the two, along the track, is what is represented in Figure~\ref{f:full_lib_L_pole_sep_tracks}.

%%%%%%%%%%%%%%%%%%%%%%%%%%%%%%%%%%%%%%%%%%%%%%%%%%

% Don't change these lines
\bsp	% typesetting comment
\label{lastpage}
\end{document}